\documentclass[journal,onecolumn]{IEEEtran}
\usepackage{cite}
\ifCLASSINFOpdf
\usepackage[pdftex]{graphicx}
\DeclareGraphicsExtensions{.pdf,.jpeg,.png}
\else
\fi
\usepackage[cmex10]{amsmath}
\usepackage{amsfonts,amssymb}
\usepackage{array}
\usepackage{url}
\usepackage[utf8]{inputenc}  
\usepackage{citesort}
\usepackage[T1]{fontenc} 
\usepackage{mathtools}
\usepackage{amsfonts}
\usepackage{algpseudocode} 

\def \({\left(}
\def \){\right)}
\def \[{\left[}
\def \]{\right]}
\def \√ï{\'}
\newcommand{\defeq}{\vcentcolon=}
\newcommand{\eqdef}{=\vcentcolon}

\newcommand{\bq}{{\textbf {q}}}
\newcommand{\bm}{{\textbf {m}}}
\newcommand{\br}{{\textbf {r}}}
\newcommand{\bF}{{\textbf {F}}}
\newcommand{\bff}{{\textbf {f}}}
\newcommand{\bG}{{\textbf {G}}}

\newcommand{\bh}{{\textbf {h}}}
\newcommand{\bw}{{\textbf {w}}}
\newcommand{\bv}{{\textbf {v}}}
\newcommand{\bA}{{\textbf {A}}}
\newcommand{\bB}{{\textbf {B}}}
\newcommand{\bx}{{\textbf {x}}}
\newcommand{\bhx}{\hat {\textbf {x}}}

\newcommand{\btau}{{\boldsymbol{\tau}}}
\newcommand{\by}{{\textbf {y}}}
\newcommand{\bz}{{\textbf {z}}}

\newcommand{\bR}{{\textbf {R}}}
\newcommand{\beps}{{\boldsymbol {\epsilon}}}
\newcommand{\bxigma}{{\boldsymbol{\Sigma}}}
\newcommand{\bxi}{{\boldsymbol{\xi}}}
\newcommand{\bzero}{{\boldsymbol{0}}}
\newcommand{\ba}{{\textbf {a}}}

\graphicspath{{plots/}}
\begin{document}
\title{Approximate Message-Passing Decoder and Capacity Achieving Sparse Superposition Codes}
\author{\IEEEauthorblockN{Jean Barbier$^\dag$ and Florent Krzakala$^{*}$}\\
  \IEEEauthorblockA{$\dag$ Laboratoire de Théorie des Communications, Facult\'e Informatique et Communications,\\
  Ecole Polytechnique F\'ed\'erale de Lausanne, Suisse. \\
  $*$ Laboratoire de Physique Statistique, UMR 8550 CNRS \& UPMC, 
  Ecole Normale Supérieure\\ \& Universit\'e Pierre et Marie Curie, Sorbonne Universit\'es, Paris, France.\\     
    jean.barbier@epfl.ch, florent.krzakala@ens.fr} }
\markboth{Approximate message-passing decoder for sparse superposition
  codes}%
{Approximate message-passing decoder for sparse superposition
  codes}
\maketitle
\IEEEpeerreviewmaketitle
\begin{abstract}
  We study the approximate message-passing decoder for sparse
  superposition coding on the additive white Gaussian noise channel
  and extend our preliminary work \cite{barbier2014superposition}. We
  use heuristic statistical-physics-based tools such as the cavity and
  the replica methods for the statistical analysis of the scheme. While superposition codes asymptotically reach
  the Shannon capacity, we show that our iterative decoder is limited
  by a phase transition similar to the one that happens in Low
  Density Parity check codes. We consider two solutions to this
  problem, that both allow to reach the Shannon capacity: $i)$ a power
  allocation strategy and $ii)$ the use of spatial coupling, a novelty
  for these codes that appears to be promising. We present in
  particular simulations suggesting that spatial coupling is more
  robust and allows for better reconstruction at finite code
  lengths. Finally, we show empirically that the use of a fast
  Hadamard-based operator allows for an efficient reconstruction, both
  in terms of computational time and memory, and the ability to deal
  with very large messages.
\end{abstract}
\IEEEpeerreviewmaketitle
\tableofcontents
\section{Introduction}
Sparse superposition codes have originally been introduced and studied
by Barron and Joseph in
\cite{barron2010sparse,barron2011analysis,joseph2012least}. They
proved the scheme to be capacity achieving for error correction over
the additive white Gaussian noise (AWGN) channel when power allocation
and (intractable) maximum-a-posteriori (MAP) decoding are used. In
\cite{barron2010sparse,barron2011analysis,joseph2012least}, a low-complexity iterative decoder called {\it adaptive successive decoder} was
presented, which was later improved in
\cite{barron2012high,choapproximate} by soft thresholding methods. The
idea is to decode a sparse vector with a special block structure over
the AWGN channel, represented in Fig.~\ref{fig_AWGN}. Using these
decoders together with a wise use of power allocation, this tractable scheme was
proved to be capacity achieving. However, both asymptotic and finite blocklength performances were far from ideal. In fact, it seemed that the asymptotic results were poor for any reasonable input alphabet (or \emph{section}) size, a fundamental parameter of the code. Furthermore, these asymptotic results could not be reproduced at any reasonable finite blocklengths.

We have proposed the Approximate Message-Passing (AMP) decoder
associated with sparse superposition codes in
\cite{barbier2014superposition}. This decoder was shown to have much
better performances. In fact it allows better decoding performance for
reasonable finite blocklengths than the asymptotic results of
\cite{barron2010sparse,barron2011analysis,joseph2012least}, and this
{\it even} without power allocation. The goal of the present
contribution is to complete and extend the short presentation in
\cite{barbier2014superposition}. In particular, we present two
modifications of sparse superposition codes that allow AMP to be
asymptotically capacity achieving as well, while retaining good finite
blocklength properties. The first strategy, new in the context of AMP
decoding but already known for sparse superposition codes
\cite{barron2010sparse}, is the use of power allocation. Without this,
the scheme of Barron and Joseph is not able to reach the Shannon capacity and
it appears that the same is true for the AMP decoder when used with homogeneous coding matrices. The second one, a novelty in the
context of sparse superposition codes, is the use of spatial coupling
which we find even more promising. We also present extensive numerical
simulations and a study of a practical scheme using Hadamard-based
operators. The overall scheme allows to practically reach near-to-capacity rates.
\subsection{Related works}
The phenomenology of these codes under AMP decoding, in particular the
sharp phase transitions different between MAP and AMP decoding, has
many similarities with what appears in low density parity check codes
(LDPC) \cite{RichardsonUrbanke08}. It is actually in the context of
LDPC codes that spatial coupling has been introduced
\cite{KudekarRichardson12,kudekar2011threshold} in order to deal with
this phase transition phenomenon that blocks the convergence of low-complexity message-passing based decoders. These similarities are not a priori
trivial because LDPC codes are codes over finite fields, the sparse
superposition codes work in the continuous framework. Furthermore LDPC codes are decoded by loopy belief-propagation (BP) whereas sparse
superposition codes are decoded by AMP which is a Gaussian
approximation of loopy BP. However, they arise due to a deep
connection to compressed sensing, where these phenomena (phase
transition, spatial coupling, etc) have been studied as well
\cite{KudekarPfister10,KrzakalaPRX2012,KrzakalaMezard12,DonohoJavanmard11}
and we shall make use of this connection extensively.

The AMP algorithm, which stands at the roots of our approach, is a
simple relaxation of loopy BP. While the principle behind AMP has been
used for a long time in physics under the name of
Thouless-Anderson-Palmer equations \cite{ThoulessAnderson77}, the
present form has been originally derived for compressed sensing
\cite{DonohoMaleki10,Rangan10b,montanari2012graphical} and is
naturally applied to sparse superposition codes as this scheme can be
interpreted as a compressed sensing problem with structured
sparsity. The state evolution technique \cite{BayatiMontanari10} is
unfortunately not yet fully rigorous for the present AMP approach, due
to the structured sparsity of the signal, but in spite of that, we
conjecture that it is exact.

Note that reconstruction of structured signals is a new trend in
compressed sensing theory that aims at going beyond simple sparsity by
introducing more complex structures in the vector that is to be
reconstructed. Other examples include group sparsity or tree structure
in the wavelet coefficients in image reconstruction
\cite{modelBasedCSBaraniuk2008,SomPotter2020}.
%
%
%
%FK: Added ref on recent relevant works
Finally, we report that upon completion of this manuscript, we became
aware of the very recent work of Rush, Greig and Venkataramanan
\cite{rush2015capacity} who also studied AMP decoding in superposition
codes using power allocation. Using the same techniques as in
\cite{BayatiMontanari10}, they proved rigorously that AMP was
capacity achieving if a proper power allocation is used by extending the state evolution analysis to power allocated sparse superposition codes~\cite{rush2015capacity} (which further supports our conjecture that state evolution indeed tracks AMP in general for sparse superposition codes, power allocated or not). This
strengthen the claim that AMP is the tool of choice for the present
problem. We will see, however, that spatial coupling leads to even
better results both asymptotically and at finite size. Another recent result giving
credibility to our physics-inspired approach is the rigorous
demonstration of the validity of the replica approach for compressed sensing \cite{reeves2016replica,barbier2016mutual}.
\subsection{Main contributions of the present study}
The main original results of the present study are listed below. In
particular, we shall also extend and give a detailed presentation of
the previous short publications by the authors
\cite{barbier2014superposition,barbier2013compressed}.
\begin{itemize}
\item A detailed derivation of the AMP decoder for sparse
  superposition codes, which was first presented in
  \cite{barbier2014superposition}, and this for a generic power
  allocation. The derivation is self-contained and starts all the way
  from the canonical loopy BP equations.
\item An analysis of the performance of the AMP decoder using the state
  evolution analysis, again presented without derivation in
  \cite{barbier2014superposition}. Here this is done in full
  generality with and without power allocation, and with and without
  spatial coupling. Note that, while we do not attempt to be mathematically
  rigorous in this contribution, the state evolution approach has been
  shown to be rigorously exact for many similar estimation problems
  \cite{BayatiMontanari10,javanmard2013state}. The present approach does not 
  verify the hypothesis required for the proofs to be valid because of the structured sparsity of the signal, but nevertheless
  we conjecture that the analysis remains exact. It is shown in
  particular that AMP, for sparse superposition codes without power allocation, suffers from a
  phenomenon similar to what happens with LDPC codes decoded with BP: there is a sharp
  transition ---different from the optimum one of the code itself---
  beyond which its performance suddenly decays.
\item An analysis of the optimum performance of sparse superposition
  codes using the non-rigorous replica method, a powerful heuristic
  tool from statistical
  physics\cite{MezardParisi87b,MezardMontanari09}. This leads in
  particular to a single-letter formulation of the minimum mean-square-error (MMSE) which we conjecture to be exact. The connection and
  consistency with the results obtained from the state evolution
  approach is also underlined. Again, this was only partially
  presented in \cite{barbier2014superposition}.
\item We present an analysis of the large section limit (partial results were only stated in \cite{barbier2014superposition}) for the
  behavior of AMP, and compute its asymptotic rate, the so-called BP threshold $R_{\rm BP}<C$ where $C$ is the Shannon capacity of the channel. As a
  by-product, we reconfirm, using the replica method, that these codes
  are Shannon capacity achieving.
\item We also show that, with a proper power allocation, the BP threshold that was blocking the AMP decoder
  disappears so that AMP becomes capacity achieving over the
  AWGN in a proper asymptotic limit.
\item Building on the connection with compressed sensing in
  \cite{KrzakalaPRX2012,KrzakalaMezard12,DonohoJavanmard11} we also
  show that the use of spatial coupling \cite{KudekarRichardson12} for
  sparse superposition codes is an alternative way to obtain capacity
  achieving performances with AMP.
\item We also present an extensive numerical study at finite
  blocklength, showing that despite improvements of the scheme thanks
  to power allocation, a properly designed spatially coupled coding
  matrix seems to allow better performances and robustness to noise for decoding over finite size messages.
\item We also discuss a more practical scheme where the i.i.d Gaussian random coding operators of the sparse superposition codes are replaced by fast
  operators based on an Hadamard construction. We show that this allows
  a close to linear time decoder able to deal with very large
  message lengths, yet performing very well at large rate for finite-length
  messages. These results were only hinted at in
  \cite{barbier2013compressed}. We study the efficiency of these
  operators combined with sparse superposition codes, with or without
  spatial coupling.
\end{itemize}

Finally, we note that our work differs from the mainstream of the existing literature. While a large part of the coding theory literature provides theorems, part of our work ---that using the
replica method--- is based on statistical physics methods that are
conjectured to give exact results. While many results obtained with
these methods on a variety of problems have indeed been proven later
on, a general proof that these methods are rigorous is not yet known. Note, however, that the state evolution technique (also called
cavity method in statistical physics) has been turned into a rigorous
tool under control in many similar cases
\cite{BayatiMontanari10,javanmard2013state}, though not yet in the
vectorial case discussed in the present contribution. We thus expect
that both the replica analysis and the state evolution results are
exact and believe it is only a matter of time before they are fully proven as already done for compressed sensing \cite{reeves2016replica,barbier2016mutual}.
\subsection{Outline}
The present paper is constructed as follows. We start by introducing
the setting of sparse superposition codes in sec.\ref{sec:Suc}. The
third section is dedicated to the AMP decoder for sparse superposition
codes and the fast Hadamard spatially coupled operator
construction. Sec.~\ref{sec:SEseeded} gives the state evolution recursions
for AMP in the simplest setting of constant power allocation case
without spatial coupling. Then follow the results of the state evolution analysis in the spatially coupled
case. We mention that this analysis is actually also valid for codes
with non constant power allocation. Some numerical experiments are
performed to confirm the approximate validity of the state evolution
analysis for predicting the behavior of the decoder with Hadamard
based operators. Sec.~\ref{sec:replica} presents the main results
extracted from the replica analysis, that provides a potential
function containing the information on the optimality of the scheme
and the transitions that can block the convergence of the decoder. In
sec.\ref{sec:AMP-PowA}, we show that a proper power allocation makes sparse superposition codes capacity achieving without
spatial coupling. Finally, sec.\ref{sec:numerics} summarizes the
results of our numerical studies of the scheme, such as the efficiency
of spatial coupling with Hadamard-based operators, and the comparisons
between power allocation and spatial coupling strategies. Finally the
last section concludes and gives some interesting open questions from
our point of view, followed by the acknowledgments and related
references.

In appendix~\ref{app:BPtoAMP}, a step-by-step derivation of the AMP
decoder for sparse superposition codes starting from the BP algorithm
is presented. Furthermore, we show how one can recover the AMP in its
original form \cite{montanari2012graphical}. Appendix~\ref{app:SE}
presents a detailed derivation of the state evolution analysis with or
without spatial coupling. The derivation is performed starting from
the AMP algorithm. Appendix~\ref{app:replicaAnalysis} details the full
derivation of the replica analysis.
\section{Sparse superposition codes}
\label{sec:Suc}
\begin{figure}[t]
\centering
\includegraphics[width=8cm]{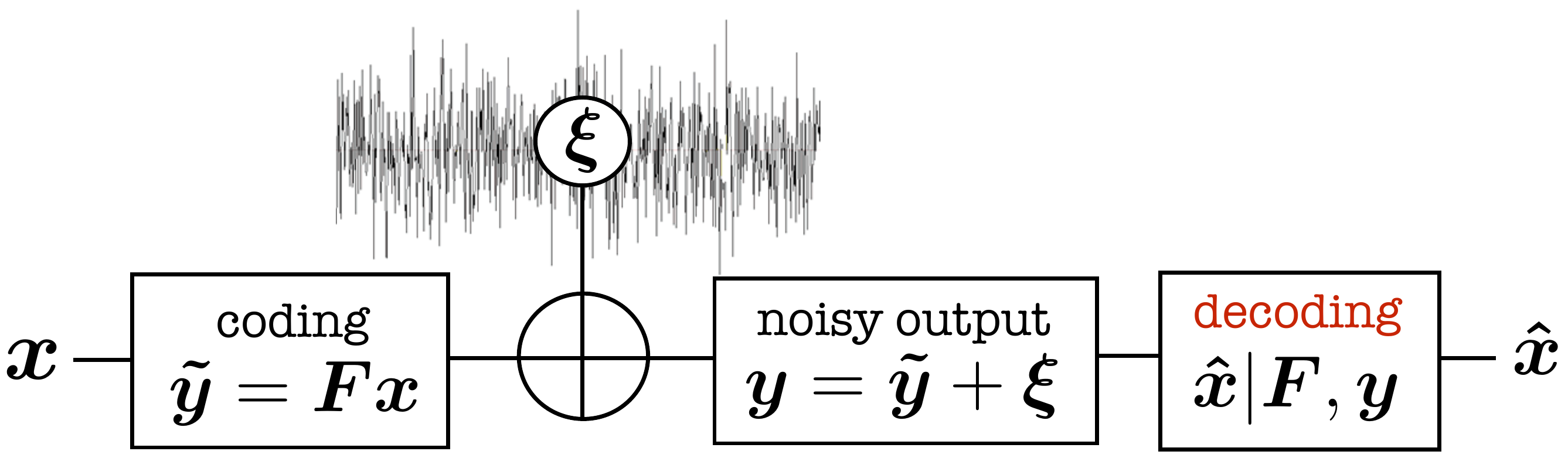}
\caption{Sending information through the AWGN channel with
  superposition codes. The message $\bx$, created such that it has a single non-zero component in each of its $L$ sections, is 
  coded by a linear transform from which we obtain the codeword $\tilde \by = \bF\bx$. The codeword is then sent
  through the AWGN channel that adds an i.i.d Gaussian noise $\bxi$
  with zero mean and a given variance $\sigma^2$ to each components. The
  receiver gets a corrupted version of the codeword, that is $\by$, and must estimate
  $\hat \bx$ as close as possible from $\bx$ from the knowledge of
  $\bF$ and $\by$. Perfect decoding happens if $\hat \bx = \bx$.}
\label{fig_AWGN}
\end{figure}
All the vectors will be denoted with bold symbols,
the matrices with capital bold symbols. Any sum or product index
starts from 1 if not specified. The notation
$x \sim P(x|\boldsymbol \theta)$ means that $x$ is a random variable with
distribution $P(x|\boldsymbol \theta)$ that can depend on some
hyperparameters $\boldsymbol \theta$. $\mathcal{N}(x|u, \sigma^2)$ is a Gaussian probability density with mean $u$ and variance $\sigma^2$.
\subsection{Sparse superposition codes: setting}
Suppose you want to send through an AWGN channel a generic message ${\tilde \bx} \defeq \{\tilde x_l : \tilde x_l \in \{1,\ldots,B\} \ \forall \ l\in \{1,\ldots,L\}\}$ made of $L$ symbols, each symbol belonging to an alphabet of $B$ letters. Starting from a standard binary representation of $\tilde \bx$, it is of course trivial to encode it in this form.

An alternative and highly {\it sparse} representation is given by the
sparse superposition codes scheme. In this scheme, the equivalent representation
$\bx$ of this message ${\tilde \bx}$ is made of $L$ sections of size
$B$, where in each section a {\it unique} component is $\not = 0$ at the location corresponding to the original symbol. We consider this
value positive as it can be interpreted as an input energy in the
channel, or power. The amplitude of the positive values, that can
depend on the section index, is given by the {\it power
  allocation}. If the $i^{th}$ component of the original message
$\tilde \bx$ is the $k^{th}$ symbol of the alphabet, then the $i^{th}$
section of $\bx$ contains only zeros except at the position $k$, where
there is a positive value.

Let us give an example in the simplest setting where the power
  allocation is constant, i.e $c_l = 1 \ \forall \ l\in\{1,\ldots,L\}$ (where $c_l$ is
the positive constant appearing in the $l^{th}$ section). If
${\tilde \bx} = [a,c,b,a]$ where the alphabet has only three symbols
$\{a,b,c\}$, i.e $L=4, B=3$ then ${\bx}=[[100],[001],[010],[100]]$ is a valid message for sparse superposition codes, where $[\ ]$ is the concatenation operator from which we obtain a vector. The $l^{th}$ section of
$\bx$ will be denoted $\bx_l \defeq [x_i : i \in l]$ where by some abuse of notation, we denote with $i \in l$ the set of components of the message $\bx$ that compose the $l^{th}$ section.

In sparse superposition codes, $\bx$ is then encoded through a linear
transform by application of an operator $\bF$ of dimension $M\times N$ (with the total number of scalar components of $\bx$ being $N=LB$) to obtain a codeword $\tilde \by \defeq  \bF \bx\in \mathbb{R}^M$. Borrowing vocabulary of compressed sensing, the ``measurement ratio'' is $\alpha\defeq M/N$. This codeword is then sent through an AWGN channel. This is summarized in Fig.~\ref{fig_AWGN}. The dimension of the operator is linked to the size $B$ of a
section and the coding rate in bits per-channel use $R$. Defining $K:=\log_2(B^L)$ as the number of information bits carried by the signal $\bx$ made of $L$ sections of size $B$, we have
\begin{align}
R&\defeq \frac{K}{M}=\frac{L\log_2(B)}{\alpha N} =\frac{\log_2(B)}{\alpha B} \Leftrightarrow \alpha \defeq \frac{M}{N}= \frac{\log_2(B)}{RB}. \label{eq_alpha}
\end{align}

Note from this relation that at fixed communication rate $R$, the codeword blocklenght $M=L\log_2(B)/R$ is proportional to the original message lenght $L$ up to a \emph{logarithmic factor in} $B$, so despite the message $\bx$ might be highly sparse when increasing $B$, this does not have a strong computational or memory cost.

In what follows we will concentrate on coding operators with independent and identically distributed (i.i.d) Gaussian entries of $0$ mean and variance fixed by a proper power constraint on the codeword. This choice is made in order to obtain analytical results. We fix the total
power sent through the channel
$P \defeq ||\tilde \by||_2^2/M = \sum_{\mu}^M \tilde y_\mu^2/M$ to
$P=1$. This is done in practice using a proper rescaling of the variance of the entries of
$\bF$. The only relevant parameter is thus the signal-to-noise ratio ${{\rm snr}}\defeq P/{\sigma^2}=1/{\sigma^2}$, where $\sigma^2$ is the variance of the AWGN in the channel. According to the celebrated Shannon formula \cite{shannon48}, the
capacity of the power constrained AWGN channel is $C=\log_2(1 + {\rm snr})/2$.
\subsection{Bayesian estimation and the decoding task}
The codeword $\tilde \by$ is sent through the AWGN channel
which outputs a corrupted version $\by$ to the receiver, see  Fig.~\ref{fig_AWGN}. Thus the linear model of interest is simply
\begin{equation} \label{eq_yfx}
\by = \tilde{\by} + \bxi = \bF \bx + \bxi \quad \Leftrightarrow \quad y_\mu = \sum_{l}^L \bF_{\mu l}^\intercal \bx_l + \xi_\mu,
\end{equation}
with $\xi_\mu \sim \mathcal{N}(\xi_\mu|0, \sigma^2) \ \forall\  \mu \in \{1,\ldots,M\}$. 

We place ourselves in a Bayesian setting and, in order to perform estimation of the message, we associate a posterior probability $P(\hat \bx|\by)$ to the signal estimate given the corrupted codeword. The memoryless AWGN of ${\rm snr}$ is modeled by the likelihood
\begin{equation}
P(\by|\hat \bx) = \prod_{\mu}^M \bigg[\sqrt{\frac{{{\rm snr}}}{2\pi}} e^{-\frac{{{\rm snr}}}{2}(y_\mu - \sum_{l}^L \bF_{\mu l}^\intercal\hat\bx_l)^2}\bigg].
\end{equation}
For the rest of the paper, we consider that the true ${{\rm snr}}$ is accessible to the
channel users, and is used by the decoder. Then the posterior distribution given by the Bayes formula is 
\begin{align}
P(\hat\bx|\by) &= \frac{P_0(\hat \bx) P(\by|\hat\bx)}{\int d\hat \bx P_0(\hat \bx) P(\by|\hat\bx)} = \frac{P_0(\hat \bx) P(\by|\hat\bx)}{P(\by)}, \label{eq:posterior}\\
P(\by)&= Z(\bF,\bxi,\bx) = \int \bigg[\prod_{l}^L d\hat \bx_l P_0^l(\hat\bx_l)\bigg] \prod_{\mu}^M \bigg[\sqrt{\frac{{{\rm snr}}}{2\pi}} e^{-\frac{{{\rm snr}}}{2}(\sum_{l}^L \bF_{\mu l}^\intercal(\bx_l-\hat\bx_l) + \xi_\mu)^2}\bigg], \label{eq_fullZ}
\end{align}
where $\by$ depends on the \emph{quenched random variables} (or disorder) $\bF,\bxi,\bx$ through the linear model \eqref{eq_yfx}. The codeword distribution, or \emph{partition function} noted $Z$ that we wrote explicitely as a function of the quenched disorder, plays the role of a normalization. The proper prior for sparse superposition codes, that enforces each section to have only one value $c_l>0$ per section, is in the present continuous framework given by $P_0(\hat\bx) = \prod_{l}^L P_0^l(\hat\bx_l)$ with
\begin{align}
P_0^l(\hat\bx_l) &\defeq \frac{1}{B} \sum_{i \in l}^B \delta(\hat x_i-c_l)\prod_{j \in l : j \neq i}^{B-1}\delta(\hat x_j) \label{eq_prior}.
\end{align}
It is designed so that it gives uniform weight, for the section $l$, to any permutation of the $B$-d vector $[c_l,0,\ldots,0]$ (most of the paper will focus on constant power allocation $c_l=1 \ \forall \ l$). 

\begin{figure}[t]
\centering
\includegraphics[width=8cm]{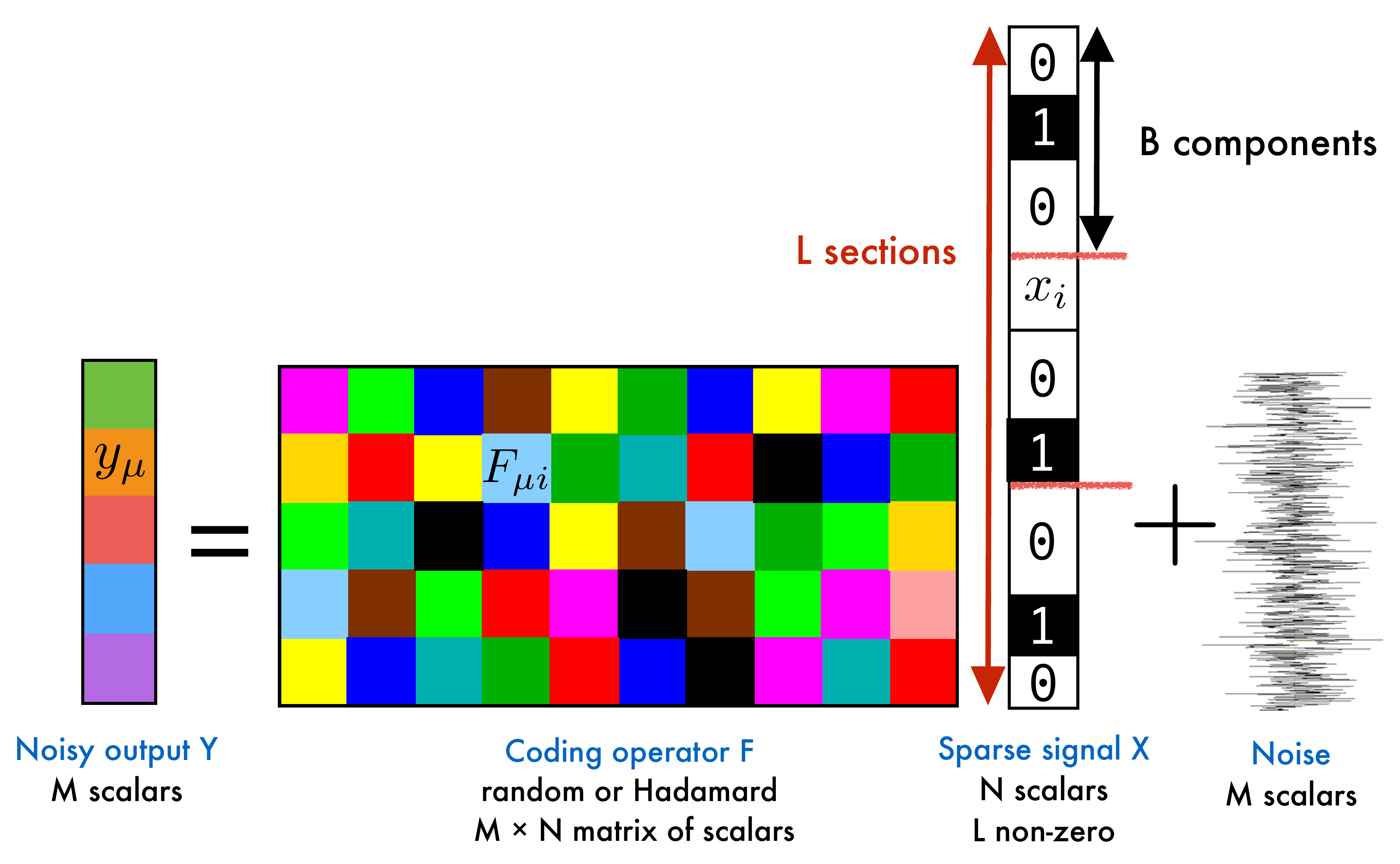}
\includegraphics[width=8cm]{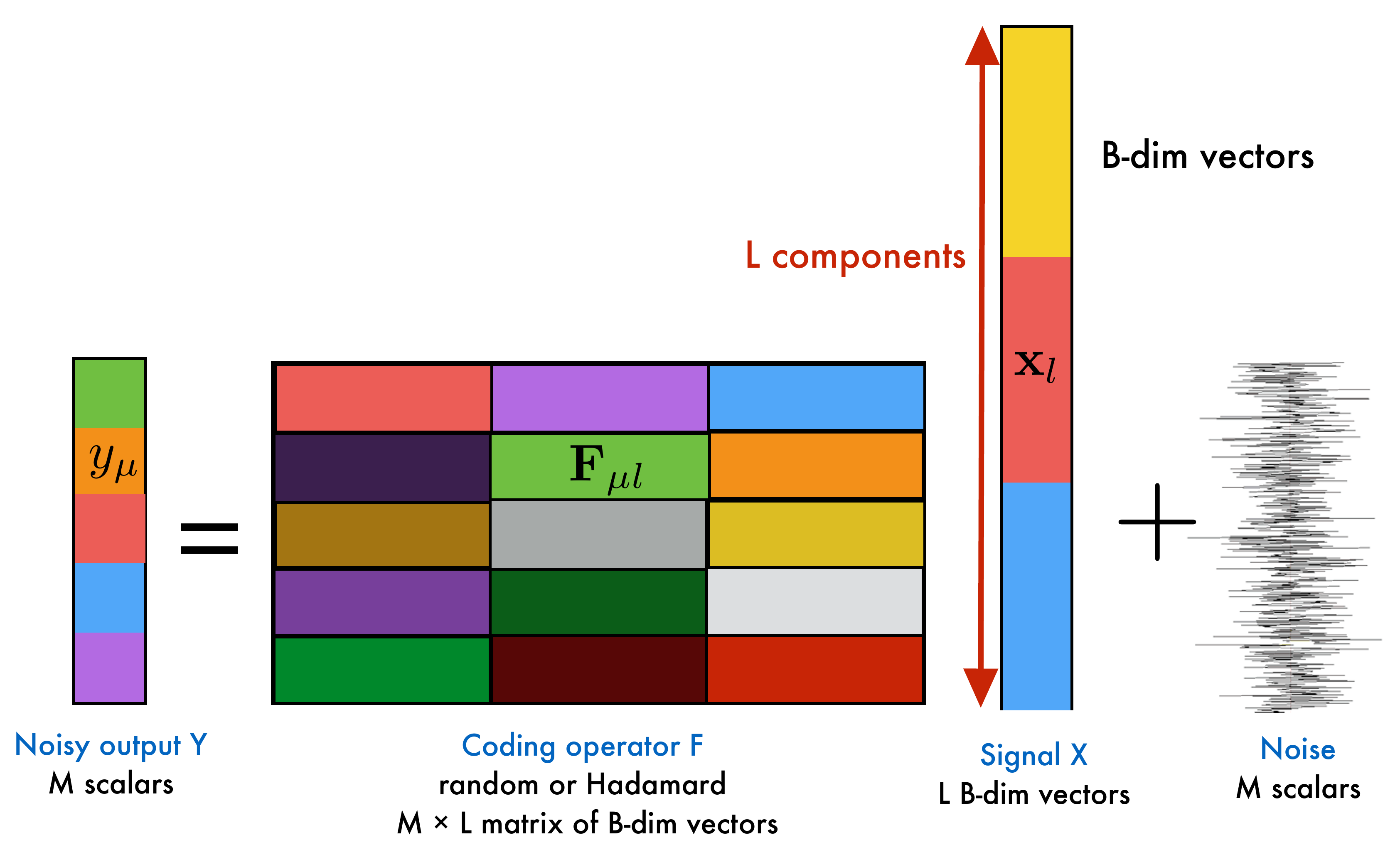}
\caption{$\textbf{Left}$: Representation of the estimation
  problem associated to the decoding of the sparse message over the
  AWGN channel. All variables in the same section
  $\{x_i : i \in l\}$ are strongly correlated due to the hard
  constraint that only one of them can be positive ($1$ in this
  example). The matrix entries are scalars as the message
  components. $\textbf{Right}$: Re-interpreting the same
  problem in terms of $B$-d variables. Now, the matrix elements of the
  previous figure are grouped to form $B$-d vectors that are applied
  (using the usual scalar product for vectors) on the associated $B$-d
  vectors representing the new components of the message. In this setting, all the message vectorial components are uncorrelated.}
\label{fig_1dOp}
\end{figure}

Let us turn now our attention to the decoding task, which we discuss
in Fig.~\ref{fig_1dOp}. It is essentially a sparse linear estimation
problem where we know $\by$ and need to estimate a sparse solution of
$\by = \bF \bx + \bxi$. However the problem is different from the canonical compressed
sensing problem \cite{Donoho:06} in that the components of $\bx$ are
correlated by the constraint that only a single component in each
section is non-zero.
We thus prefer to think of the problem as a multidimensional one,
as discussed in \cite{barbier2014superposition}. Each section
$l \in \{1,\ldots,L\}$ made of $B$ components in $\tilde \bx$ is interpreted
as {\it a single} $B$-dimensional ($B$-d) variable for which we have a
strong prior information: it is zero in all dimensions {\it but} one
where there is a fixed positive value. Given its length, we thus know
the vector must point in only one of the directions of the $B$-d hypercube. In this setting, instead of dealing with a $N$-d
vector with scalar components, we deal with a $L$-d vector
$\bx$ whose components $\{\bx_l\}$ are $B$-d vectors. We define
$\bF_{\mu l}\defeq [F_{\mu i} : i \in l]$ as the vector of
entries of the $\mu^{th}$ row of the matrix $\bF$ that act on $\bx_l$, see
Fig.~\ref{fig_1dOp}. 

The decoding task is thus exactly of the kind considered in the Bayesian approach to compressed sensing, see e.g. \cite{Rangan10b,montanari2012graphical,KrzakalaPRX2012,KrzakalaMezard12,BarbierKrzakalaAllerton2012}
and we can thus directly apply these techniques to the present
problem. Other analogies between compressed sensing and error correction over the AWGN exist 
in the litterature such as \cite{barbierRobustErrorCorrection13}. 

We are interested in two error estimators, namely the mean-square
error per section (${\rm MSE}$) $E$ and the section error rate ${\rm{SER}}$. They
are defined respectively as the ${\rm MSE}$ associated to the sections and the fraction of wrongly reconstructed sections
\begin{align}
&E = \frac{1}{L} \sum_{i}^N (x_i - \hat x_i)^2, \quad {\rm SER} = \frac{1}{L} \sum_{l}^L \mathbb{I}(\bx_l \neq \hat\bx_l), \label{eq_MSE} 
\end{align}
where $\mathbb{I}\left(A\right)$ is the indicator function of the
event $A$ which is one if $A$ occurs, zero else and
$\hat \bx\defeq [\hat \bx_1,\dots, \hat \bx_L]=[\hat x_i, \dots, \hat x_N]$ is the estimate of the signal obtained using the decoder. 
\section{The approximate message-passing decoder and spatial coupling}
\label{sec:AMP}
\subsection{Why belief-propagation is not an option for decoding with sparse superposition codes}
In sparse superposition codes, as the message $\bx$ has discrete
components, one could think about BP as a good decoder, that is a proper algorithm to sample the posterior \eqref{eq:posterior} and perform estimation from it. Indeed, it is
numerically easier to perform discrete sums than the numerical
integrations one would have to perform in the continuous setting where
the variables to infer are real numbers. Let us discuss why a direct
approach with BP is nevertheless intractable in the present
setting. We define
$\mathcal{S}_k \defeq \{[c_k, 0, \ldots, 0], [0,c_k, 0, \ldots,
0],\ldots, [0, \ldots, 0, c_k]\}$.
Thus $\mathcal{S}_k$ is the ensemble of the $B$ authorized sections
for position $k\in\{1,\ldots,L\}$ in the context of sparse superposition codes. Let's write the canonical BP
equations associated to the factor graph Fig.~\ref{fig_factorSC} for
the $B$-d variables in order to understand why it is not appropriate
here:
\begin{align}
&\hat m_{\mu l}(\hat \bx_l) = \frac{1}{\hat z_{\mu l} } \sum_{\{\hat\bx_k\in\mathcal{S}_k:k\not = l\}}^{B^{L-1}} e^{-\frac{{\rm snr}}{2}\left(\sum_{k\neq l}^{L-1} \bF_{\mu k}^\intercal \hat\bx_k + \bF_{\mu l}^\intercal \hat\bx_l - y_\mu\right)^2}\prod_{k\not = l}^{L-1} m_{k\mu}(\hat\bx_k),\label{eq:BP1}\\
&m_{l\mu}(\hat \bx_l)= \frac{1}{z_{l\mu} } \prod_{\nu \not = \mu}^{M-1}\hat m_{\nu l}(\hat\bx_l),\label{eq:BP2}
\end{align}
where the Greek letters are associated to the soft factors enforcing $\hat\bx$ to verify the system (\ref{eq_yfx}) up to some error. These factors, that take into account the deviation of the transmitted codeword due to the AWGN are Gaussian densities. The Roman letters correspond to the variable nodes, that is the sections to decode. 

The basic objects in this approach are the so-called cavity messages, that are the usual BP messages: $\{\hat m_{\mu l}(\hat \bx_l), m_{l\mu}(\hat \bx_l)\}$ is the set of factor-to-node and node-to-factor messages respectively. The messages associated to $\hat \bx_l$ are probability distributions from $\mathcal{S}_l\to[0,1]$, i.e the joint probability distribution of the components inside a given section, but in \emph{modified graphical models} with respect to Fig.~\ref{fig_factorSC}. Indeed, $\hat m_{\mu l}(\hat \bx_l)$ is the distribution of $\hat \bx_l$ in a graphical model where the variable node associated to $\hat \bx_l$ is only connected to the $\mu^{th}$ factor node. Instead $m_{l\mu}(\hat \bx_l)$ is its probability in a graph where $\hat \bx_l$ is connected to all the factor nodes except the $\mu^{th}$ one. These distributions can be computed iteratively in an exact way on a graphical model which is a tree, or approximately on a generic graph. In the latter case, the procedure is called loopy BP because of the loops present in a generic graph, which is therefore not a tree. 

The terminology of cavity messages, referring to the procedure of ``removing'' factors of the original graph when computing the messages, comes from the physics vocabulary. This is because the BP algorithm can be understood as the cavity method of statistical physics of disordered systems, an asymptotic statistical analysis originally developed in the context of spin-glasses \cite{MezardParisi87b,MezardMontanari09}, but applied to single instances of a problem defined by a graphical model. The cavity method is referred to as the state evolution analysis in the present context, and more generally in the context of dense linear estimation such as compressed sensing. When dealing with codes with a low density coding matrix such as in LDPC codes, the method is called density evolution analysis.

The problem with loopy BP for sparse superposition codes is now clear: the sum that has to be performed is over an combinatorial number of terms, which comes from the fact that the underlying factor graph is densely connected. In addition, there are $2ML$ messages to deal with ($2$ per edge), which is way too many. It would become quickly intractable even for small signals. BP is efficient only when the factor graph defining the inference problem to be solved has a low average connectivity like in LDPC codes (was speak in this case of tree-like graphs).
\subsection{The approximate message-passing decoder for sparse superposition codes}
\begin{figure}[t!]
\centering
\includegraphics[width=4.2cm]{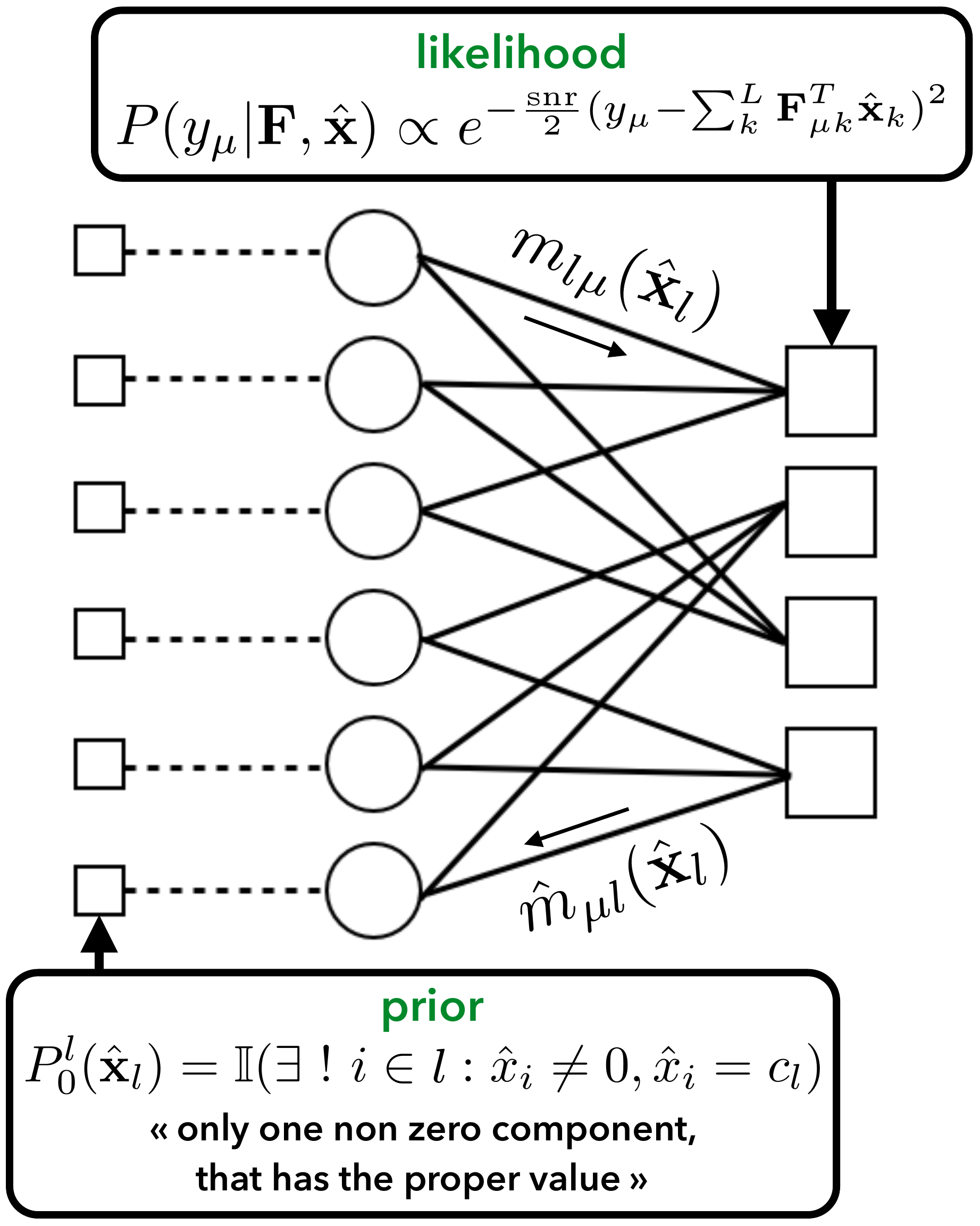}
\caption{Factor graph associated to sparse
  superposition codes, i.e to the posterior \eqref{eq:posterior}. It is a bipartite graph where the variables estimates $\{\hat{\bx}_l:l\in \{1,\ldots,L\}\}$ are
  represented by circles, the constraints (or factors) by squares. The
  variables are constrained by two kind of factors. The $M$ factors on the right side represent the soft
  constraints called the likelihood factors. These enforce the system $\by = \bF \hat\bx$ to be fulfilled up to some error controlled by the ${\rm snr}$, due to the precense of AWGN in the communication channel. On the left side are the prior hard constraints, enforcing each section to have only one non-zero value fixed by the power allocation. In the non spatially coupled operator case, the
  variable nodes are connected to all the likelihood factors
  and vice versa, whereas in the spatially coupled case, they are connected to a finite fraction of the factors which depends on the spatial coupling design. The factor-to-node $\hat m_{\mu l}(\hat\bx_l)$ and node-to-factor $m_{l \mu}(\hat\bx_l)$ BP messages are represented. The two messages should sit on the same edge as they depend on the same variable and factor but we put them on distinct edges for readibility purpose.}
\label{fig_factorSC}
\end{figure}
The purpose of AMP is to go beyond BP, in order to solve efficiently inference problems defined on dense graphs, which is here to compute the posterior marginal means $\{\ba_l : l\in\{1,\ldots,L\}\}$ of \eqref{eq:posterior} for each section
\begin{equation}\label{eq_aiTrue}
\ba_l = [a_i : i\in l] \quad \text{with}\quad a_i = \int d\hat \bx\, \hat x_i\, P(\hat \bx | \by).
\end{equation}
It is a message-passing algorithm originally derived in its modern form for compressed sensing \cite{BayatiMontanari10,KrzakalaMezard12,KrzakalaPRX2012,rangan2011generalized} where one writes the BP equations on a densely connected factor graph with linear constraints ((\ref{eq:BP1}), (\ref{eq:BP2}) for sparse superposition codes), see Fig.~\ref{fig_factorSC} for the factor graph associated with \eqref{eq:posterior}. One then expands them up to the second order in the interaction terms, as we will derive it in appendix~\ref{app:BPtoAMP}. This step gives what is sometimes referred as the Gaussian approximation of BP, or the relaxed-BP algorithm \cite{KrzakalaMezard12}. A second step is then required to lower the number of messages, from which AMP is obtained. AMP can also be seen as a special case of the non-parametric BP algorithm \cite{sudderth2010nonparametric} in the case where one takes only one Gaussian density per message in the parametrization, see~\cite{phdBarbier} for details on this point. 

In the AMP algorithm, even if the variables to infer are discrete, they are estimated using continuous representations. This is why when deriving AMP, we start from the BP equations (\ref{eq_bp1}), (\ref{eq_bp2}) written in the continuous framework. The marginals associated to the variables become densities. It is difficult to store these distributions if not properly parametrized. AMP being a second order Gaussian approximation of the original equations, these densities are Gaussian distributions simply parametrized by a mean and a variance. But other mean and variances of different quantities naturally appear in the derivation appendix~\ref{app:BPtoAMP}. 

The final AMP decoder is given in Fig.~\ref{algo_AMP}. Let us give some meaning to the various quantities appearing. $\bw^t$ is an estimation of the codeword $\by$ at iteration $t$, $\boldsymbol\Theta^t$ is its vector of associated variance per component (an estimation of how much AMP is ``confident'' in its estimate $\bw^t$). $\bR^t$ is the estimation of the message before the prior has been taken into account, which is thus the average with respect to the likelihood. $(\bxigma^t)^2$ are the associated variances. Finally $\textbf{a}^t$ and $\textbf{v}^t$ are the posterior estimate and variance of $\bx$, that takes into account all the information available at iteration $t$. As such, $\bv^t$ should vanish in the successful decoding case. These are obtained thanks to the so-called \emph{denoisers} $f_{a_i}$ and $f_{c_i}$, given respectively by \eqref{eq_meani}, \eqref{eq_vari}.

In dense linear estimation with AWGN, when the noise variance and the
prior are known (the prior is always known in the context of coding
theory), this algorithm is Bayes-optimal and asymptotically performs
minimum mean-square estimation as long as the communication rate is
below the so-called BP threshold (or transition) $R_{\rm BP}$, if it exists
\cite{KrzakalaMezard12}. The transition location depends on the noise
variance. This transition that prevents the algorithm to reach the
MMSE estimate is inherent to the problem: we believe that its presence
does not depend on the decoding algorithm. It can be, however,
overcomed by a slight change in the code either by power allocation or by the use of spatial coupling that we present now (we shall see in sec.\ref{sec:numerics} that the later solution seems to give better results in practice).
\begin{figure}[!t]
\centering
\begin{minipage}{.45\textwidth}
\centering
\begin{algorithmic}[1]
\State $t\gets 0$
\State $\delta \gets \epsilon + 1$
\While{$t<t_{\rm max} \ \textbf
{and} \ \delta>\epsilon$} 
\State $\Theta^{t+1}_\mu \gets \sum_{c}^{L_c}\tilde O_\mu(\textbf{v}_c^t)$
\State $w^{t+1}_\mu \gets \sum_{c}^{L_c}O_\mu(\textbf{a}_c^t) - \Theta^{t+1}_\mu\frac{y_\mu-w^t_\mu}{{1/{{\rm snr}}} + \Theta^t_\mu}$
\State $\Sigma^{t+1}_i \gets \left[\sum_{r}^{L_r}\tilde O_i\left([{1/{{\rm snr}}} + \boldsymbol{\Theta}_r^{t+1}]^{-1}\right)\right]^{-1/2}$
\State $R^{t+1}_i \gets a^t_i + (\Sigma^{t+1}_i)^2 \sum_{r}^{L_r} O_i\left(\frac{\textbf{y}_r - \textbf{w}^{t+1}_r}{{1/{{\rm snr}}} + \boldsymbol{\Theta}^{t+1}_r}\right)$
\State $v^{t+1}_i \gets f_{c_i}\left((\bxigma_{l_i}^{t+1})^2,\bR_{l_i}^{t+1}\right)$
\State $a^{t+1}_i \gets f_{a_i}\left((\bxigma_{l_i}^{t+1})^2,\bR_{l_i}^{t+1}\right)$
\State $t \gets t+1$
\State $\delta \gets 1/N\sum_i^N(a_i^t - a_i^{t-1})^2$
\EndWhile
\State \textbf{return} $\{a_i\}$
\end{algorithmic}
\end{minipage}
\begin{minipage}{.45\textwidth}
\centering
\begin{algorithmic}[1]
\State $t\gets 0$
\State $\delta \gets \epsilon + 1$
\While{$t<t_{\rm max} \ \textbf
{and} \ \delta>\epsilon$} 
\State $\Theta^{t+1}_\mu \gets \sum_{i}^{N}F_{\mu i}^2v_i^t$
\State $w^{t+1}_\mu \gets \sum_{i}^N F_{\mu i}a_i^t - \Theta^{t+1}_\mu\frac{y_\mu-w^t_\mu}{{1/{{\rm snr}}} + \Theta^t_\mu}$
\State $\Sigma^{t+1}_i \gets \left[\sum_{\mu}^{M}\frac{F_{\mu i}^2}{{1/{{\rm snr}}} + \Theta_\mu^{t+1}}\right]^{-1/2}$
\State $R^{t+1}_i \gets a^t_i + (\Sigma^{t+1}_i)^2 \sum_{\mu}^{M} F_{\mu i}\frac{y_\mu - w^{t+1}_\mu}{{1/{{\rm snr}}} + \Theta^{t+1}_\mu}$
\State $v^{t+1}_i \gets f_{c_i}\left((\bxigma_{l_i}^{t+1})^2,\bR_{l_i}^{t+1}\right)$
\State $a^{t+1}_i \gets f_{a_i}\left((\bxigma_{l_i}^{t+1})^2,\bR_{l_i}^{t+1}\right)$
\State $t \gets t+1$
\State $\delta \gets 1/N\sum_i^N(a_i^t - a_i^{t-1})^2$
\EndWhile
\State \textbf{return} $\{a_i\}$
\end{algorithmic}
\end{minipage}
\caption{The AMP decoder for sparse superposition codes written in two different equivalent forms. The functions $f_{a_i}$ and $f_{c_i}$ refered as the \emph{denoisers} are given by \eqref{eq_meani}, \eqref{eq_vari} respectively. The first form underlines how AMP is operating when spatially coupled operators are used instead of matrices and takes advantage from this structure. The second form is more easy to read and explicits the operations done by the operators \eqref{eq_fastOpDefs2} in the first form. This form can be less efficient than the first if many blocks are only zeros as it is the case with spatially coupled operators due to their sparsity. $l_i$ is the index of the section to which the $i^{th}$ $1$-d variable belongs to. $\epsilon$ is the accuracy for convergence and $t_{\rm max}$ the maximum number of iterations. A suitable initialization for the quantities is ($a_i^{t=0}=0$, $v_i^{t=0}=\rho \sigma^2 $, $w_\mu^{t=0}=y_\mu$). Once the algorithm has converged, i.e the quantities do not change anymore from iteration to iteration, the estimate $\hat{\bx}_l^t$ of the $l^{th}$ section at iteration $t$ is the projection of the AMP estimate of the posterior marginal means \eqref{eq_aiTrue} given by $\textbf{a}_l^t$ (that is made of real numbers) on the closest authorized section.}  
\label{algo_AMP}
\end{figure}
\subsection{Spatial coupling}
\label{susec:fastHad}
We now discuss how the phase transition encountered by message-passing decoding is overcomed using spatially coupled
codes. The term ``spatially coupled codes'' was first used in
\cite{kudekar2011threshold}, in the context of LDPC codes. Their aim
was to show that this ``coupling'' of graphs leads to a remarkable
change in the algorithmic performances and that ensembles of codes designed in
this way combine the property that they are capacity achieving under
low complexity decoding, with the practical advantages of sparse graph
codes: this is referred as \emph{threshold saturation}\footnote{Since the first version of this work, one of the authors have rigorously proven with his collaborators that this threshold saturation phenomenon indeed occurs for spatially coupled sparse superposition codes, and this whatever memoryless channel used for communication \cite{barbier2016proof,barbier2016threshold}.}. Spatially coupled codes require, however, a very specific underlying graph, or in our case, a very specific coding matrix.
Following these breakthrough, spatial coupling has been extensively
used in the compressed sensing setting as well
\cite{KudekarPfister10,KrzakalaPRX2012,KrzakalaMezard12,DonohoJavanmard11,DonohoMaleki10}. It
rigorously allows to reach the information theoretical bound in LDPC
\cite{kudekar2011threshold} and in compressed sensing in the random
i.i.d Gaussian measurement matrix case \cite{DonohoJavanmard11}. We thus naturally apply this technique here, using a properly designed coding
operator: the sparse superposition codes scheme, being a structured compressed sensing problem, spatial coupling is expected to work.

In a nutshell, spatially coupled coding (or sensing) matrices are simply (almost) random band-diagonal matrices (see Fig.~\ref{fig_seededHadamard} for the general strucuture). More precisely, they represent a one
dimensional chain of different systems that are ``spatially coupled''
across a finite window along the chain. A fundamental ingredient for spatial coupling to work is to introduce a \emph{seed} at the boundary: the matrix is designed such that the first system on the chain lives into the ``easy region'' of the
phase diagram, while the other ones stay in the ``hard region''. Consider first a collection of different, independent sub-systems,
where the first one has a low rate, so that a perfect decoding is
easy, while all the other ones have a large rate where the naive
decoder fail to reconstruct the message. In a spatially coupled
matrix, additional measurements are coupling all these systems in a
very specific way. Initially, we expect these couplings to be, at
first, neglectible, so that the variables corresponding to the first
system will be decoded, but not the other ones. As the algorithm is
further iterated, however, the coupling from the first sub-system will
help the algorithm to decode the second one, and so and so forth: this triggers a \emph{reconstruction wave} starting from the seed and propagating inwards the signal. This
is the basis of the construction in the LDPC case.  Alternatively, one
can also provide generic ``statistical physics-type'' argument on why these codes work (see
\cite{HassaniMacris10,hassani2012chains,KrzakalaPRX2012,caltagirone2014dynamics} for more on this subject)).

We study spatially coupled coding operators constructed as in
Fig.~\ref{fig_seededHadamard}, see the caption for the details. The
operator has a block structure, i.e it is decomposed in
$L_r \times L_c$ blocks, each of them being either only zeros or a
given sub-matrix. We focus on two particular constructions: the
sub-matrices are made of random selections of modes of an Hadamard
operator or are Gaussian i.i.d matrices. The Hadamard-based
construction is predominantly used in this study for computational and
memory efficiency purpose, and is presented in more details in the
next section. In both cases, the matrix elements are always rescaled
by some constant which enforces the power of the codeword to be one,
so that the ${\rm snr}$ is the only relevant channel parameter.

We shall not prove in this paper that these coding matrices allow to reach threshold saturation (i.e that they allow to reach capacity under low complexity message-passing decoding) and let the study of this theorem for further work\footnote{See the previous footnote.}, but we nevertheless conjecture that this is true. Indeed, we will show explicit examples where we construct codes that are going as close as needed to the desired threshold.

The structure of the coding operator induces a spatial structure in
the signal, which becomes the concatenation of sub-parts
$\bx=[\bx_1,\dots,\bx_{L_c}]$. One has to be careful to ensure that
these sub-parts remain large enough for the assumption behind AMP to
be valid, essentially $L\gg L_c$. Concurrently, however, the larger $L_c$,
the better it is to get closer to the optimal treshold. This is due to
the relation (\ref{eq_alphaRest}) between the communication rate and
the effective rate of the seed block $R_{{\rm seed}}$ and that of the
remaining ones $R_{{\rm rest}}$. Indeed, in the construction of
Fig.~\ref{fig_seededHadamard}, the link between the overall
measurement rate $\alpha$ defined in (\ref{eq_alpha}), that of the
seed (the first block on the left upper corner on
Fig.~\ref{fig_seededHadamard}) $\alpha_{{\rm{seed}}}$ and that of the
bulk $\alpha_{{\rm{rest}}}$ is
\begin{equation}
\alpha_{{\rm{rest}}}= \frac{\alpha L_c -\alpha_{{\rm{seed}}}}{L_r - 1} = \alpha\left(\frac{L_c - \beta_{{\rm{seed}}}}{L_r - 1}\right) \Leftrightarrow R = \frac{L_c R_{{\rm rest}} R_{{\rm seed}} }{(L_r-1)R_{{\rm seed}} + R_{{\rm rest}} } \underset{L_c, L_r \gg 1}{\longrightarrow} R_{{\rm rest}},
\label{eq_alphaRest}
\end{equation}
where $R_{{\rm rest}}$ can be asymptotically as large as the Bayes optimal rate. This optimal rate $R_{{\rm opt}}(B)$, defined precisely in sec.\ref{sec:replica}, is the highest rate until which the superposition codes allow to decode (up to an inherent error floor) the input message for a given section size $B$ under MAP (or equivalently MMSE) decoding. 

In practice, $\alpha$ is fixed by choosing the rate $R$ thanks to
(\ref{eq_alpha}) and
$\alpha_{{\rm{seed}}}\defeq \alpha\beta_{{\rm{seed}}}$ as well by
fixing $\beta_{{\rm{seed}}}> 1$. $\alpha_{{\rm{rest}}}$ is then deduced
from (\ref{eq_alphaRest}).  In the rest of the paper, we will define
the spatially coupled ensemble of coding operators by
$(L_c,L_r,w,\sqrt{J},R, \beta_{{\rm{seed}}})$ instead of
$(L_c,L_r,w,\sqrt{J},\alpha_{{\rm{seed}}},\alpha_{{\rm{rest}}})$.
\subsection{The fast Hadamard-based coding operator}
In order to get a practical decoder able to deal with very large messages, we combine the spatial coupling technique with the use of a structured Hadamard operator, i.e the standard fast Hadamard transform which is as efficient as the fast Fourier transform from the computational point of view. These operators have been empirically shown to be as efficient in terms of reconstruction error (or even \emph{better}) as the random i.i.d Gaussian ones in the context of compressed sensing \cite{barbier2013compressed,JavanmardMontanari12,do2008fast}. This is confirmed by the replica analysis for orthogonal operators done in \cite{WenW14}.

\begin{figure}[t]
\centering
\includegraphics[width=8cm]{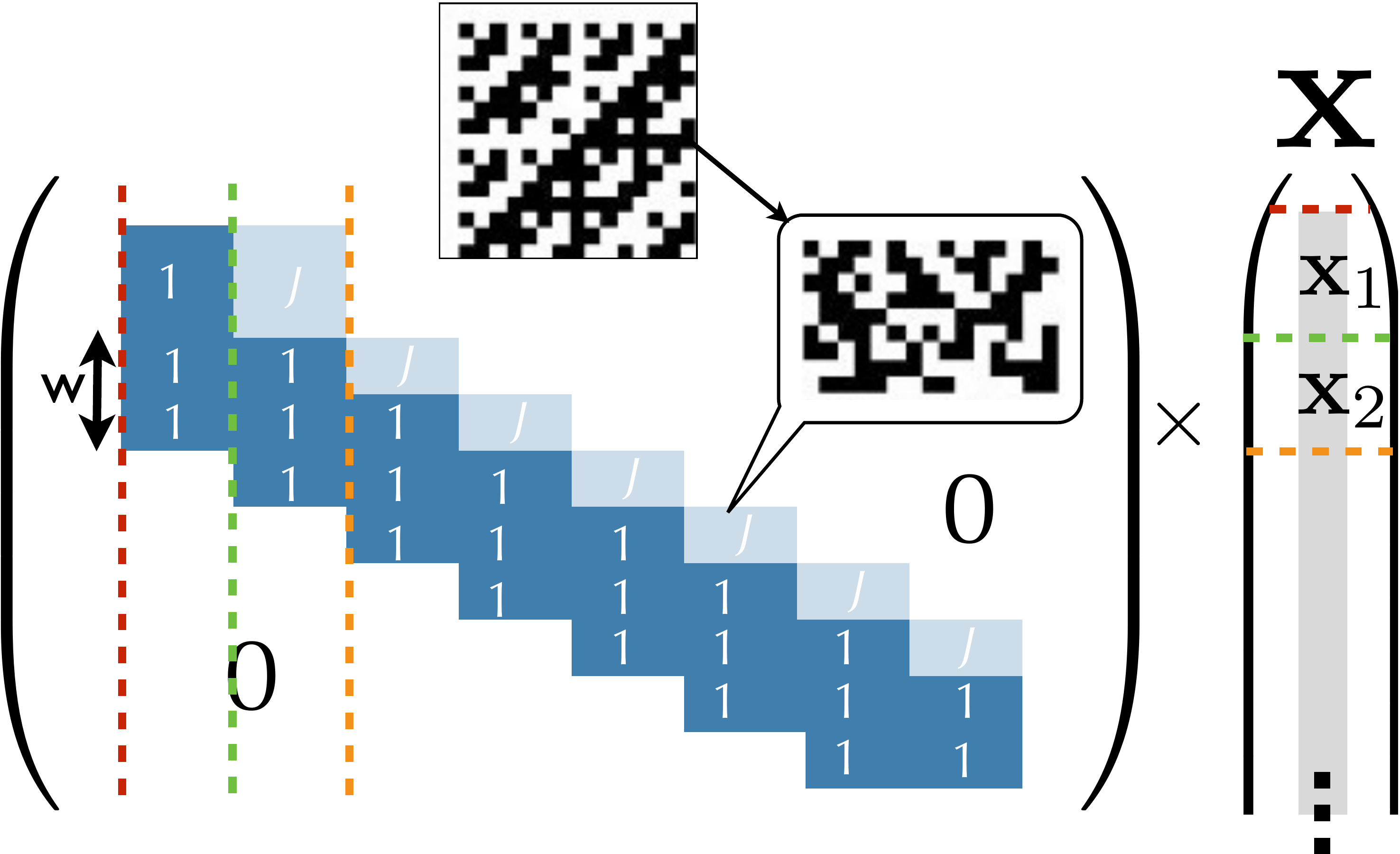}
\caption{The spatially coupled coding operator used in our study. It is decomposed in
  $L_r\times L_c$ blocks, each being made of $N/L_c$ columns and
  $\alpha_{\rm seed}N/L_c$ lines for the blocks of the first
  block-row, $\alpha_{\rm rest}N/L_c$ lines for the following
  block-rows. Futhermore $\alpha_{\rm seed}>\alpha_{\rm rest}$ and $\alpha_{\rm seed}+ (L_r-1)\alpha_{\rm rest} = L_c\alpha$. We focus on two possible ways to construct the non-zero blocks of the operator that are colored (the white parts are only zeros). The first is using sub-matrices that are i.i.d Gaussian, the second, that is represented here, is using Hadamard operators. The figure shows how the
  lines of the original Hadamard matrix (of size $N/L_c\times N/L_c$)
  are randomly selected and re-ordered to form a
  block of the final operator. There is a number $w$ (the coupling
  window) of lower diagonal blocks. In the case of an Hadamard construction, these blocks have entries $\in \{\pm 1\}$ as
  the diagonal blocks, the upper diagonal blocks have entries
  $\in \{\pm\sqrt{J}\}$ where $\sqrt{J}$ is the coupling strength. In the case of a Gaussian construction, it is the variance of the i.i.d entries inside the blocks that changes: the variance is $1$ in the dark blue blocks, $J$ in the light blue ones. In full generality, in the random operator case, the structure of the coding matrix is encoded through a matrix of variances with entries $J_{r,c}\ge 0$, the block $(r,c)$ being a Gaussian matrix with i.i.d entries $\sim \mathcal{N}(\cdot |0, J_{r,c})$. The colored dotted lines help
  to visualize the block decomposition of the signal induced by the
  operator structure: each block of the signal will be decoded
  at different times (see Fig.~4 of
  \cite{barbier2013compressed}). The parameters that define the
  spatially coupled coding operator ensemble are
  $(L_c,L_r,w,\sqrt{J},\alpha_{{\rm{seed}}},\alpha_{{\rm{rest}}})$. The matrix elements are all multiplied by $\sqrt{C/L}$, where $C$ is chosen such that the power of the codeword is equal to one.}
\label{fig_seededHadamard}
\end{figure}

All the blocks are thus constructed from the same Hadamard operator of size $N/L_c \times N/L_c$ with the constraint that $N/L_c$ is a power of two, intrinsic to the Hadamard construction (simple numerical tricks such as $0$-padding allow to relax this constraint at virtually no computational cost). The difference between blocks is the random selection of modes and their order, see Fig.~\ref{fig_seededHadamard}.

The decoder requires four operators in order to work. We define $\textbf{e}_c$, with $c\in\{1,\ldots,L_c\}$, as the vector of size $N/L_c$ which is the $c^{th}$ block of $\textbf{e}$, itself of size $N$. For example, in Fig.~\ref{fig_seededHadamard}, the signal
$\bx$ is naturally decomposed as $[\bx_1,\bx_2, \ldots,\bx_{L_c}]$ due to the block structure of the coding operator. We define similarly $\textbf{f}_r$, with $r\in\{1,\ldots,L_r\}$, as the vector of size $\alpha_rN/L_c$  which is the $r^{th}$ block of $\textbf{f}$, itself of size $M$. We call $\alpha_r$ the measurement rate of all the blocks at the $r^{th}$ block-row and $\sum_r^{L_r} \alpha_r/L_c = \alpha$ from \eqref{eq_alphaRest}. In Fig.~\ref{fig_seededHadamard}, $\alpha_1 = \alpha_{{\rm{seed}}}$ and $\alpha_{j} = \alpha_{{\rm{rest}}} \ \forall \ j\ge 2$.
The notation $i\in c$ (resp. $\mu\in r$) means all the components of $\textbf{e}$ that are in $\textbf{e}_c$ (resp. all the components of $\textbf{f}$ that are in $\textbf{f}_r$). Using this, the operators required by the decoder are defined as following
\begin{align}
&\tilde O_\mu(\textbf{e}_c) \defeq  \sum_{i\in c}^{N/L_c} F_{\mu i}^{2} e_i, \  O_\mu(\textbf{e}_c) \defeq  \sum_{i\in c}^{N/L_c} F_{\mu i} e_i, \ \tilde O_i(\textbf{f}_r) \defeq  \sum_{\mu\in r}^{\alpha_rN/L_c} F_{\mu i}^2 f_\mu, \ O_i(\textbf{f}_r) \defeq  \sum_{\mu\in r}^{\alpha_rN/L_c}F_{\mu i} f_\mu. \label{eq_fastOpDefs2}
\end{align}
In the case of an Hadamard-based operator, $F_{\mu i}^2=1$ or $J$ depending on the non-zero block to which the indices $(\mu,i)$ belongs to. It implies that these four operators  are implemented as fast transforms ($O_\mu$ and $O_i$) or simple sums ($\tilde O_\mu$ and $\tilde O_i$) and do not require any costly direct matrix multiplications. This is the advantage of using Hadamard-based operators: it reduces the cost of the matrix multiplications required by the decoder from $O(N^2)$ (the cost with non structured matrices) to $O(N\ln N)$ and the matrix has never to be stored in the memory, which allows to decode very large messages in a fast way without memory issues. The AMP decoder written in terms of these operators and that makes explicit the operator block structure is given in Fig.~\ref{algo_AMP} \cite{Schniter2012compressive,barbier2013compressed}.

The practical implementation of the operator $\bF$ requires caution: the necessary ``structure killing'' randomization of the Hadamard modes inside each non-zero block is obtained by applying a permutation of lines after the use of the standard fast Hadamard transform ${\rm H}$. For each block $(r,c)$, we choose a random subset of modes $\Omega^{r,c} = \{ \Omega^{r,c}_1 ,\ldots, \Omega^{r,c}_{\alpha_rN/L_c} \} \subset \{ 1,\ldots, N/L_c \}$.
The definition of 
$O_\mu(\textbf{e}_c)$ using ${\rm H}$ is
\begin{equation}
 O_\mu(\textbf{e}_c) :=   {\rm H}(\textbf{e}_c)|_{\Omega^{r_{\mu},c}_{\mu - \mu_{r_{\mu}} + 1}},
\end{equation}
where $r_{\mu}$ is the index of the block-row that includes $\mu$, $\mu_{r_{\mu}}$ is the index of the first line of the block-row $r_{\mu}$ and $\lambda|_{\mu}$ is the $\mu^{th}$ component of $\lambda$. For $O_i(\textbf{f}_r)$ instead,

\begin{equation}
 O_i(\textbf{f}_r):={\rm H}^{-1}( \tilde{\textbf{f}}_r )|_{i - i_{c_i}+1},
\end{equation}
where $c_{i}$ is the index of the block column that includes $i$, $i_{c_{i}}$ is the index of the first column of the block column $c_{i}$, ${\rm H}^{-1}$ is the standard Hadamard fast inverse operator of ${\rm H}$ (which is actually ${\rm H}$ itself) and $\tilde{\textbf{f}}_r$ is defined in the following way
\begin{equation}
  \forall \ \gamma \in \{ 1,\ldots,\alpha_r N/L_c \}, \quad \tilde{\textbf{f}}_r|_{\Omega_{\gamma}^{r,c}} = \textbf{f}_r|_{\gamma} \quad {\text{and}} \quad  \forall \ i \notin \Omega^{r,c}, \quad \tilde{\textbf{f}}_r|_i = 0. 
\end{equation}

\begin{figure}[!t]
\centering
\includegraphics[width=0.4\textwidth]{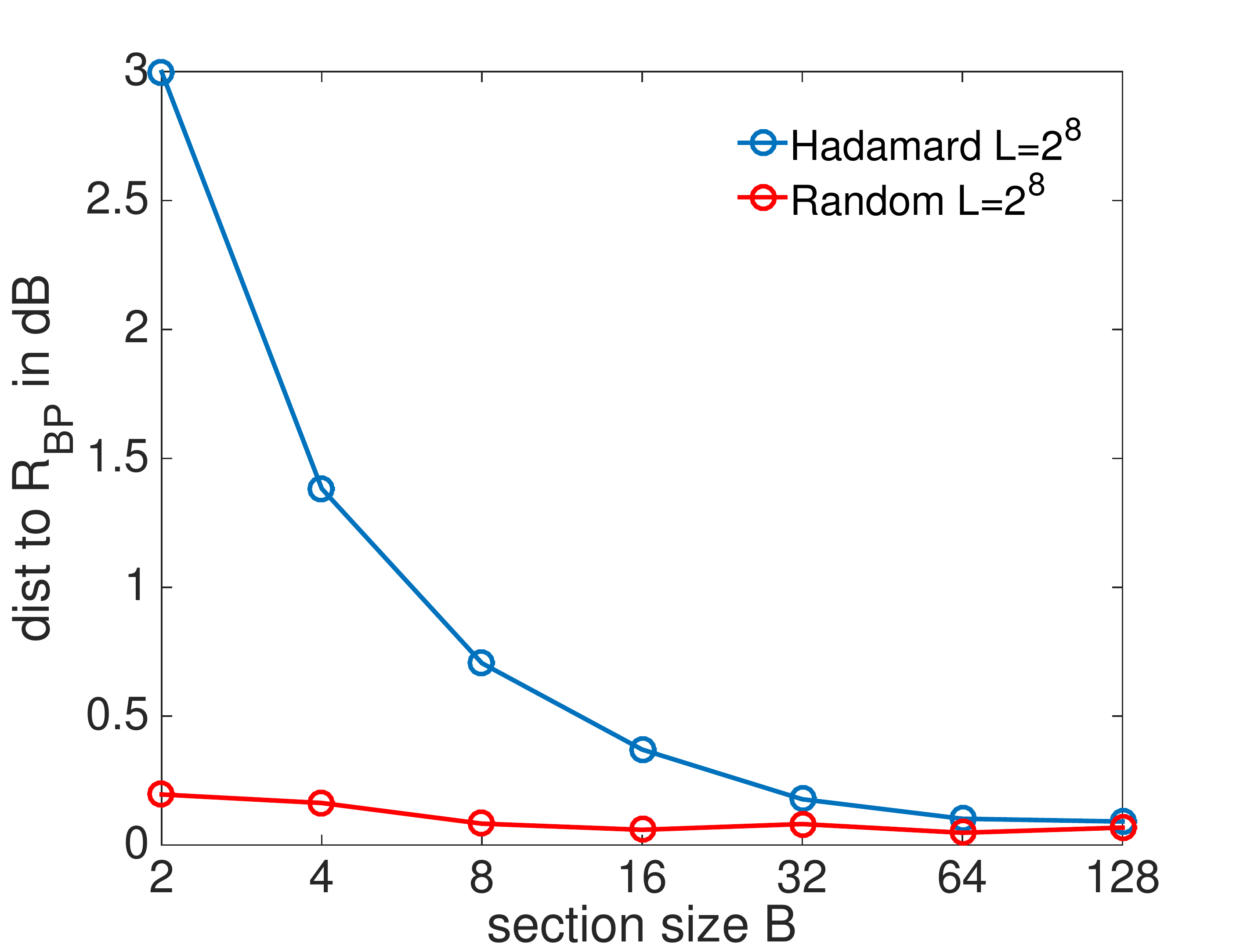}
\caption{Comparison between the distance in dB to the asymptotic $L\to\infty$ BP threshold $R_{\rm BP}(B)$ at which the AMP decoder with homogeneous (i.e non spatially coupled) Hadamard coding operators (blue line) or with random i.i.d Gaussian matrices (red line) starts to reach an ${\rm{SER}} <10^{-5}$ (which is then almost always strictly $0$). The experiement is for a fixed number of sections $L=2^8$ and ${{\rm snr}}=100$. The points have been obtained by averaging over $100$ random instances. The BP threshold is obtained by state evolution analysis. The Hadamard operator works poorly when the signal density increases (i.e when $B$ decreases), but reaches quickly performances close to the random matrix ones as it decreases. The random Gaussian i.i.d matrices have a performance that is close to constant as a function of $B$ at fixed $L$.}\label{fig_distToRbp}
\end{figure}

Fig.~\ref{fig_distToRbp} shows that when the signal sparsity increases, i.e when the section size $B$ increases, using Hadamard-based operators becomes equivalent to random i.i.d Gaussian ones in terms of performances (this point is studied in more details in \cite{barbier2013compressed}). We fix the ${\rm snr}=100$ and plot the distance in dB to the BP threshold $R_{\rm BP}(B)$ (computed for $L\to\infty$) at which the decoder starts to decode perfectly with Hadamard or random i.i.d Gaussian operators. Recall that $R_{\rm BP}(B)$ is defined as the highest rate until which AMP decoding is optimal \emph{without} the need of non constant power allocation nor spatial coupling. It appears that at low section size, it is advantageous to use random operators but as $B$ increases, structured operators quickly reach the random operator performances. The BP threshold is predicted by the state evolution analysis presented in the next section.
\section{Results of the state evolution analysies}
\label{sec:SEseeded}
Most of the following empirical results are given for Hadamard-based operators for practical and computationnal reasons. In contrast, the state evolution analysies are derived (in appendix~\ref{app:SE}) for i.i.d Gaussian matrices, which remains quite accurate when Hadamard-based operators are employed.
\subsection{State evolution for homogeneous coding operators and constant power allocation}
\begin{figure}[!t]
\centering
\includegraphics[width=0.32\textwidth, trim=20 30 70 5, clip=true]{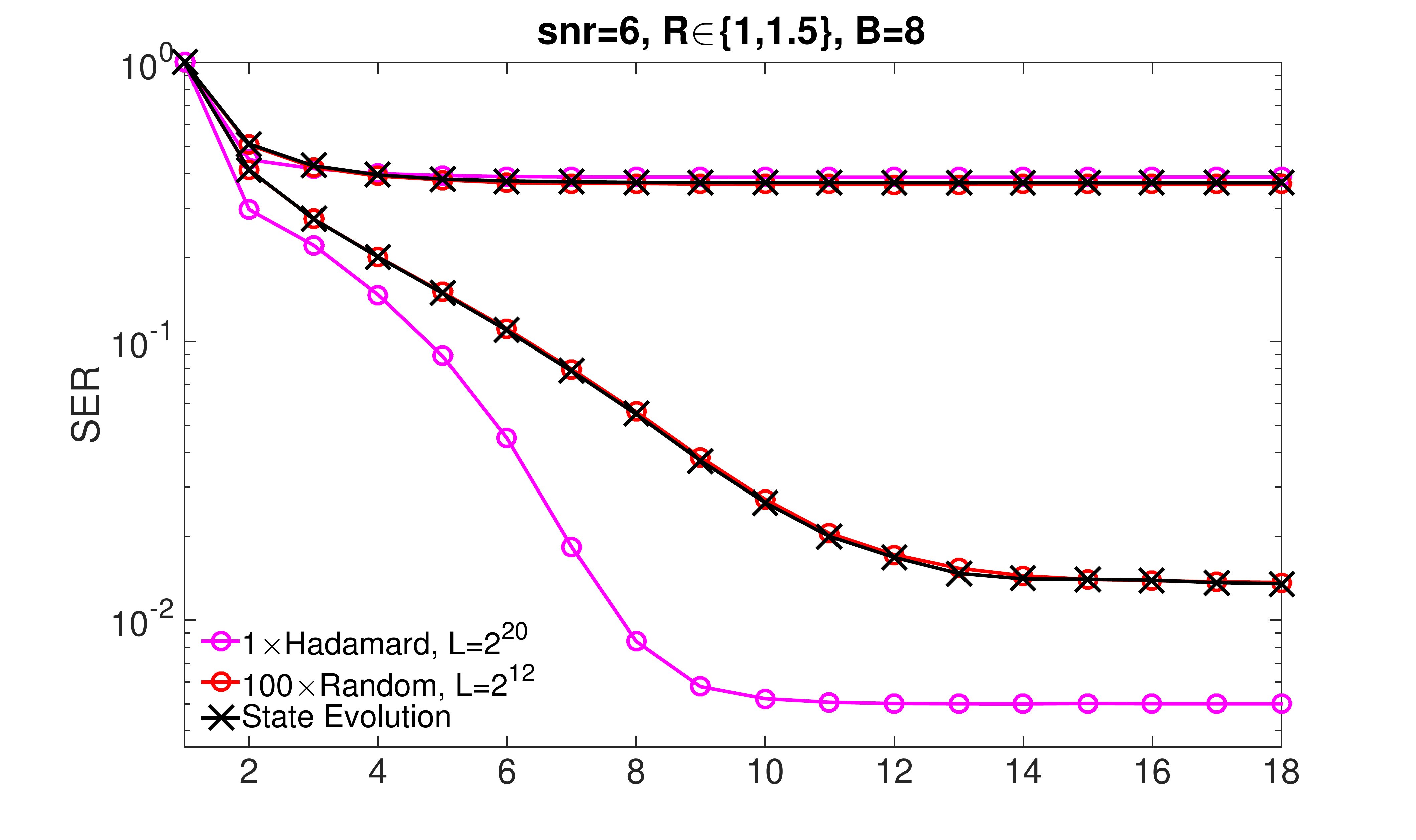}
\includegraphics[width=0.32\textwidth, trim=20 30 70 5, clip=true]{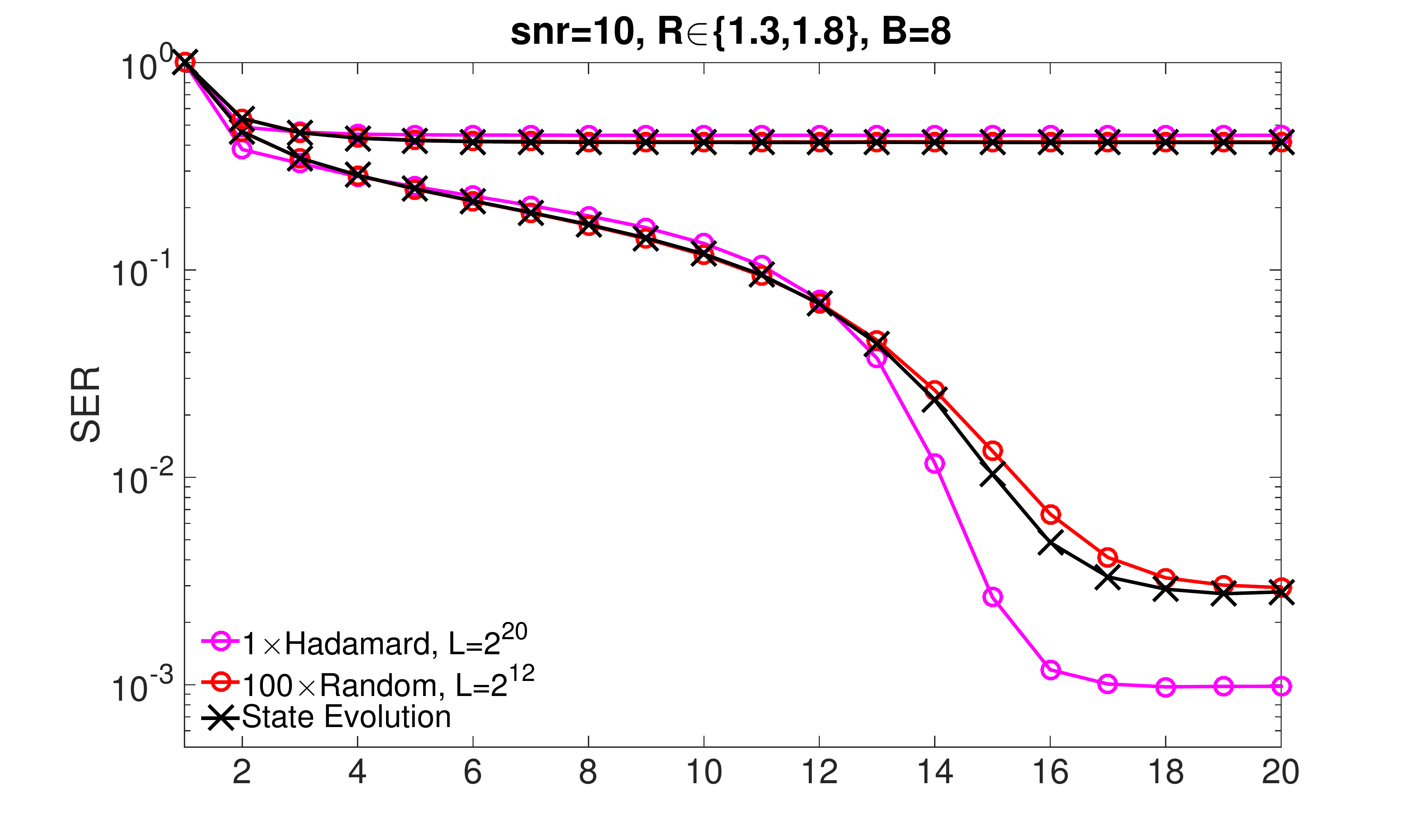}
\includegraphics[width=0.32\textwidth, trim=20 30 70 5, clip=true]{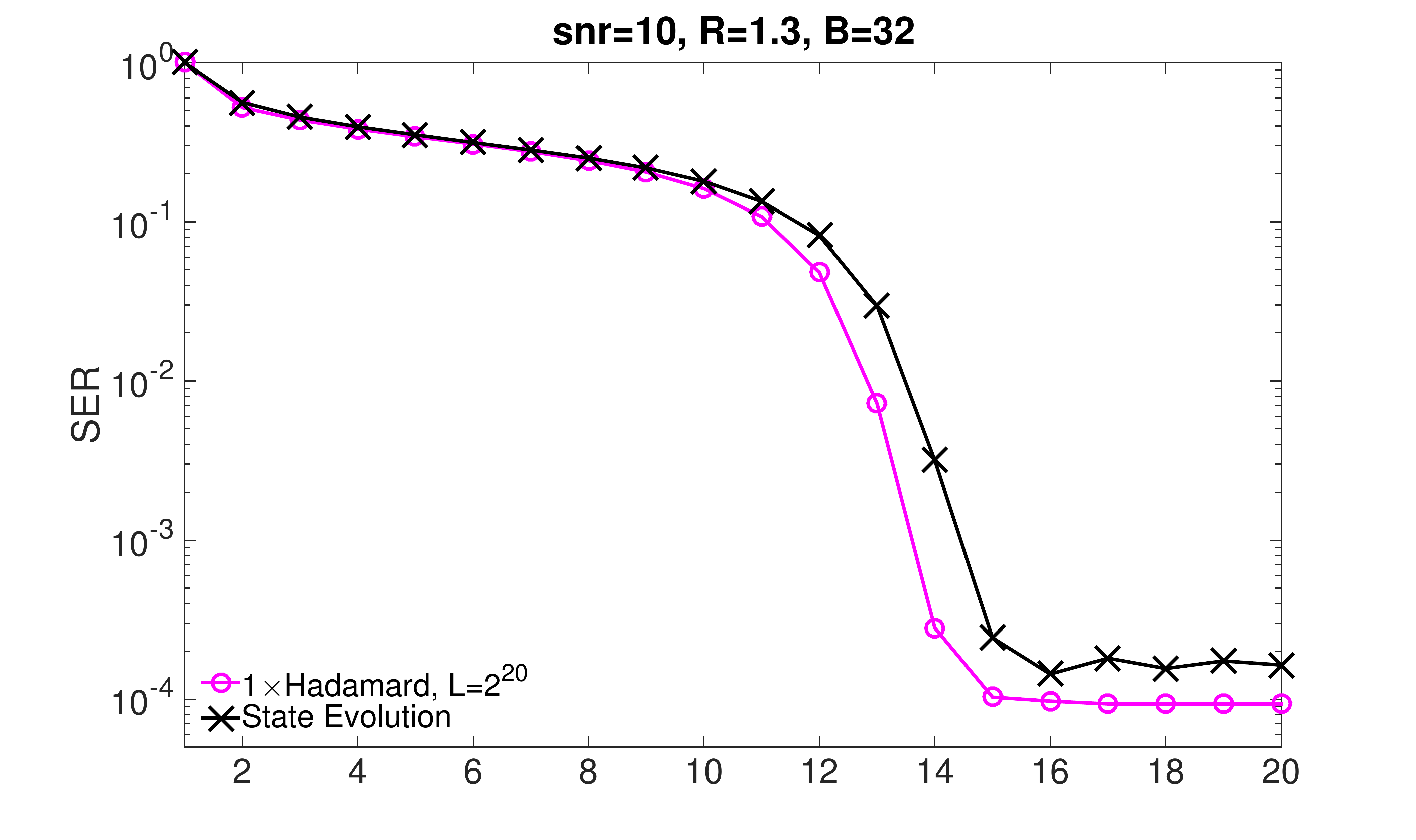}
\caption{The state evolution prediction of the section error rate ${\rm{SER}}^t$ as a function of time (black curves), compared to the actual ${\rm{SER}}^t$ obtained with the AMP decoder for sparse superposition codes. We use constant power allocation and various ${{\rm snr}}$, rates $R$ (one above and one below the BP threshold $R_{\rm BP}$), a section size $B=64$ and using both homogeneous Hadamard-based (pink curves) and random i.i.d Gaussian operators (red curves). The figure shows how close is the theoretical prediction from the true AMP behavior at finite sizes. The integrals appearing in (\ref{eq_SE_MSE}), (\ref{eq_SE_SER}) are computed by monte carlo with a sample size of $10^6$. We observe that when the signal size $L$ is large and the use of an Hadamard-based operator is mendatory due to speed and memory issues, the decoder behavior follows the theoretical predictions very accurately despite not being exact. We also observe that the final ${\rm{SER}}$ reached with Hadamard-based operator is lower than with Gaussian i.i.d ones and convergence is generally faster in terms of number of decoding iterations. Note that due to the memory issue, the curves for the Gaussian operators cases are obtained averaging over $100$ random instances, while with Hadamard-based ones, a single instance is sufficient as we can take a very large $L$ making the instance-to-instance fluctuations negligible.}
\label{fig_SEfull}
\end{figure}
The state evolution technique (referred to as the cavity method in
physics) is a statistical analysis that allows to monitor the AMP
dynamics and performance in the limit of decoding infinitely large
signals \cite{BayatiMontanari10}. In the present case, we consider the
matrix $\bF$ to be i.i.d Gausian with zero mean, for which state
evolution has been originally derived
\cite{BayatiMontanari10}. Extension to more general ensembles such as
row-orthogonal matrices could be considered \cite{WenW14} but it is
out of the scope of the present paper. In addition, the present
authors have numerically shown in \cite{barbier2013compressed} that
the state evolution analysis derived in the Gaussian i.i.d case is a
good predictive tool of the behavior of the AMP decoder with
structured operators such as the Hadamard one, despite not perfect nor
rigorous.

The complete (but heuristic) derivation of the state evolution
recursions is done in appendix~\ref{appSec:SE_Nonseeded}. We define
\begin{align}
E^t\defeq \lim_{L\to\infty} \frac{1}{L} \mathbb{E}_{\bF, \bx, \bxi}[||\hat{\bx}^t - \bx||_2^2]
\end{align}
as the asymptotic average ${\rm MSE}$ (\ref{eq_MSE}) per section of the AMP estimate $\hat{\bx}^t$ at iteration $t$, where the average is over the model \eqref{eq_yfx}. The state evolution should be initialized with initial condition $E^0 = 1$, corresponding
to no prior knowledge of the sent message. A convenient form of the
state evolution recursion is
\begin{align}
E^{t+1} &= \int_{\mathbb{R}^B} \mathcal{D} \bz \left([f_{a_{1|1}}((\Sigma^{t+1})^2,\bz )-1]^{2}+(B-1)f_{a_{2|1}}((\Sigma^{t+1})^2,\bz )^{2}\right), \label{eq_SE_MSE} \\
{\text{with}}  \quad \Sigma^{t+1}(E^t) &= \sqrt{R\ln(2)({1/{{\rm snr}}} + E^t) }, \label{eq_SE_var}
\end{align}
where $\mathcal{D} \bz \defeq \prod_{i}^{B} \mathcal{D} z_i = \prod_{i}^{B} \mathcal{N}(z_i|0,1) dz_i $ is a $B$-d unit centered Gaussian measure, and the following functions are used
\begin{align}
f_{a_{i|i}}(\Sigma^2,\bz ) &\defeq  \Big[1 + e^{-\frac{\ln(B)}{\Sigma^2}}  \sum_{1\le j\le B : j\ne i}^{B-1}  e^{\frac{\sqrt{\ln(B)}(z_j-z_i)}{\Sigma}} \Big]^{-1}, \label{eq_fa1fun} \\
f_{a_{j|i}}(\Sigma^2,\bz )  &\defeq  \Big[1 + e^{\frac{\ln(B)}{\Sigma^2} + \frac{\sqrt{\ln(B)}(z_j - z_i)}{\Sigma}}  +  \sum_{1\le k\le B : k \ne i,j }^{B-2} \!\!\!\!\!\!\!\!\!\!e^{\frac{\sqrt{\ln(B)}(z_k-z_i)}{\Sigma}} \Big]^{-1}. \label{eq_fa2fun}
\end{align}
The interpretation of $E$ is the following: it is the MMSE associated with the estimation of a single section sent through an \emph{effective AWGN channel} with noise variance $\Sigma^2/\ln(B)$, where $\bz$ plays the role of this effective AWGN. Here $f_{a_{i|i}}(\Sigma^2, \bz)$ outputs the asymptotic estimate by the AMP decoder of the posterior probability that the $i^{th}$ component is the unique $1$ in the section, given that is is actually the $1$ in the transmitted section (all section permutations are equivalent). Instead, $f_{a_{j|i}}(\Sigma^2, \bz)$ outputs the asymptotic posterior probability estimate by the AMP decoder that the $j^{th}$ component is the $1$ given that is is actually the $i^{th}$ component that is the true $1$, thus of an error. With this interpretation in mind, there is a simple correspondance between the ${\rm{MSE}}$ and the ${\rm{SER}}$ given by
\begin{equation}
{\rm{SER}}^{t+1} = \int_{\mathbb{R}^B}  \mathcal{D}\bz \ \mathbb{I}\left(\exists \ j \in \{2,\ldots,B\} : f_{a_{j|1}}((\Sigma^{t+1})^2,\bz) > f_{a_{1|1}}((\Sigma^{t+1})^2,\bz) \right).
\label{eq_SE_SER}
\end{equation}

From this equation, we can predict the asymptotic time evolution of the decoder performance measured by the ${\rm{SER}}$, such as in Fig.~\ref{fig_SEfull}. Recall that the state evolution predictions are asymptotically exact when AMP is used with i.i.d Gaussian coding matrices, and approximate but yet accurate for Hadamard operators. The black curves on this figure represent the iteration of \eqref{eq_SE_MSE}, \eqref{eq_SE_var}, \eqref{eq_SE_SER} for different parameters $({{\rm snr}},R)$ and fixed section size $B=64$, using randomized Hadamard-based or random Gaussian i.i.d operators. \eqref{eq_SE_SER} and (\ref{eq_SE_MSE}) are computed at each step by monte carlo. 

We restrict these experiments to relatively low values of ${{\rm snr}}$, because if these are too high, the experimental and theoretical curves would stop at some iteration without reaching an error floor and decoding ``seems'' perfect. For the experimental curves, this is due to the fact that in order to observe an ${\rm{SER}} = O(\epsilon)$, the message must be at least made of $L\approx 1/\epsilon$ sections, which is not the case for messages of reasonnable sizes when the asymptotic ${\rm{SER}}$ is very small. In fact, when the rate is below the BP threshold, the decoding is usually perfect and is found to reach with high probability ${\rm{SER}}=0$. The black theoretical curves should anyway always reach a positive error floor but they would not because of the same reason: this error floor is so low at high ${{\rm snr}}$ that the minimal sample size required to observe it when computing the integrals present in (\ref{eq_SE_MSE}), (\ref{eq_SE_SER}) by monte carlo should be way too large to practically deal with.

We also naturally observe, from the definition of the state evolution technique as an asymptotic analysis, that the theoretical and experimental results match better for larger messages. At rate $R>R_{\rm BP}$ (the curves converging to an high ${\rm{SER}}$ for the first two cases on Fig.~\ref{fig_SEfull}), we see that AMP decoding does not reconstruct the messages and converges to an error precisely predicted by the state evolution. On the contrary, below the threshold, the reconstruction succeeds up to an error floor dependent on the parameters $(B,{{\rm snr}},R)$. We also observe, as in \cite{barbier2013compressed}, that the state evolution, despite being derived for random i.i.d Gaussian matrices, predicts well the behavior of AMP with Hadamard-based operators, especially for large $B$.

Let us discuss a bit more the error floor. The replica analysis from which we borrow some results now will be discussed in details in sec.\ref{sec:replica}, but let us just consider now the potential (\ref{eq_freeEnt2}) obtained from this analysis as \emph{a function which extrema correspond to the fixed points of the state evolution} (\ref{eq_SE_MSE}).
\begin{figure}[!t]
\centering
\includegraphics[width=0.35\textwidth, trim=20 30 70 5, clip=true]{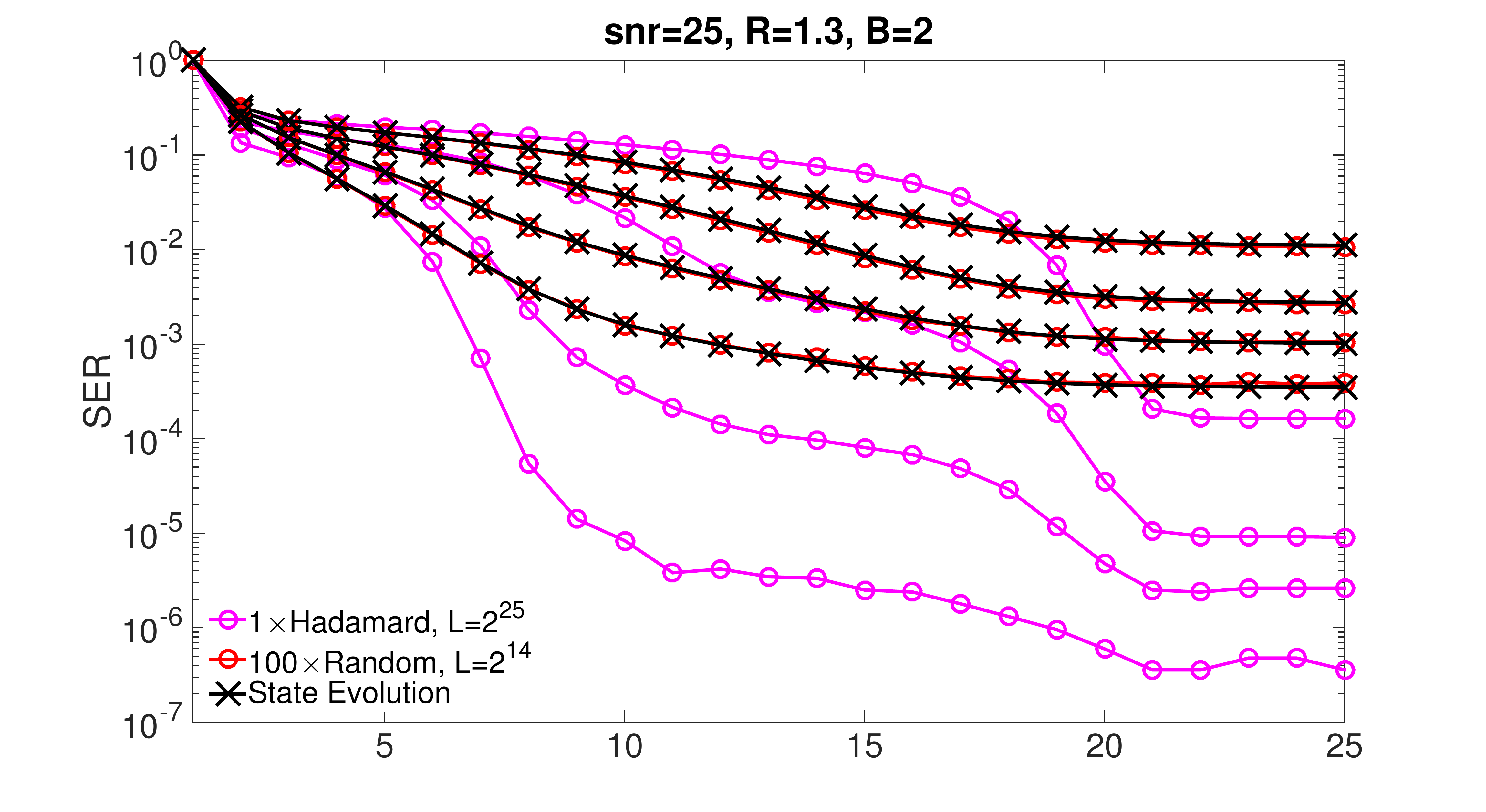}
\includegraphics[width=0.35\textwidth, trim=20 30 70 5, clip=true]{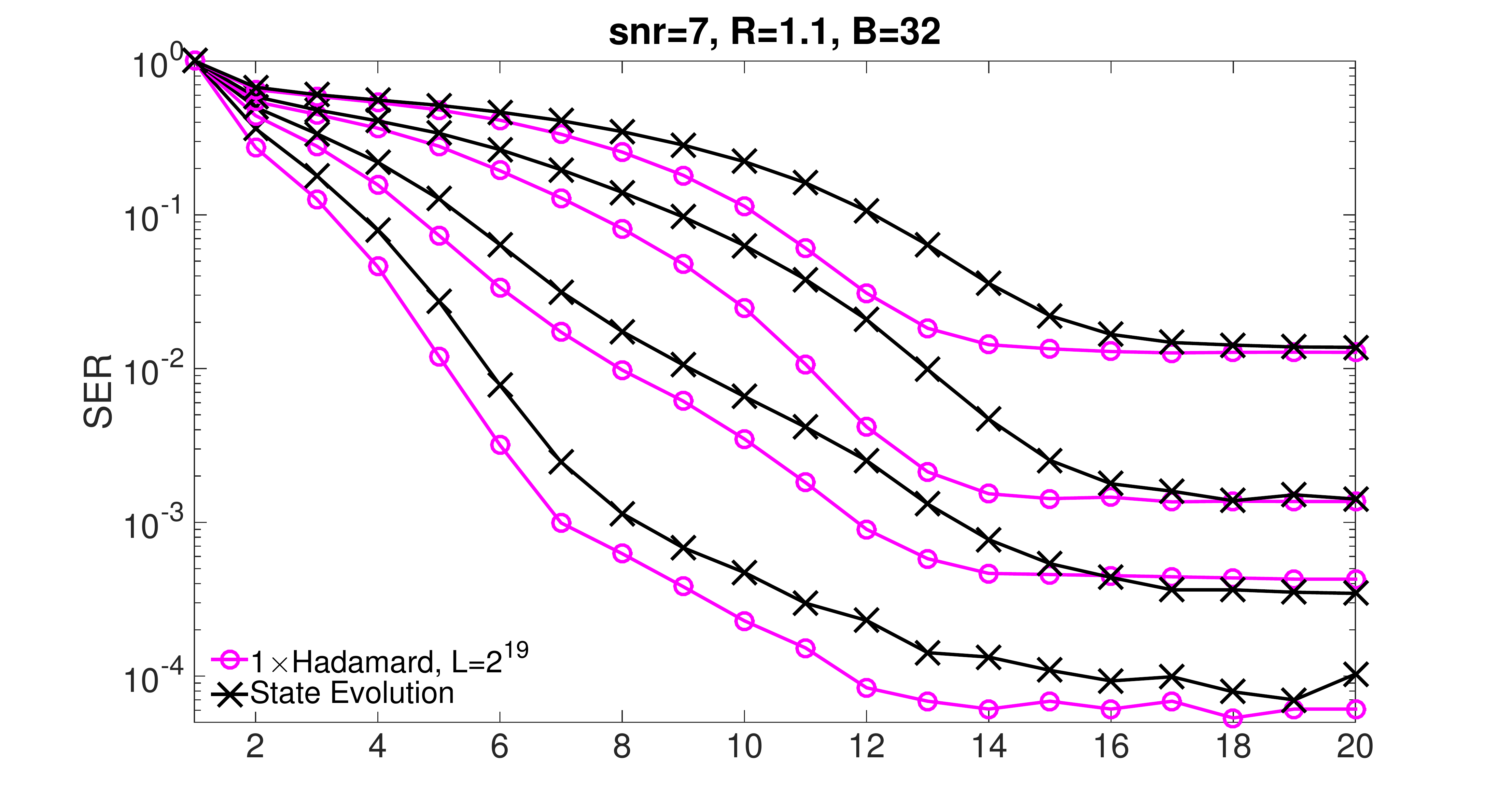}
\caption{The state evolution prediction (black curves) of the section error rate $\{{\rm{SER}}^t_c:c\in\{1,\ldots,L_c=4\}\}$ for each of the four blocks of the message, induced by the block structure of the spatially coupled operator, see Fig.~\ref{fig_seededHadamard}, as a function of the decoder iterations. This is for sparse superposition codes with constant power allocation. The state evolution curves are compared to the actual $\{{\rm{SER}}^t_c\}$ of the AMP decoder for two different settings. The spatially coupled Hadamard-based operator (pink curves) is drawn from the ensemble $(L_c=4,L_r=5,w=2,\sqrt{J}=0.6,R,\beta_{{\rm{seed}}}=1.5)$. In the low ${{\rm snr}}=7$ case, the error floor that is different in each block is well predicted by state evolution while for higher ${{\rm snr}}=25$, the results of the Hadamard-based operator is way better than the i.i.d Gaussian operator performance (red curves), perfectly predicted by state evolution. As for Fig.~\ref{fig_SEfull}, the finite-size performance in the Gaussian operator case is obtained averaging over $100$ random instances.}\label{fig_SEspc}
\end{figure}

In sparse superposition codes, there exists a inherent error floor, and this independently of the finite size effects. Indeed, for any finite section size $B$, the asymptotic $L\to\infty$ state evolution and replica analysies show that this error floor is present, but is in general very small and quickly decreasing when $B$ or the ${\rm snr}$ increase. See for example the Fig.~\ref{fig_optSER}, obtained from the replica analysis, that shows how the SER associated with the MMSE estimator (i.e the optimal ${\rm SER}$) of sparse superposition codes with constant power allocation falls with a power law decay as a function of $B$ or the ${\rm snr}$. The right part of Fig.~\ref{fig_freeEnt} also illustrates the ${\rm MSE}$ error floor decaying when the ${\rm snr}$ increases. It shows the potential function, which maxima indicate the stable fixed points ${\rm MSE}$ of the state evolution, at fixed $B=2$, $R=1.8$ and for values of the ${\rm snr} \in \{28,30,32,\ldots,46\}$ (the top black curve is for ${\rm snr}=46$, the bottom one for ${\rm snr}=28$). On each curve, there is a red point at a relatively low ${\rm MSE}$ value, which corresponds to the MMSE of the code (as long as $R<R_{\rm opt}$, otherwise the MMSE corresponds to the high error maximum). This is the $\neq 0$ ${\rm MSE}$ error floor. 

This phenomenology is also present in low density generator matrix (LDGM) codes. These codes also present an error floor, but a very important difference between sparse superpostion codes and LDGM ones is that \emph{for sparse superposition codes, the error floor can be made arbitrarily small for a fixed ${\rm snr}$ by increasing $B$ while maintaining low-complexity AMP decoding}. Indeed, in the case of an i.i.d Gaussian operator (spatially coupled or not), the decoding complexity of AMP scales as $O((BL)^2)$, or $O(BL\ln(BL))$ with Hadamard-based operators. In the case of LDGM codes, the error floor can be decreased as well by increasing the generator matrix density, but it has a large computational cost: the BP decoder used for LDGM codes which is very similar to the one used for LDPC codes \cite{RichardsonUrbanke08} has to perform a number of operations which scales exponentially with the average degree of the factor nodes in the graph. Thus the reduction of the error floor in LDGM codes becomes quickly intractable due to this computational barrier that is not present in sparse superposition codes, where the cost is at worst quadratic with the section size $B$.

Finally let us stress that the rapid decrease of the error floor observed in Fig.~\ref{fig_optSER} when the ${\rm snr}$ increases is a generic scenario, in the sense that the very same phenomenon happens when $B$ increases or the rate $R$ decreases. This can be easily understood. Looking at the state evolution recursion for the effective noise variance (\ref{eq_SE_var}) together with (\ref{eq_SE_MSE}) and the functions (\ref{eq_fa1fun}), (\ref{eq_fa2fun}), wee observe the following: it is perfectly equivalent to decrease the rate or increase $B$ by the proper amount, and increasing the ${\rm snr}$ has a similar effect to reduce the effective noise variance (but not in a simple multiplicative way as $R$ and $\ln(B)$).
\subsection{State evolution for spatially coupled coding operators and constant power allocation} \label{sec:SE_spc}
In the spatially coupled case, the interpretation of state evolution is similar to the homogeneous operator case: $E_c^{t+1}$ tracks the asymptotic average ${\rm{MSE}}$ of the AMP decoder in the block $c$ of the reconstructed signal, see Fig.~\ref{fig_seededHadamard}. It is further interpreted as the MMSE associated with an effective AWGN channel which noise variance (\ref{eq_SEsigmaSeeded}) now depends on the block index, and which is coupled to the other blocks. The derivation of the analysis for the spatially coupled operators is presented in details in appendix~\ref{appSec:SE_seeded}. The final recursion for the average ${\rm{MSE}}$ asymptotically attained by AMP for the block $c\in\{1,\dots,L_c\}$ is 
\begin{align}
E_c^{t+1} &= \int_{\mathbb{R}^B} \mathcal{D} \bz \left([f_{a_{1|1}}((\Sigma_c^{t+1})^2,\bz )-1]^{2}+(B-1)f_{a_{2|1}}((\Sigma_c^{t+1})^2,\bz )^{2}\right), \label{eq_SESeeded}\\
{\text{with}}\quad \Sigma_c^{t+1}(\{E_{c'}^t\}) &= \left[\frac{B}{\ln(B)} \sum_{r}^{L_r} \frac{\alpha_{r} J_{r,c}}{{L_c/{{\rm snr}}} + \sum_{c'}^{L_c} J_{r,c'}E_{c'}^t}\right]^{-1/2}, \label{eq_SEsigmaSeeded}
\end{align}
where the $f_{a_{1|1}},f_{a_{2|1}}$ functions are given by (\ref{eq_fa1fun}), (\ref{eq_fa2fun}). The relation linking the $E_c$ and ${\rm{SER}}_c\ \forall\ c\in\{1,\dots,L_c\}$ is similar to the homogenous operator case 
\begin{equation}
{\rm{SER}}^{t+1}_c = \int_{\mathbb{R}^B}  \mathcal{D}\bz \ \mathbb{I}\left(\exists \ j \in \{2,\ldots,B\} : f_{a_{j|1}}((\Sigma_c^{t+1})^2,\bz) > f_{a_{1|1}}((\Sigma_c^{t+1})^2,\bz) \right).
\label{eq_DE_E}
\end{equation}

Fig.~\ref{fig_SEspc} shows a comparison of $\{{\rm{SER}}_c^t:c\in\{1,\ldots,L_c\}\}$ predicted by state evolution (black curves) with the actual reconstruction ${\rm{SER}}$ per block of messages transmitted using sparse superposition codes with Hadamard-based spatially coupled operators and AMP. Again, the discrepancies between the theoretical and experimental curves come from that state evolution is derived for random i.i.d Gaussian matrices. The final error using these Hadamard operators is at least as good as predicted by state evolution. As already noted in the homogeneous case and \cite{barbier2013compressed}, AMP in conjunction with structured Hadamard-based operators converges slightly faster to the predicted final error than Gaussian matrices.
\begin{figure}[!t]
\centering
\includegraphics[width=8cm]{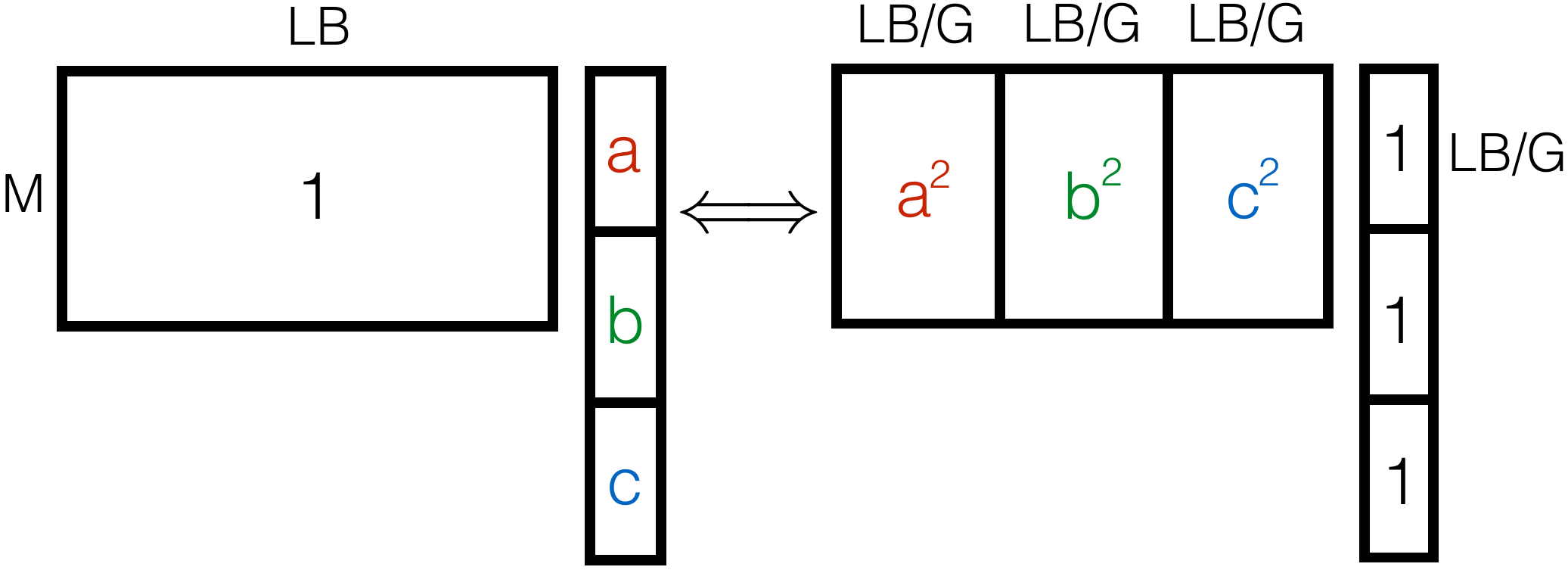}
\caption{The figure shows how to convert a non constant power allocated message encoded through an operator with homogeneous variance $=1$ into an equivalent system, from the point of view of state evolution, with a constant power allocated message encoded by a structured operator. The values on the matrix represent the variance of the entries of the matrix up to some rescaling factor used for the codeword power constraint. The values on the message represent the non-zero values inside the sections that belong to a given group: here the message is decomposed into $G=3$ groups, and all the sections inside the first group have a non-zero value equal to $a$, and so forth. The transformation is performed by structuring the operator into block columns, with as many block columns as different values in the power allocation, or groups: if a column of the original matrix acts on a component of a section where the non-zero value is $u$, then this column variance is multiplied by $u^2$ in the new structured operator (such that the entries of this column are multiplied by $u$).}
\label{fig_equivPowaSpc}
\end{figure}
\subsection{State evolution for homogeneous coding operators and non constant power allocation}
\label{sec:powA_SE}
From the previous analysis sec.\ref{sec:SE_spc}, we can trivially extract the state evolution for sparse superposition codes with non constant power allocation when an homogeneous i.i.d Gaussian matrix is used. This is done thanks to the transformation of Fig.~\ref{fig_equivPowaSpc}: starting from an homogeneous matrix and non constant power allocated message, we convert the system into an equivalent one (from the state evolution point of view) that has a structured matrix but with a constant power allocated message. 

Let us detail the procedure. Suppose the message is decomposed into $G$ groups, where inside the group $g$, the power allocation is the same for all the sections belonging to this group and equals $c_g$. Now one must create a structured operator starting from the original one, decomposing it into $LB/G$ column blocks and multiply all the elements of the column block $g$ by $c_g$, as shown in Fig.~\ref{fig_equivPowaSpc}. This new operator acting on a constant power allocated message is totally equivalent to the original system from the state evolution point of view, and fortunately, we already have the state evolution for this new system from the previous section. Using (\ref{eq_SEsigmaSeeded}) in the present setting, one has to be careful with the value of $\alpha_r$ defined as the number of lines over the number of columns of the $r^{th}$ block-row. Here there is a unique value that equals $M/(N/G)=G\alpha$ where $\alpha$ is the measurement rate of model (\ref{eq_alpha}). Given that, $L_c=G$ we obtain for all $g'\in\{1,\dots,G\}$
\begin{align}
E_g^{t+1} &= \int_{\mathbb{R}^B} \mathcal{D} \bz \left([f_{a_{1|1}}((\Sigma_g^{t+1})^2,\bz
  )-1]^{2}+(B-1)f_{a_{2|1}}((\Sigma_g^{t+1})^2,\bz )^{2}\right),
\label{eq:SE_POWA_E}\\
\text{with}\quad \Sigma_g^{t+1}(\{E_{g'}^t\}) &= 
  \left[\frac{B\alpha c_g^2}{\ln(B)({1/{{\rm snr}}} + 1/G\sum_{g'}^{G}
  c_{g'}^2E_{g'}^t)}\right]^{-1/2},
\label{eq:SE_POWA_S}
\end{align}
and where the $f_{a_{1|1}},f_{a_{2|1}}$ functions are again given by (\ref{eq_fa1fun}), (\ref{eq_fa2fun}).
\section{Results of the replica analysis}
\label{sec:replica}
\begin{figure}[!t]
\centering
\includegraphics[width=0.45\textwidth]{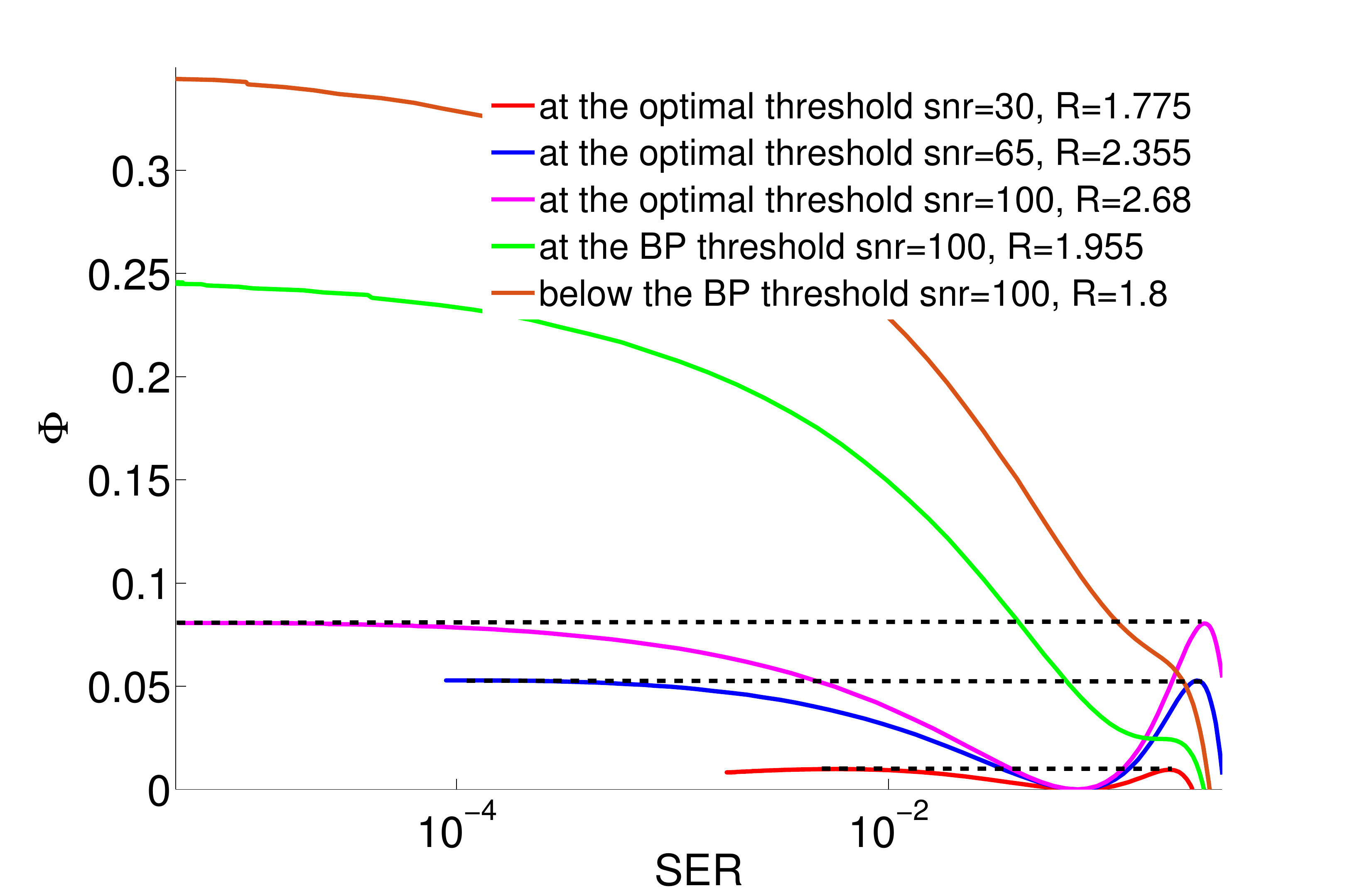}
\includegraphics[width=0.4\textwidth]{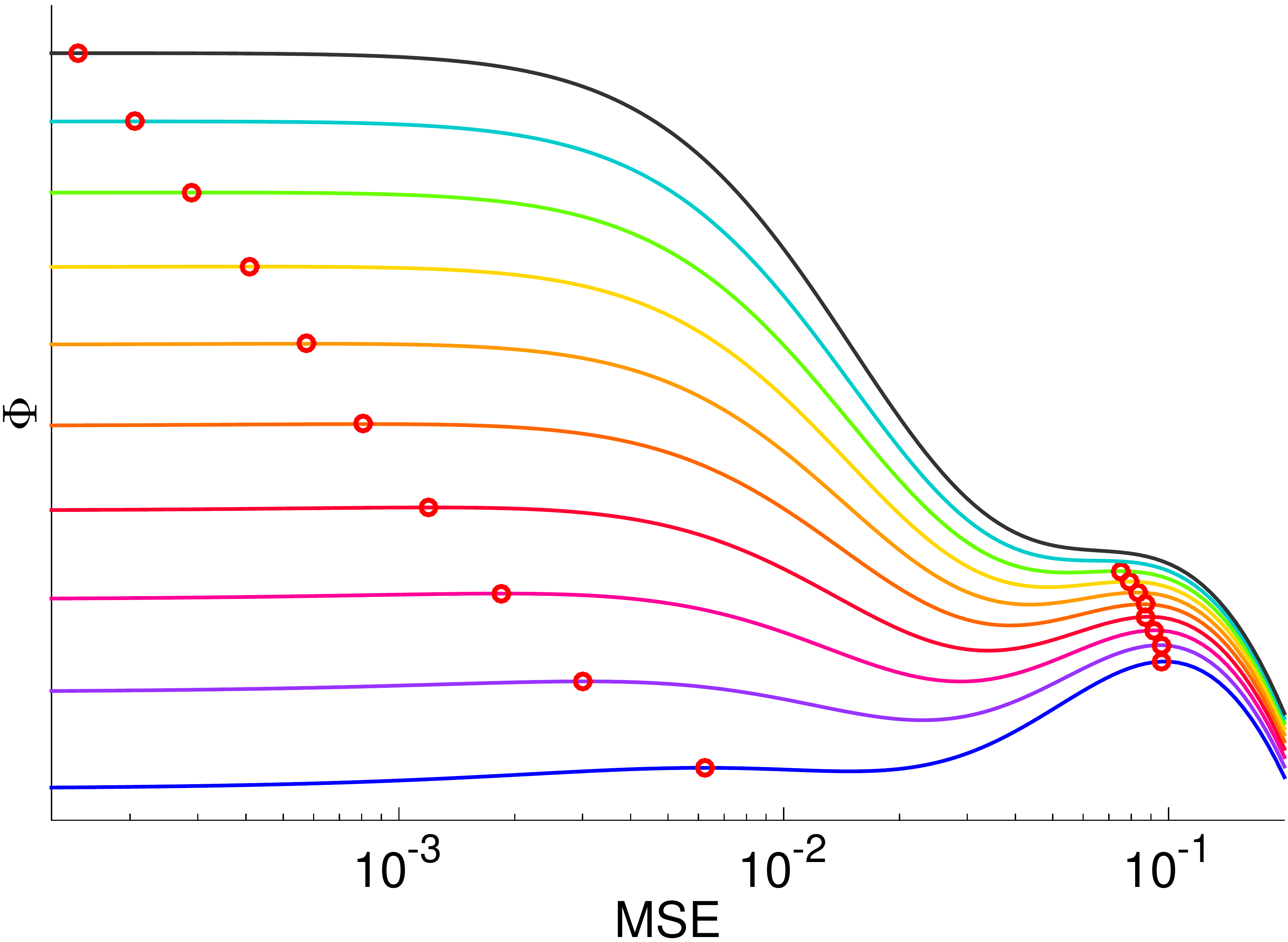}
\caption{\textbf{Left:} The free entropy (or potential) $\Phi({\rm SER})$ for $B=2$, different rates and ${{\rm snr}}$. The
maxima of the curves correspond to the ${\rm{SER}}$ which are fixed points of the state evolution recursion (\ref{eq_SE_SER}) for a given set of parameters $(R,B,{{\rm snr}})$. The global maximum is the \emph{equilibrium state}, corresponding to the optimal ${\rm SER}$. The curves are obtained by numerical integration of (\ref{eq_freeEnt2}). The optimal threshold $R_{\rm opt}(B,{{\rm snr}})$ is the rate where the high and low error maxima have same height, see pink, blue and red curves. The BP threshold $R_{\rm BP}(B,{{\rm snr}})$ is the rate at which the metastable local maximum at high error (that blocks the convergence of AMP) appears, i.e the appearance of the first horizontal inflexion point when increasing $R$, see green curve. The plot illustrates how the potential gap between the two maxima at the optimal threshold increases with the ${{\rm snr}}$. $\textbf{snr}\boldsymbol{= 100}\textbf{:}$ Here for rates larger than $R>2.68$, the
optimal ${\rm SER}$ jumps discontinuously from a low value to a large $O(1)$ one (pink curve). This defines the maximum possible rate (to compare here to
$C=3.3291$) below which acceptable performance can be obtained with AMP combined with spatial coupling or non constant power allocation. For
$R<2.68$, the optimal ${\rm SER}$ is much lower (and decay with $R$). The AMP decoder Fig.~\ref{algo_AMP} asymptotic error can be thought as performing an ascent of this function. As long as the maximum is unique (i.e. for
$R<1.955$, see green curve), it achieves the predicted optimal performance without the need of spatial coupling nor non constant power allocation, as in the case of the brown curve. \textbf{Right:} We plot the potential for $R=1.8, B=2, {\rm snr} \in \{28, 30, 32, \ldots, 46\}$ (from blue to black) and observe the displacement of the maxima, illustrating how that the error floor decays with increasing ${\rm snr}$.}
\label{fig_freeEnt}
\end{figure}
\subsection{The replica symmetric potential of sparse superposition codes with constant power allocation}
The replica analysis is an heuristic asymptotic $L\to\infty$ and \emph{static} statistical analysis (as opposed to the \emph{dynamical} state evolution analysis). It allows to compute the so-called replica symmetric free entropy $\Phi_B(E)$ (\ref{eq_freeEnt2}), a potential function of the ${\rm{MSE}}$. This potential is related to the mutual information (per section) $i(\by;\bx)$ of model \eqref{eq_yfx} between the random received corrupted codeword and the radom transmitted message through 
\begin{equation} \label{eq:true_mutual_info}
i(\by;\bx) \defeq \frac{1}{L}\mathbb{E}_{\bF, \bx, \bxi}\Big[\ln\Big( \frac {P(\by|\bx)}{P(\by)}\Big)\Big] = -\frac{\alpha B}2 + \max_{E\ge 0} \Phi_B(E).
\end{equation}
This potential contains all the information about the location of the information theoretic and algorithmic transition (blocking the decoder if $R>R_{\rm BP}$ and no spatial coupling is employed) of the problem, the MMSE performance or the attainable asymptotic ${\rm{MSE}}$ of AMP \cite{barbier2016mutual}. Indeed, AMP is deeply linked to $\Phi_B(E)$: recall that the extrema of this potential \eqref{eq_freeEnt2} match the fixed points of state evolution \eqref{eq_SE_MSE}, \eqref{eq_SE_var}. 

The replica method used to derive this potential has been developed in the context of statistical physics of disordered systems in order to compute averages with respect to some source of quenched disorder of physical observables of the system, the ${\rm{MSE}}$ and ${\rm{SER}}$ in the present case. The method has then be extended to information theoretical problems \cite{tanaka2002statistical,nishimori2001statistical} due to the close connections between the physics of spin glasses and communications problems \cite{MezardMontanari09,nishimori2001statistical}, where the sources of quenched disorder to average over are the noise, coding matrix and the transmitted message realizations. See the recent rigorous results on the validity of the replica approach for linear estimation \cite{barbier2016mutual,reeves2016replica}.

The expression of the potential for sparse superposition codes at fixed section size $B$ with constant power allocation is
\begin{align}
&\Phi_B(E) = -\frac{\ln(B)}{2R\ln(2)} \left(\ln({1/{{\rm snr}}} + E) + \frac{1 - E}{{1/{{\rm snr}}} + E} \right) + \int \mathcal{D}\bz \ln\left( e^{\frac{\ln(B)}{2\Sigma(E)^2} + \frac{\sqrt{\ln(B)}z_1}{\Sigma(E)} } + \sum_{i = 2}^B e^{-\frac{\ln(B)}{2\Sigma(E)^2} + \frac{\sqrt{\ln(B)}z_i}{\Sigma(E)} }\right), \label{eq_freeEnt2}
\end{align}
where
\begin{equation} \label{eqSIGMA2}
\Sigma(E)^2 \defeq R\ln(2)({1/{{\rm snr}}} + E). 
\end{equation}
This potential depends on $E$ that is interpreted as a mean-square error per section, but it can be implicitely expressed as a function of the section error rate thanks to the one-to-one correspondance (\ref{eq_SE_SER}) between the ${\rm{MSE}}$ and ${\rm{SER}}$.

We plot (\ref{eq_freeEnt2}) in the $(R,B=2,{{\rm snr}}=100)$ case on Fig.~\ref{fig_freeEnt}. The AMP algorithm follows a dynamics that can be interpreted as a gradient ascent of this potential, which starts the ascent from an high error (i.e from a random guess of the message). The brown curve of the left figure thus corresponds to an ``easy'' case as the global maximum is unique and correponds to a low error, which is the optimal ${\rm SER}$, i.e associated to the MMSE through \eqref{eq_SE_SER} (the optimal ${\rm SER}$ is the error associated with the \emph{global} maximum of this potential). There $R<R_{\rm BP}$ and AMP is asymptotically optimal in the sense that it performs MMSE estimation (and thus leads to optimal ${\rm SER}$). The green curve corresponds to the BP threshold, that is the appearance of the first horizontal inflexion point in $\Phi_B(E)$ when increasing $R$. This threshold marks the appearance of the \emph{hard phase}, where the inference is typically hard and AMP cannot decode (without spatial coupling or power allocation). The local maximum at high error blocks the convergence of AMP, preventing it to reach the MMSE. Still, the MMSE corresponds to an higher free entropy meaning that it has an exponentially larger statistical weight (and thus corresponds to the true \emph{equilibrium state} in physics terms). The problem is to reach it under AMP decoding despite the precense of the high error local maximum. Spatial coupling has been specifically designed to achieve this goal. The pink curve (or blue and red ones for other ${\rm snr}$) marks the appearance of the \emph{impossible inference phase} defined as the rate where the low and high error maxima have same free entropy. At this threshold, the local and global maxima swith roles which corresponds to a jump discontinuity of the MMSE and optimal ${\rm SER}$ from low values to high ones (one speaks in this case of a \emph{first order phase transition}). In this phase, even optimal MMSE estimation leads to a wrong decoding. In constrast, if $R<R_{\rm opt}$, the AMP algorithm combined with spatial coupling or well designed power allocation is theoretically able to decode and as we will see, $R_{\rm pot}$ tends to the Shannon capacity as $B\to\infty$, see sec.\ref{sec:AMP-PowA} and sec.\ref{sec:numerics}.
\subsection{Results from the replica analysis for sparse superposition codes with constant power allocation and finite section size}
\begin{figure}[!t]
\centering
\includegraphics[width=0.34\textwidth]{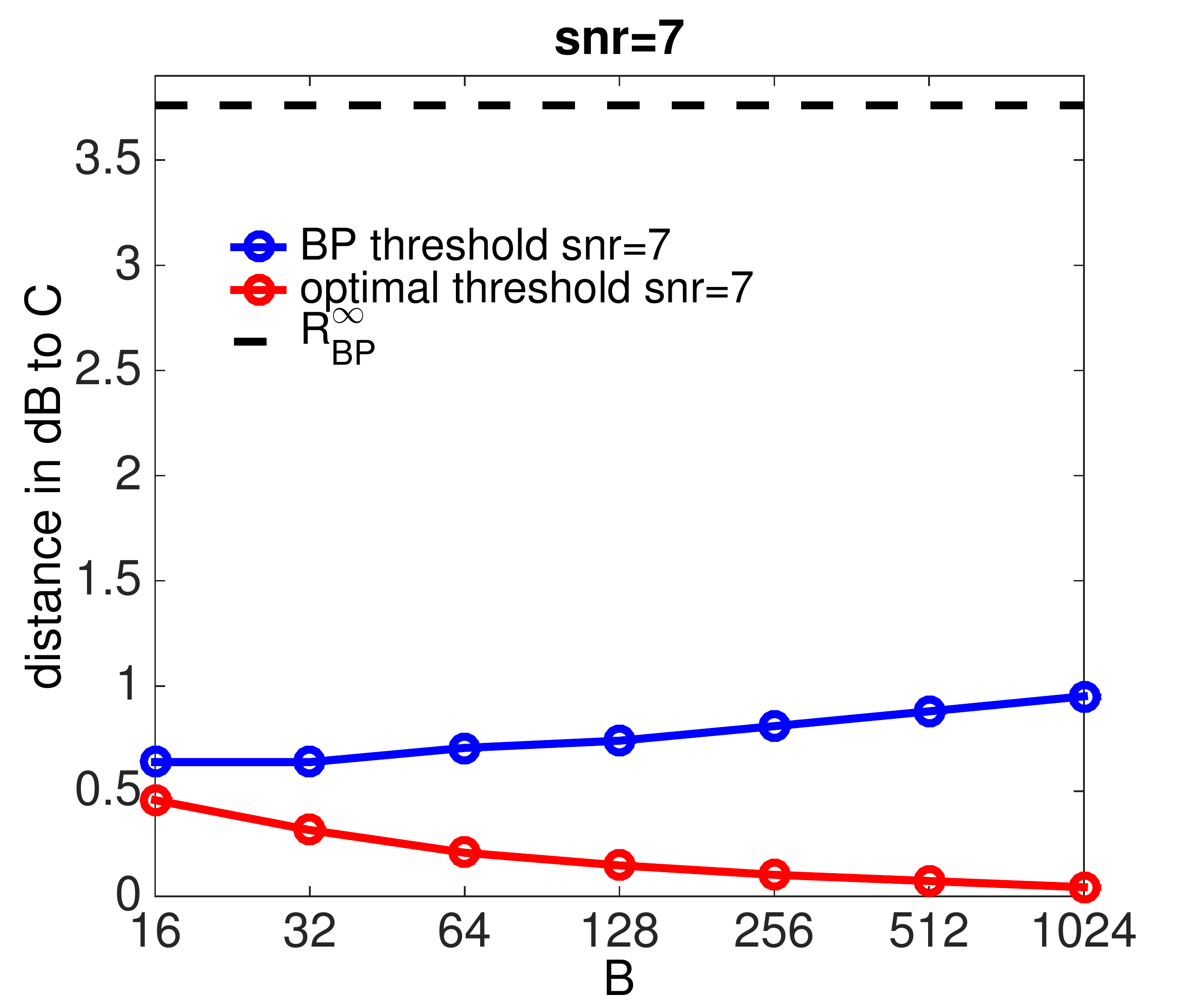}
\includegraphics[width=0.32\textwidth, trim=40 0 0 0, clip=true]{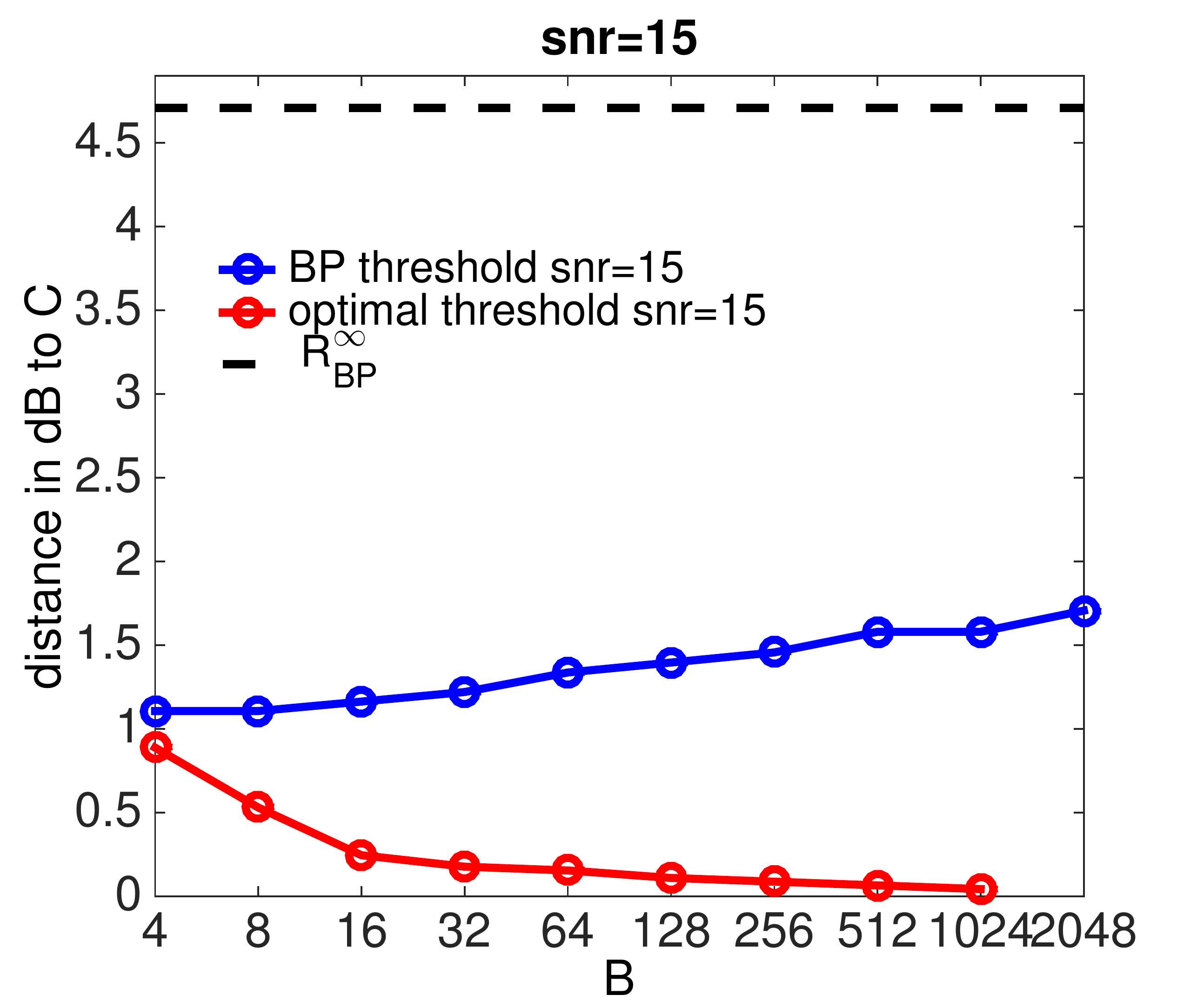}
\includegraphics[width=0.31\textwidth, trim=65 0 0 0, clip=true]{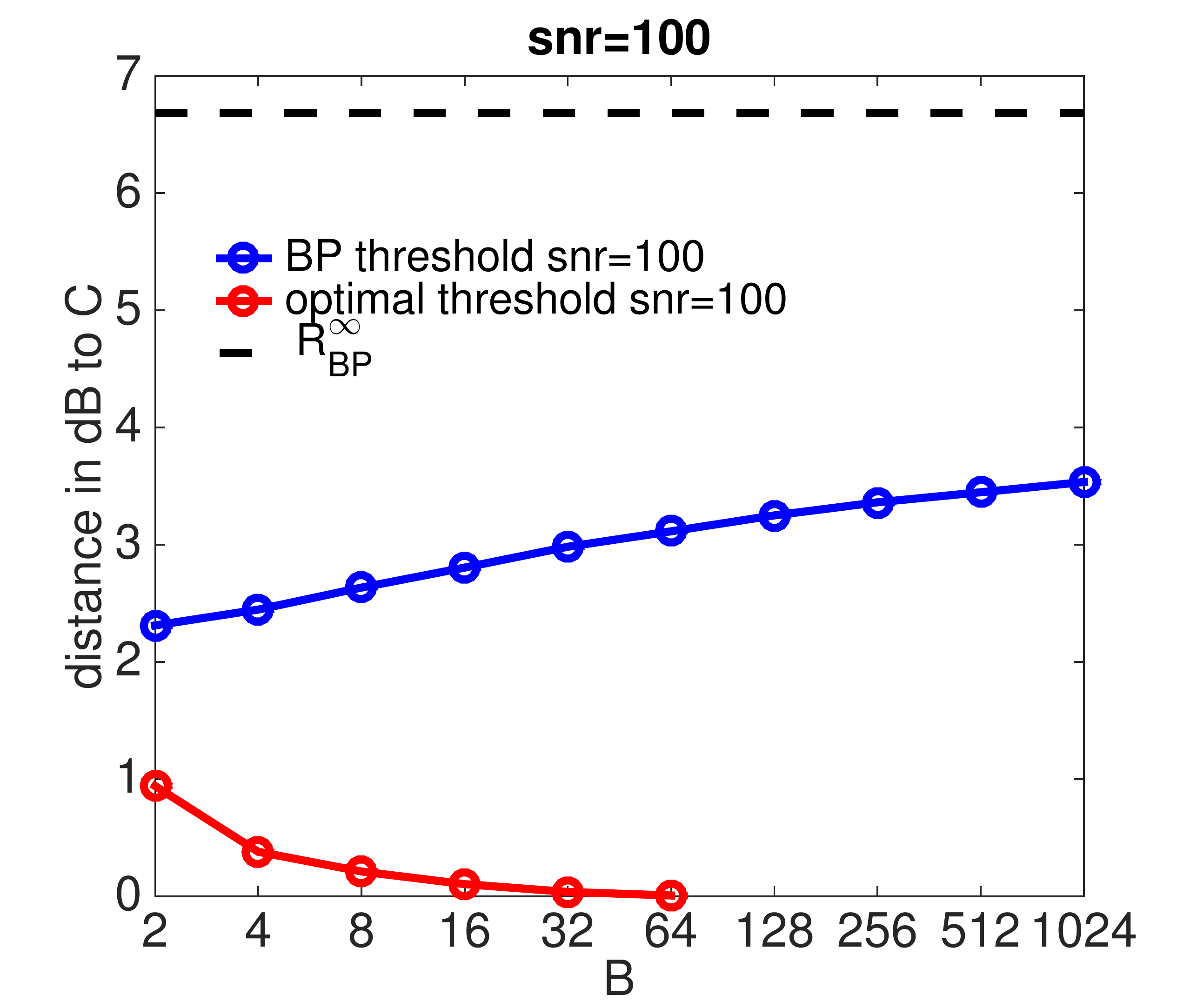}
\caption{Phase diagram of sparse superposition codes without power allocation for different ${\rm snr}$, where the $x$ axis is the section size $B$, the $y$ axis is the distance to the capacity $C$ in dB. The transitions are computed from the potential (\ref{eq_freeEnt2}) where the integral is computed by monte carlo. The blue and red curves are respectively the algorithmic BP and optimal thresholds. The black dashed line is the asymptotic value (in $B$) of the BP threshold $R_{\rm BP}^{\infty}$ (\ref{BPcrit}).}
\label{fig_diagsDist}
\end{figure}
From this analysis, we can extract the phase diagram of the superposition codes scheme. Fig.~\ref{fig_diagsDist} shows phase diagrams for different ${\rm snr}$ values, where the $x$ axis is the section size $B$ while the $y$ axis is the distance to the Shannon capacity in dB. The blue curve is the BP threshold extracted from the potential (\ref{eq_freeEnt2}) which marks the end of optimality of the AMP decoder without spatial coupling or proper power allocation. The red curve is the optimal threshold: the highest rate until decoding is information theoretically possible\footnote{also formally defined as the first non-analiticity point of the asymptotic $L\to \infty$ mutual information \eqref{eq:true_mutual_info} when increasing $R$, see \cite{barbier2016mutual}.}, also extracted from (\ref{eq_freeEnt2}). The black dashed curve is the asymptotic $B\to \infty$ BP threshold (\ref{BPcrit}), which derivation is done in sec.~\ref{subsec_largeBrep}.

\begin{figure}[!t]
\centering % 0.47
\includegraphics[width=0.38\textwidth]{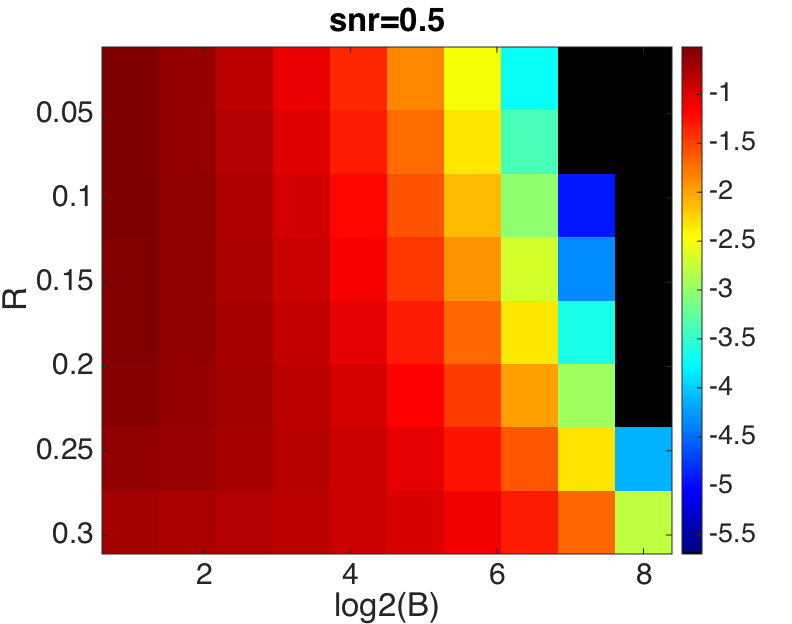}
\includegraphics[width=0.38\textwidth]{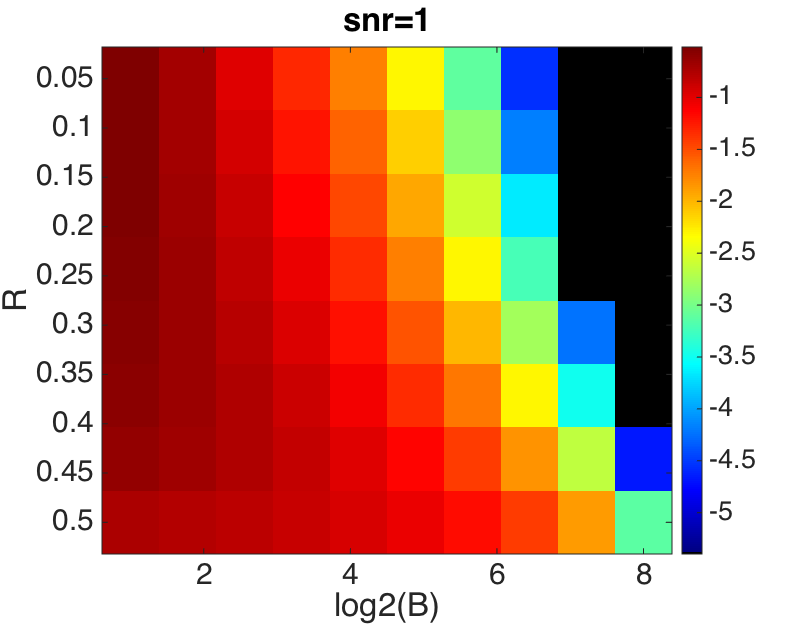}
\includegraphics[width=0.38\textwidth]{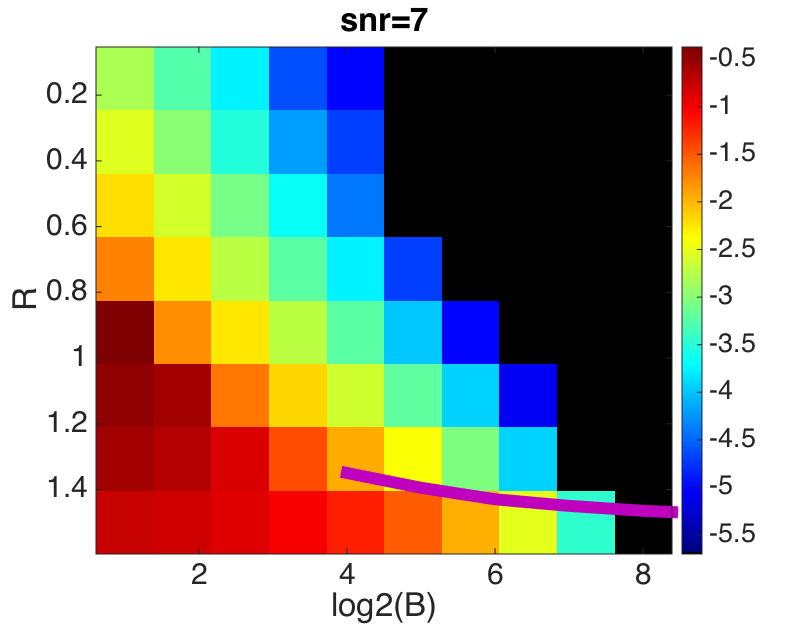}
\includegraphics[width=0.38\textwidth]{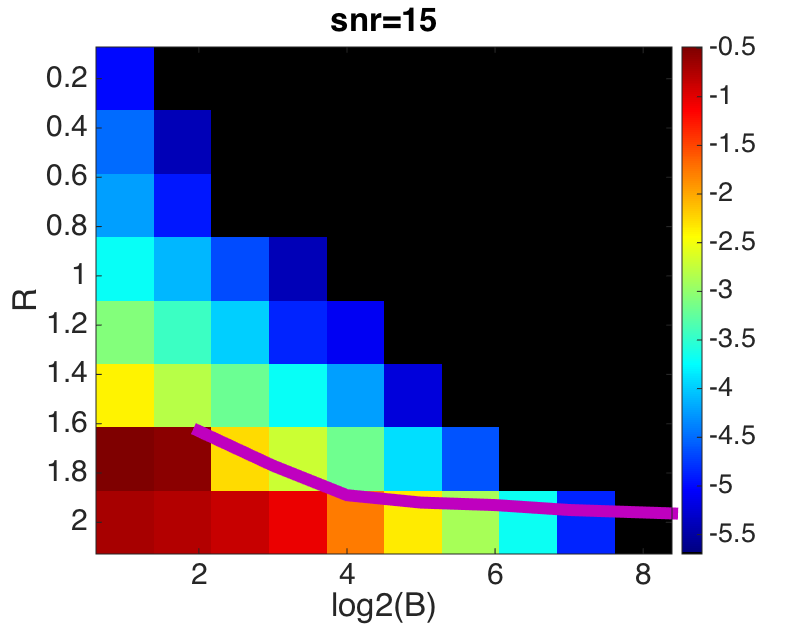}
\caption{The color code is for the logarithm in base 10 of the section error rate corresponding to the maximum of the replica potential (\ref{eq_freeEnt2}) that has lowest ${\rm SER}$, in the $(R,B)$ plane and for various ${\rm snr}$. The values are obtained from the state evolution recursion (\ref{eq_SE_SER}) starting from the solution (i.e. with an initial error equal to $0$). The recursions (\ref{eq_SE_MSE}), (\ref{eq_SE_SER}) are computed by monte carlo with a sample size of $5B\times10^5$. The black squares correspond to points where the computed value is ${\rm SER}=0$ which actually means a value that is lower to $(5\times10^5)^{-1}$ with high probability. The solid pink curve on the two lower plots correspond to the optimal rates $R_{\rm opt}(B,{{\rm snr}})$ as in Fig.~\ref{fig_diagsDist}. In the two upper plots that correspond to high noise regimes, there is no transition (the potential has a unique maximum and the AMP decoder is thus always Bayes optimal, at least for these manageable section sizes $B$) and the optimal ${\rm SER}$ is a smooth increasing function of the rate $R$ at fixed $B$; a decreasing function of $B$ at fixed $R$. The ${\rm SER}$ in the two lower plots, corresponding to low noise regimes, match the optimal ${\rm SER}$ as long as $R < R_{\rm opt}(B,{{\rm snr}})$. For higher rates, the maximum of the potential corresponding to the plotted ${\rm SER}$ is not the global maximum and thus cannot be reached, even with spatial coupling (that works asymptotically until the optimal rate). For $B$ smaller than $4$ (resp. $2$) on the ${\rm snr}=7$ (resp. ${\rm snr}=15$) plot, there is no sharp transition and the represented ${\rm SER}$ value is the optimal one, that can be reached by AMP without spatial coupling, as in the high noise regime.}\label{fig_phaseDiagsFinalSER}
\end{figure}

A first observation is that the BP threshold is converging quite slowly to its asymptotic $B\to\infty$ value $R_{\rm BP}^{\infty}$ \eqref{BPcrit} (computed in the next section) if compared to the convergence rate of the optimal threshold to the capacity. We also note that the section size where start the transitions, and thus marks the appearance of the hard phase where the AMP decoder without spatial coupling is not Bayes optimal anymore, increases as the ${\rm snr}$ decreases. When the ${\rm snr}$ is not too large, we see that the optimal and BP thresholds almost coincide at small $B$ values, such as for $B=16$ at ${\rm snr}=7$ and $B=4$ for ${\rm snr}=15$. Below this section size value, there are no more sharp phase transitions as only one maximum exists in the potential (\ref{eq_freeEnt2}) and the AMP dedoder is optimal at any rate even without spatial coupling. In this regime, the ${\rm SER}$ increases continuously with the rate. As the ${\rm snr}$ increases, the curves split sooner until they remain different for all $ B$ such as in the ${\rm snr}=100$ case. See Fig.~\ref{fig_phaseDiagsFinalSER} and Fig.~\ref{fig_optSER} for more details on the achievable values of the ${\rm SER}$. A second observation is that despite the optimal performance of the code improves and approaches capacity with increasing $B$, instead the AMP perfomance monotonously reduces (in terms of possible communication rate). But as we will see with Fig.~\ref{fig_finiteSizeSeeded} and Fig.~\ref{fig_phaseDiagSC}, spatial coupling allows to enter the hard phase (between the two transitions), making AMP to improve as well with increasing $B$.

\begin{figure}[!t]
\centering
\includegraphics[width=0.34\textwidth]{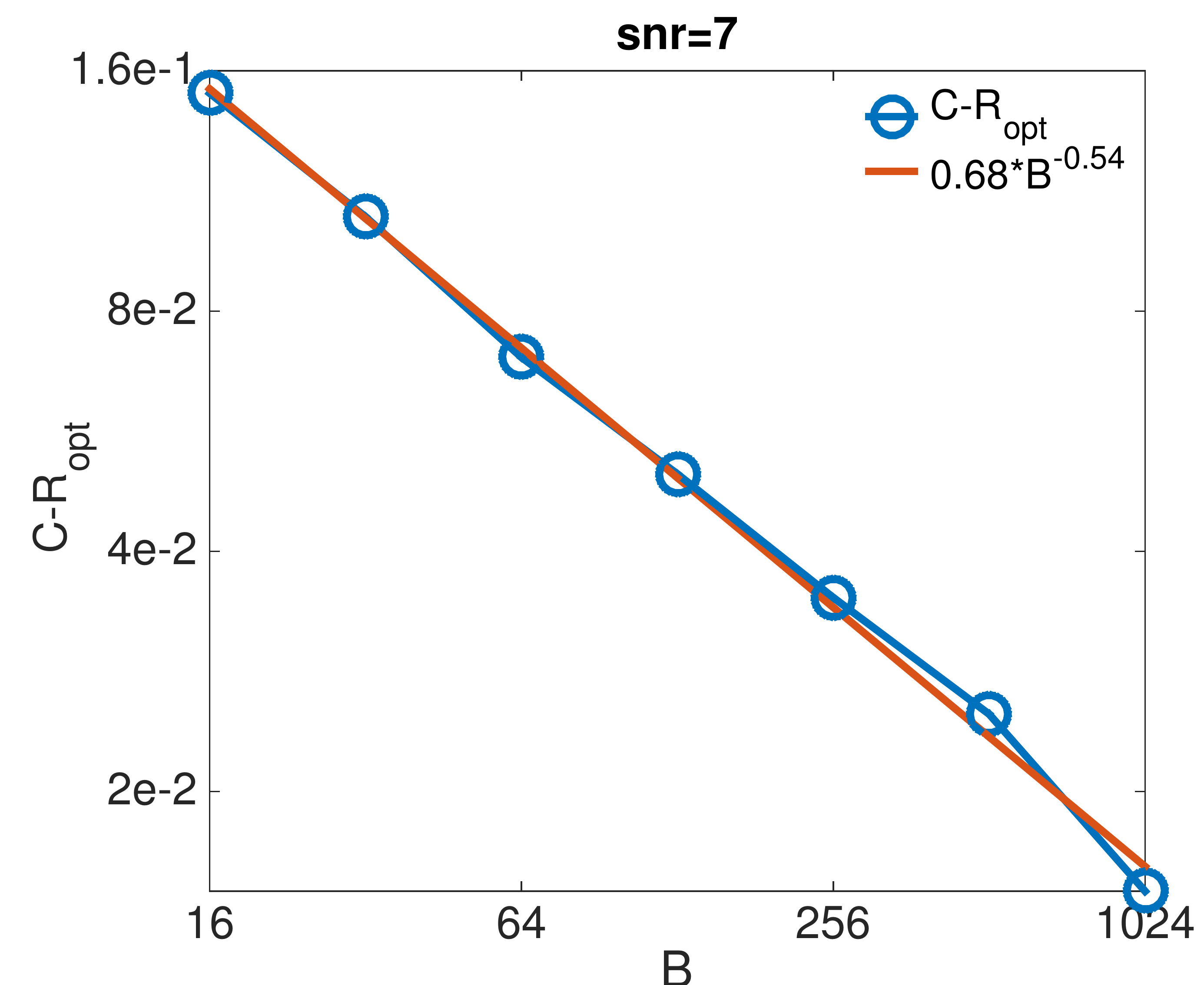}
\includegraphics[width=0.32\textwidth, trim=50 0 0 0, clip=true]{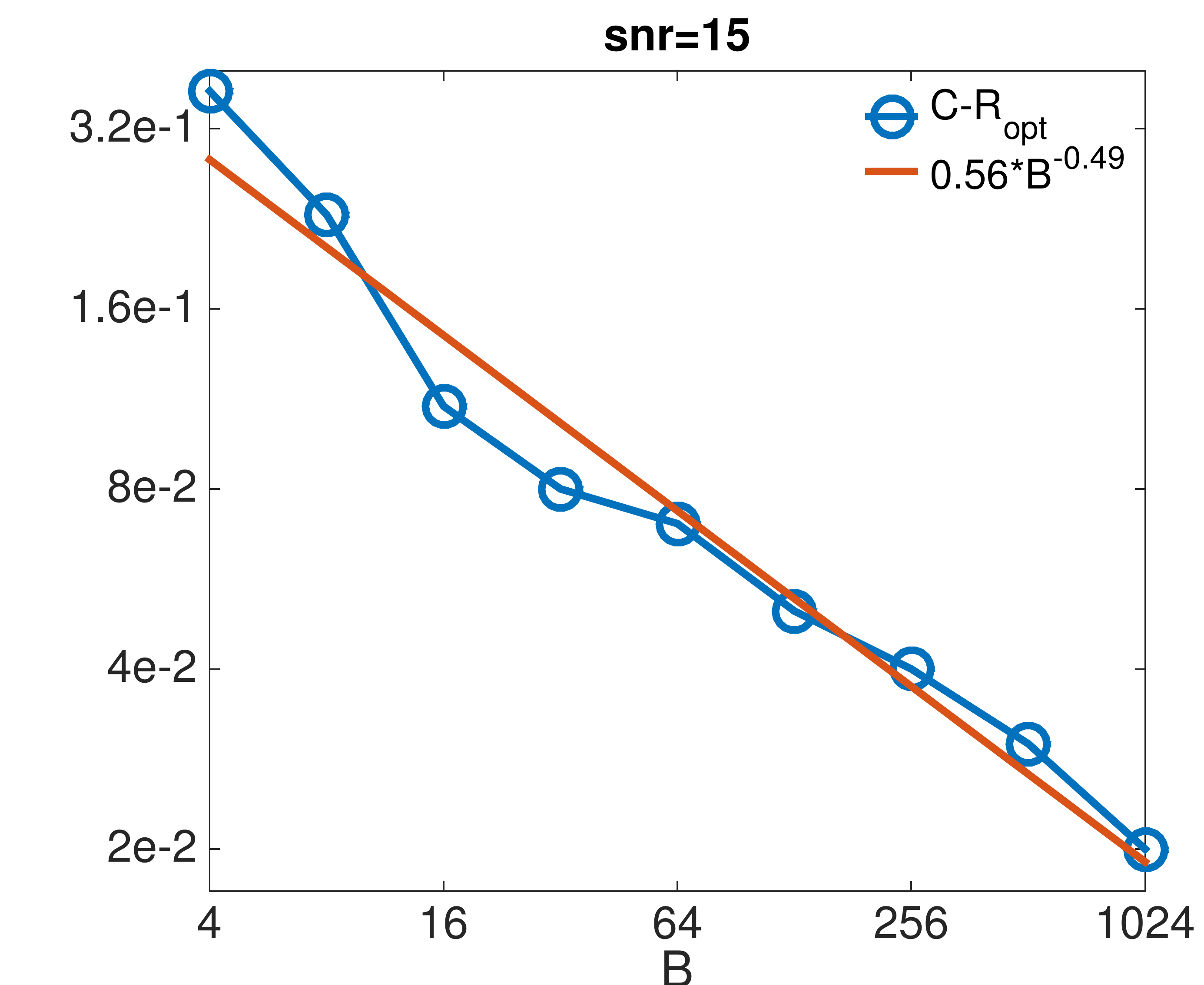}
\includegraphics[width=0.32\textwidth, trim=50 0 0 0, clip=true]{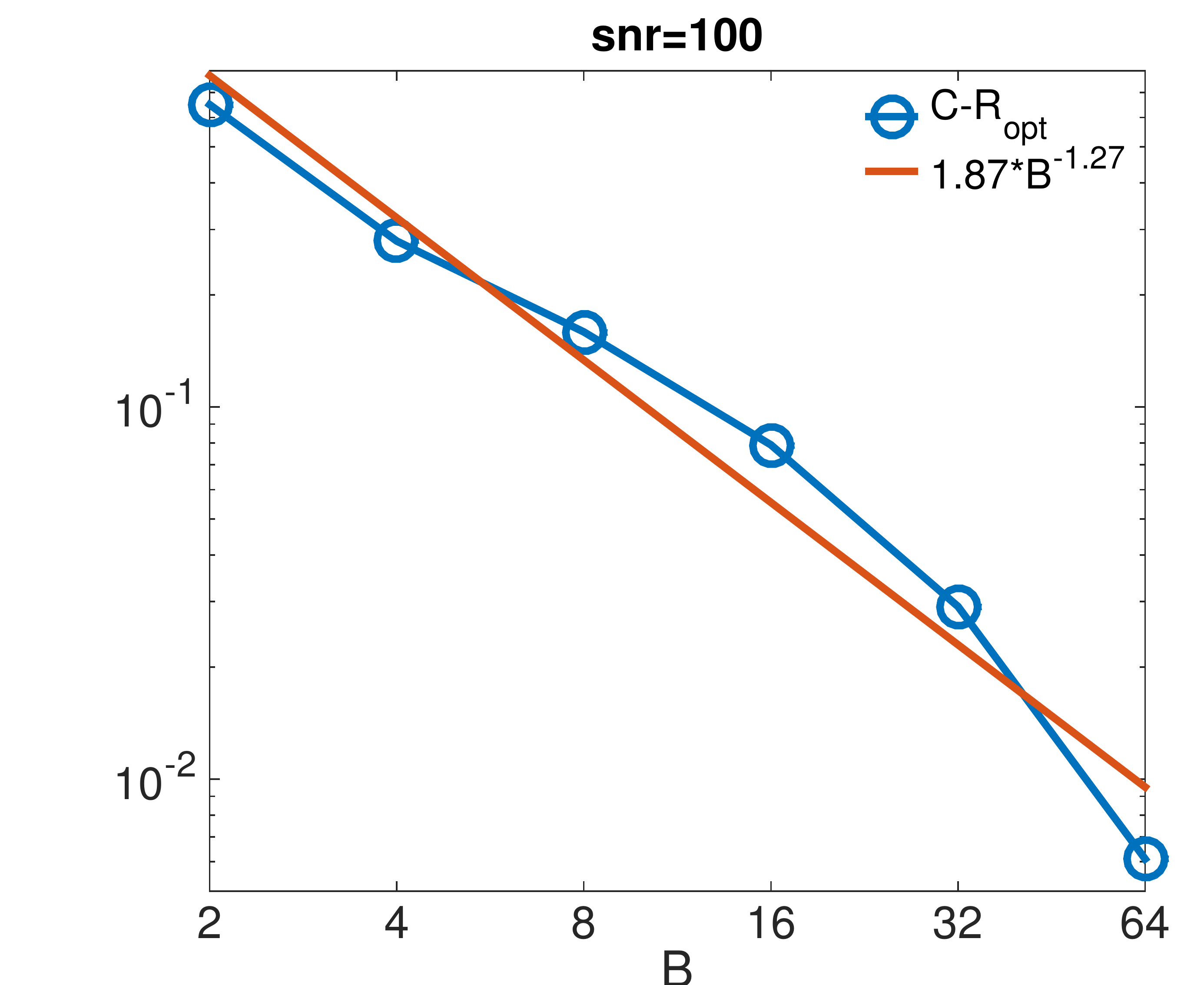}
\includegraphics[width=0.34\textwidth]{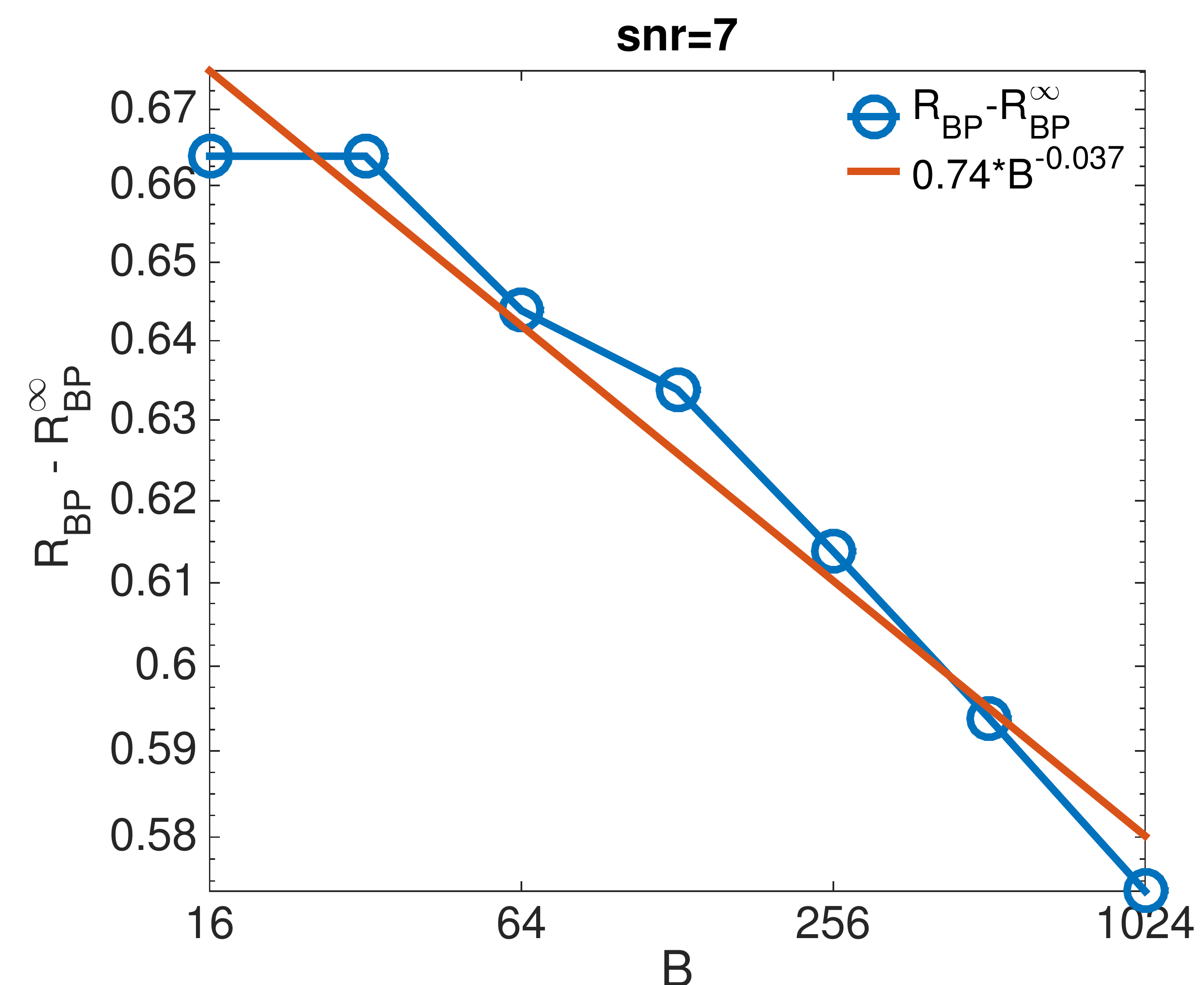} 
\includegraphics[width=0.32\textwidth, trim=45 0 0 0, clip=true]{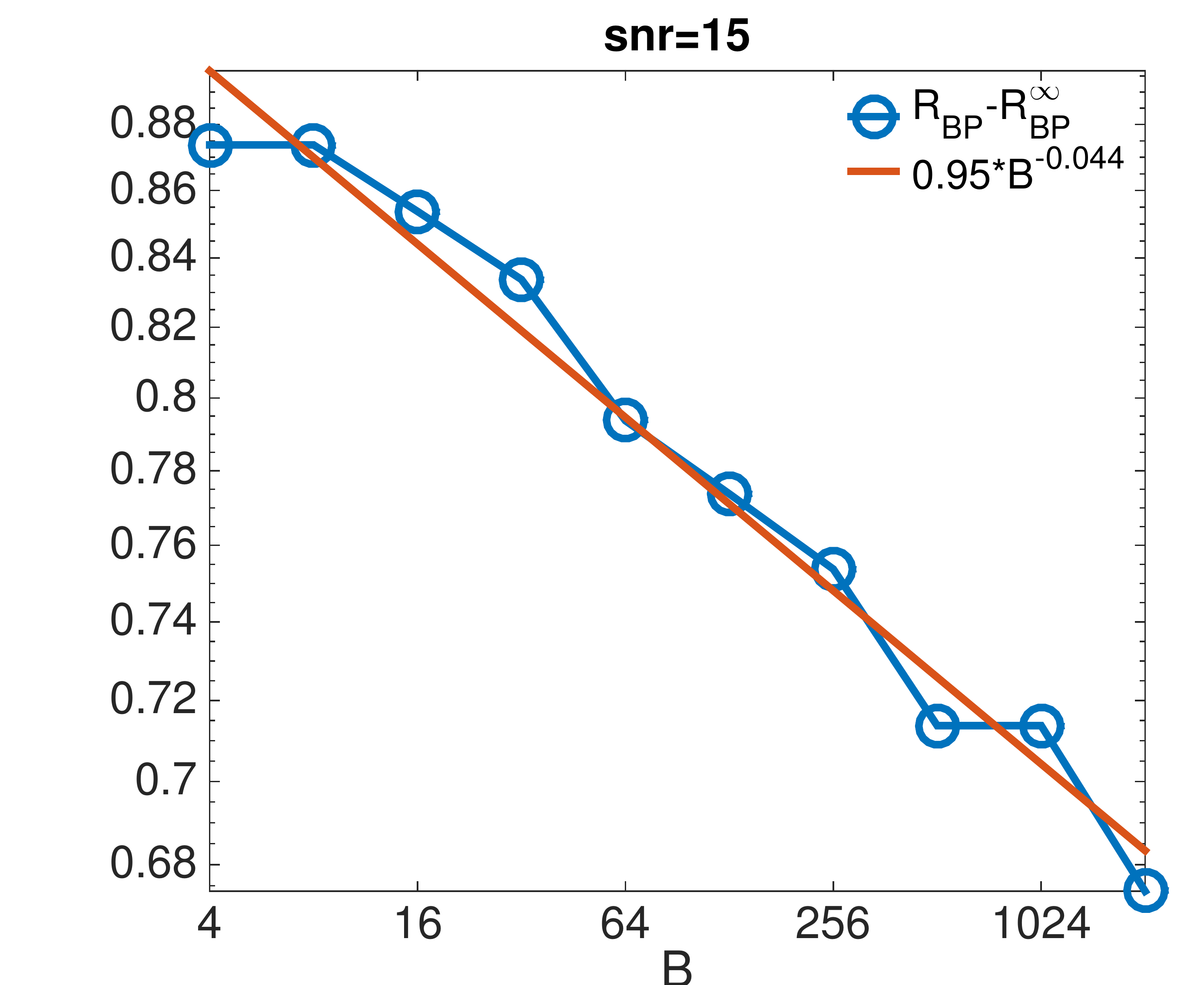}
\includegraphics[width=0.32\textwidth, trim=45 0 0 0, clip=true]{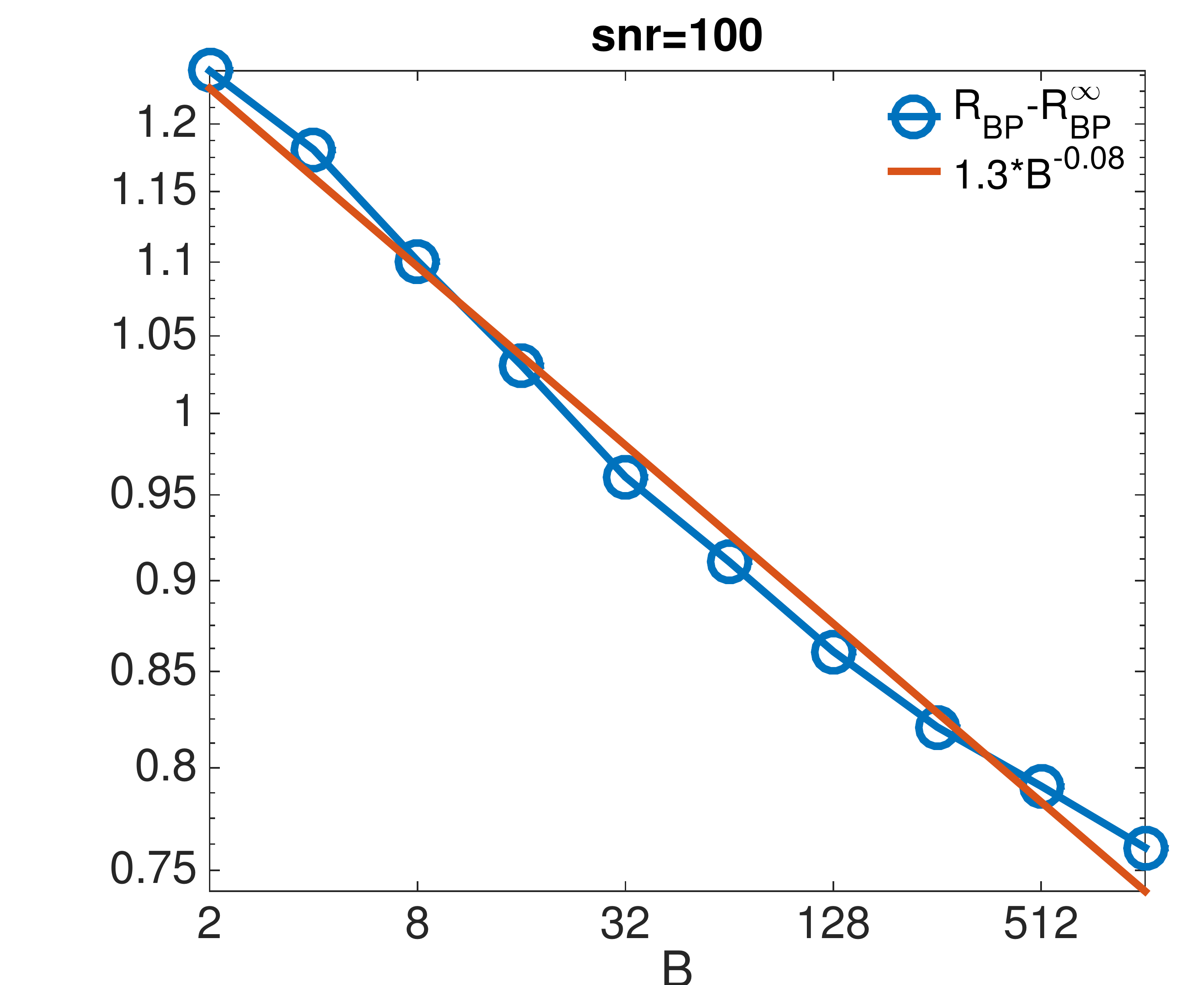}
\caption{These plots show how $R_{\rm BP}$ and $R_{{\rm opt}}$ change when
  $B$ increases according to the replica analysis. The blue points
  are computed from the potential
  (\ref{eq_freeEnt2}). \textbf{Upper plots}: These pictures show how
  fast with $B$ the optimal threshold $R_{{\rm opt}}$ is approaching the capacity for different ${{\rm snr}}$. We plot
  the difference $C-R_{{\rm opt}}$ as a function of $B$ in double
  logarithmic scale. The lines are guides for the eyes, and should not
  be taken as serious fits. They strongly suggest, however, a power
  law behavior. \textbf{Lower plots}: We did the same for
  $R_{\rm BP}$ by plotting $R_{\rm BP}-R_{\rm BP}^{\infty}$ as a function of $B$ in double logarithmic scale, where $R_{\rm BP}^{\infty}({{\rm snr}})$ is the asymptotic BP threshold (\ref{BPcrit}). In all cases, we observe a behavior quite well predicted by a power law. The low values of the exponents might also suggest a very slow logarithmic behavior for the convergence to $R_{\rm BP}^{\infty}$.}\label{fig_CminRscaling}
\end{figure}

%here
Fig.~\ref{fig_CminRscaling} gives details on the rate of convergence of the thresholds to their asymptotic value, and it seems it can be well approximated by a power law in both cases. On Fig.~\ref{fig_CminRscaling} we show the differences between the finite $B$ transitions of Fig.~\ref{fig_diagsDist} and their asymptotic (in $B$) values which are the capacity $C$ for the optimal threshold (as shown in the previous subsection) and $B_{BP}^{\infty}$ (\ref{BPcrit}) for the BP threshold. It appears that the scaling exponents increase in amplitude as the ${\rm snr}$ increases: the larger the ${\rm snr}$, the faster the convergence to asymptotics values is. 

Fig.~\ref{fig_optSER} represents how the optimal ${\rm SER}$, the ${\rm SER}$ corresponding to the MMSE, evolves with the section size $B$ at fixed rate and ${\rm snr}$ (left plot) and then as a function of the ${\rm snr}$ at fixed rate and $B$ (right plot). In both cases, the curves seem to be well approximated by power laws with exponent given on the plots. The points are extracted from the potential (\ref{eq_freeEnt2}).

Fig.~\ref{fig_phaseDiagsFinalSER} quantifies the optimal performance of the code, obtained from the state evolution analysis. We plot the base 10 logarithm of the ${\rm SER}$ corresponding to the maximum of the potential (\ref{eq_freeEnt2}) that has lower error, this as a function of $R$ and $B$ (again the state evolution and replica analysies are equivalent as shown in appendix~\ref{subsec:repIsSE}). For high noise regimes, the plotted ${\rm SER}$ is always attainable by AMP without the need of spatial coupling as there is no sharp phase transition (the potential has a single maximum). For lower noise regimes, the plotted ${\rm SER}$ matches the optimal one as long as $R<R_{{\rm opt}}$ (pink curves). When there is no transition (before the pink curves start), the ${\rm SER}$ is the optimal one too (here also, the potential has a single maximum). Fig.~\ref{fig_optSER} left plot is a cut in the ${\rm snr}=15$ plot. The information brought by the replica analysis, not explicitly included in the state evolution analysis, is the identification of the phase in which the system is (easy/hard/impossible inference) for a given set of parameters $(R,B,{{\rm snr}})$.
\subsection{Large section limit for sparse superposition codes with constant power allocation}
\label{subsec_largeBrep}
In order to access the $B\to\infty$ limit of the
potential and thus the asymptotic performance of the code, we need to compute the asymptotic value $I$ of the integral $I_B$ that appears in the potential (\ref{eq_freeEnt2}):
\begin{align}
I\defeq \lim_{B\to\infty}I_B = \lim_{B\to\infty} \int_{\mathbb{R}^B} \mathcal{D}\bz \ln\left( e^{\frac{\ln(B)}{2\Sigma^2} + \frac{\sqrt{\ln(B)}z_1}{\Sigma} } + \sum_{i = 2}^B e^{-\frac{\ln(B)}{2\Sigma^2} + \frac{\sqrt{\ln(B)}z_i}{\Sigma} }\right). \label{eq_logKB}
\end{align}
Recall $\mathcal{D}\bz$ is an standardized Gaussian measure over the i.i.d $\{z_i\}$. We present here an heuristic computation based on an analogy with the so-called random energy model \cite{derrida1980random,MezardMontanari09} of statistical physics. An alternative heuristic derivation of the following results based instead on the replica method is given in appendix~\ref{subsec:largeB_rep}. This independent analysis brings the same results, strenghtening the claim of the exactness of the analysis despite not being rigorous.

\begin{figure}[t!]
\centering
\includegraphics[width=0.43\textwidth]{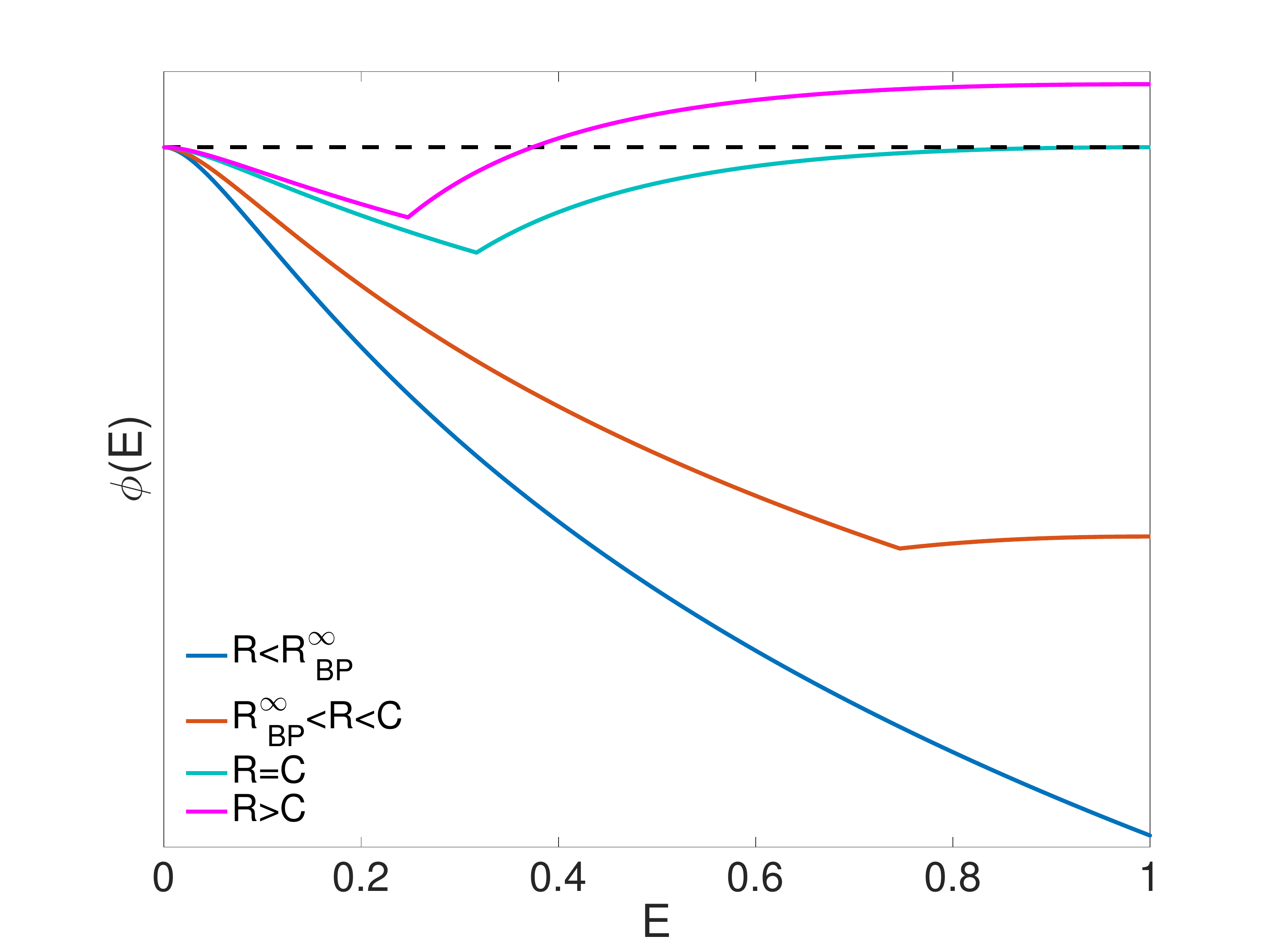}
\centering
\includegraphics[width=0.43\textwidth]{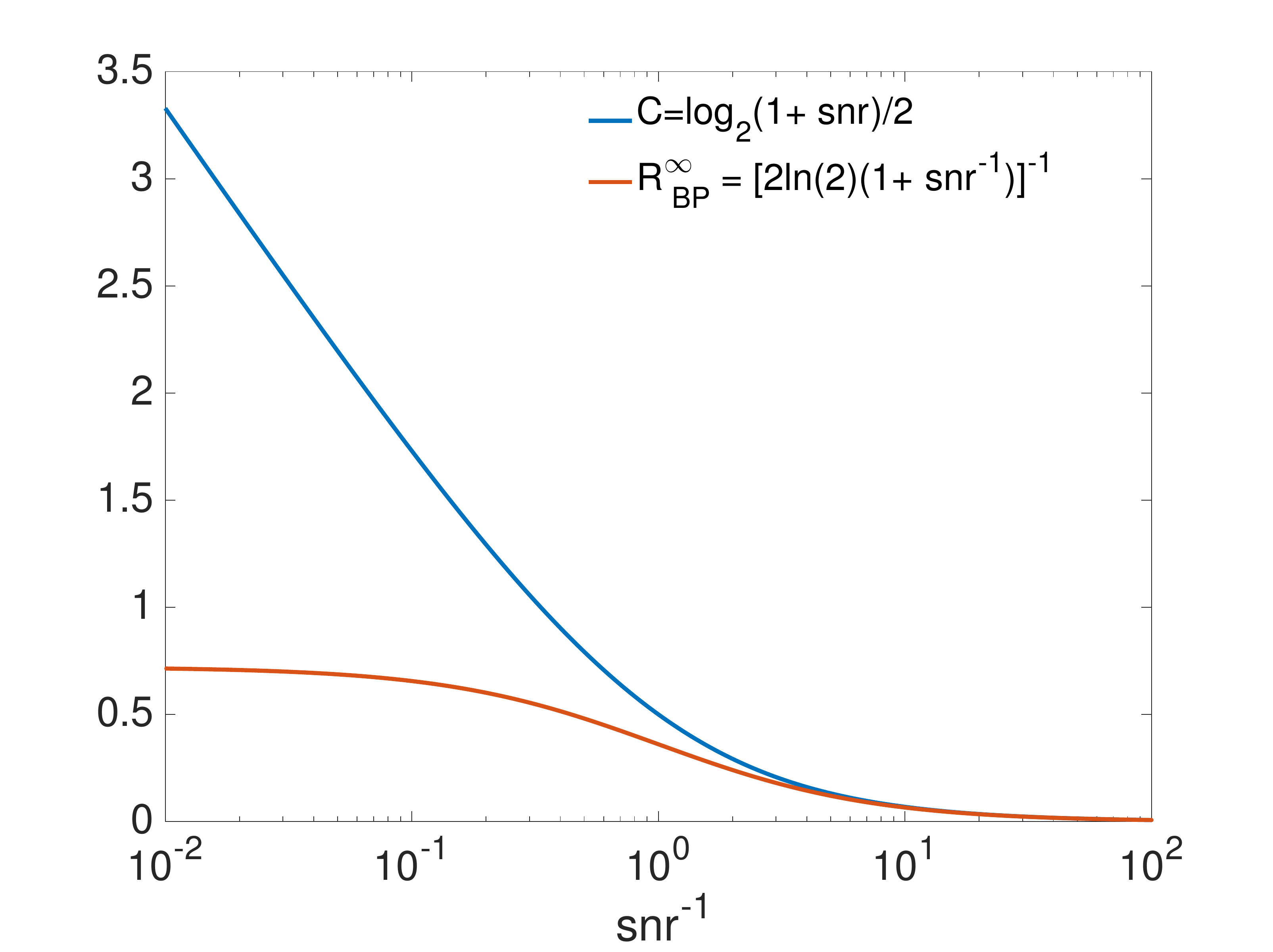}
\caption{\textbf{Left:} The large alphabet (or large section) limit of the potential, that is $\phi(E)$ \eqref{Phi_largeB}, with ${\rm snr}=10$. The potential is scaled such that $\phi(0)=0$. For $R$ below $R_{\rm BP}^{\infty}$, there is a unique maximum at $E=0$ while just above, this maximum coexists with a local one at $E=1$. At the optimal threshold of the code, that coincides with the Shannon capacity, the two maxima are equal. Then, for $R>C$ the maximum at $E=1$ becomes the global one, and thus decoding is impossible. \textbf{Right:} We plot the the Shannon capacity and the asymptotic BP threshold as a function of the ${\rm snr}^{-1}$ to illustrate the computational gap appearing between the low-complexity AMP decoding performance and the optimal performance. Note that at low ${\rm snr}$, the curves coincide and there is no gap anymore, thus AMP is optimal.}
\label{fig:largeB}
\end{figure}

We shall drop the dependency of $\Sigma$ \eqref{eqSIGMA2} in $E$ 
to avoid confusions. We adopt here the vocabulary of statistical physics \cite{MezardMontanari09}: this is formally a problem of computing the average of the logarithm of a partition function of a system with $B$ (disordered) states. Indeed, one can rewrite (\ref{eq_logKB}) as:
\begin{align}
I_B &= -\frac{\ln(B)}{2\Sigma^2} + \int \mathcal{D}\bz \ln\Big(
e^{\frac{\ln(B)}{\Sigma^2} + \frac{\sqrt{\ln(B)}z_1}{\Sigma} } +
  \sum_{i = 2}^B e^{\frac{\sqrt{\ln(B)}z_i}{\Sigma} }\Big) \\
&= -\frac{\ln(B)}{2\Sigma^2} +\int \mathcal{D}\bz \ln\Big( {\cal  Z}_1(z_1) + {\cal Z}_{2}(\{z_i:i\in\{2,\ldots,B\}\})\Big), \label{eq_32}
\end{align}
where
\begin{align}
&{\cal  Z}_1(z1)\defeq \exp\left( \ln(B)/\Sigma^2 +\sqrt{\ln(B)} z_1/\Sigma\right), \\
&{\cal Z}_2(\{z_i:i\in\{2,\ldots,B\}\})\defeq\sum_{i=2}^B \exp\left(\sqrt{\ln(B)} z_i/\Sigma\right). 
\end{align}
In fact ${\cal Z}_2$ is formally known as a random energy model
in the statistical physics literature
\cite{derrida1980random,MezardMontanari09}, a statistical physics
model where i.i.d energy levels are drawn from some given
distribution. This analogy can be further refined by writing the energy levels as $U_i=-\sqrt{\ln(B)}z_i$ and by denoting $\Sigma$ as the
temperature. In this case, a standard result
\cite{derrida1980random,MezardMontanari09,arous2005limit} is:
\begin{figure}[!t]
\centering
\includegraphics[width=0.35\textwidth]{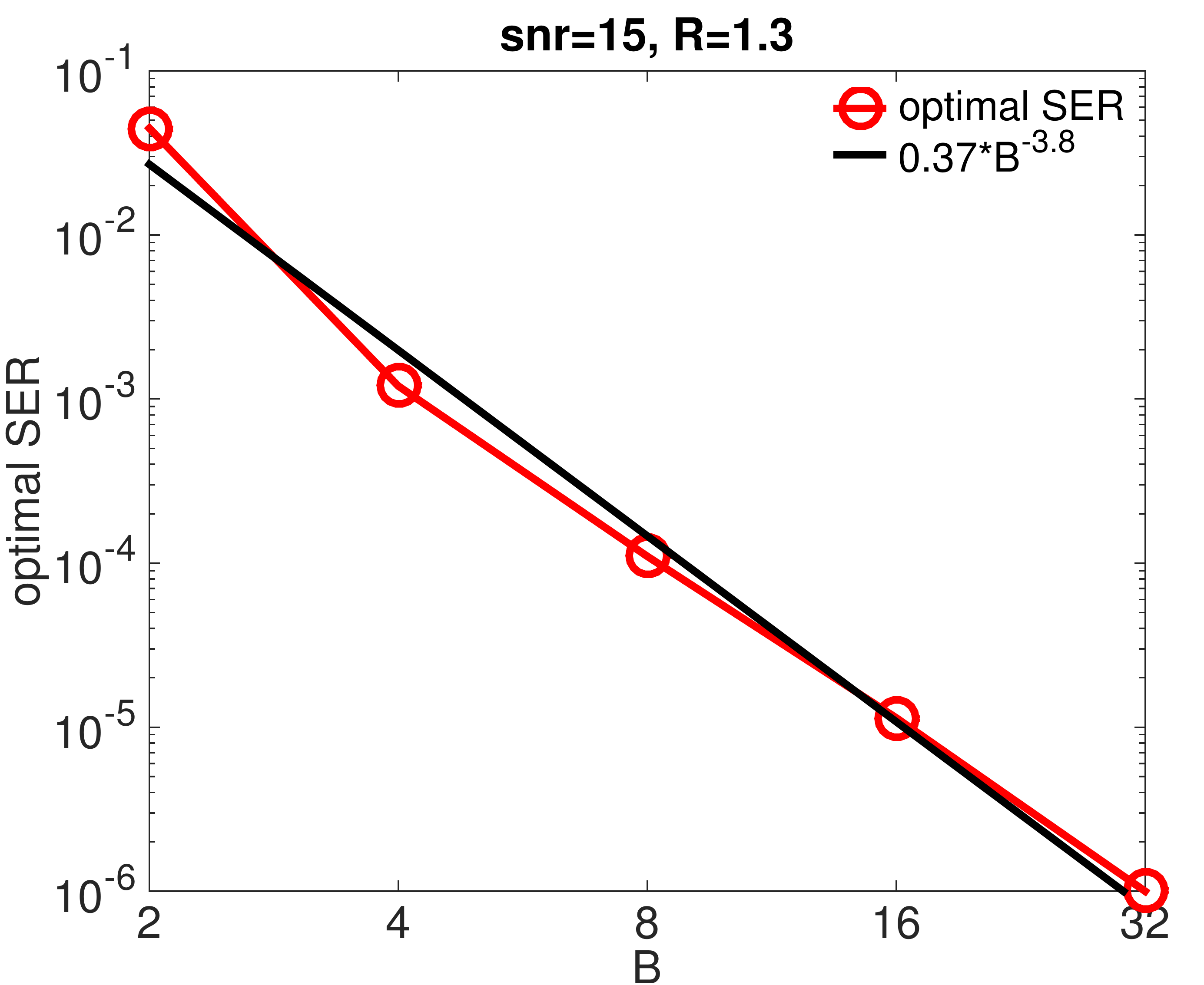}
\includegraphics[width=0.35\textwidth]{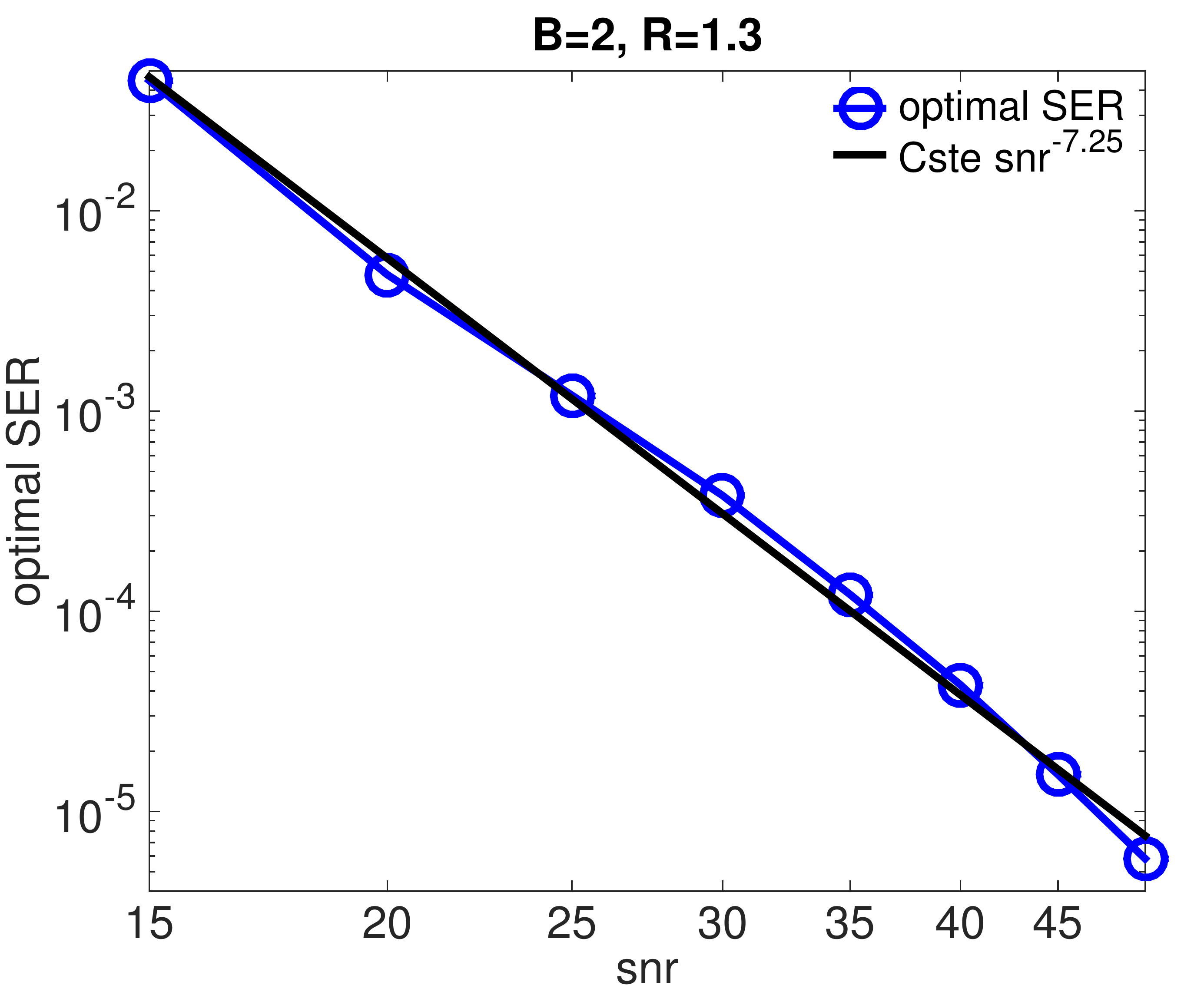}
\caption{These double logarithmic plots show how the optimal ${\rm SER}$ (the ${\rm SER}$ corresponding to the MMSE through \eqref{eq_SE_SER}) changes when $B$ increases at fixed rate $R=1.3$ and ${\rm snr}=15$ (red) or when the ${\rm snr}$ increases at fixed rate $R=1.3$ and $B=2$ (blue), according to the replica analysis. All the points are extracted from the potential (\ref{eq_freeEnt2}). The best linear
  fit is added on top of the curves (Cste is a constant). \label{fig_optSER}}
\end{figure}
\begin{itemize}
\item The asymptotic limit for large $B$ of ${\cal J}\defeq\ln({\cal Z}_2)/\ln(B)$ exists, and is concentrated (i.e it does not depend on the disorder realization, that is the ensemble of energy levels).
\item It is equal to ${\cal J} = \sqrt{2}/\Sigma \ \mathbb{I}\left(\Sigma<1/\sqrt{2}\right)+ (1/(2\Sigma^2) + 1) \ \mathbb{I}\left(\Sigma>1/\sqrt{2}\right)$.
\end{itemize}
We can thus now obtain the value of the integral by comparing
${\cal Z}_{1}$ and ${\cal Z}_{2}$ and keeping only the dominant
term. First let us consider the case where $\Sigma>1/\sqrt{2}$:
\begin{equation}
\frac 1{\ln(B)} \ln\left( {\cal  Z}_1 + {\cal Z}_{2}\right)\approx \frac{\ln ( {\cal
  Z}_{2})}{\ln(B)} ,
\end{equation}
where the approximate equality is an ansatz motivated by physical arguments: at high temperature, all the configurations have approximately same weight and thus the favored state has negligible influence. In communication terms, $\Sigma^2$ plays the role of the variance of an effective AWGN added to the transmitted section and thus when it is high, it prevents recovering the section. If, however, $\Sigma<1/\sqrt{2}$, then using again an ansatz one obtains
\begin{equation}
\frac 1{\ln(B)} \ln\left( {\cal  Z}_1 + {\cal Z}_{2}\right)\approx \frac{\ln ( {\cal
  Z}_{1})}{\ln(B)} \,.
\end{equation}
Indeed, at low temperature, the favored state should be dominant. In communication terms, the noise is low and thus one recovers the section. From \eqref{eq_32} this leads to
\begin{align}
\lim_{B\to \infty} \frac {I_B}{\ln(B)} &=\frac 1{2\Sigma^2} \ \mathbb{I}(\Sigma< 1/\sqrt{2}) +\mathbb{I}(\Sigma > 1/\sqrt{2}).
\end{align}
From these results combined with (\ref{eq_freeEnt2}), we now can give the asymptotic expression of the potential:
\begin{equation}
\phi(E) \defeq \lim_{B\to \infty} \frac{\Phi_B(E)}{\ln(B)} = -\frac{1}{2R\ln(2)}
\Big(\ln({1/{{\rm snr}}} + E) + \frac{1 - E}{{1/{{\rm snr}}} + E}
\Big) + \max\Big(1,\frac 1{2\Sigma^2(E)}\Big),
\label{Phi_largeB}
\end{equation}
with $\Sigma^2(E) = R \ln(2) (1/{\rm snr}+E)$, see (\ref{eq_defS2}). See Fig.~\ref{fig:largeB} for a graphical representation of this potential.

Let us now look at the extrema of this potential. We see that we have
to distinghish between the high error case
($\Sigma>1/\sqrt{2}$ so that $E>1/(2R\ln(2)) - 1/{\rm snr}$) and the low
error one ($\Sigma<1/\sqrt{2}$, so that $E<1/(2R\ln(2)) - 1/{\rm snr}$).

In the high error case, the derivative of the potential is zero when
\begin{equation}
\frac 1{2R\ln(2)} \Big( \frac 1{1/{\rm snr}+E} - \frac{
1/{\rm snr}+1}{(1/{\rm snr}+E)^2}\Big)=0,
\end{equation}
which happens when $E=1$. Therefore, if both the condition $E=1$ and
$E>1/2R\ln(2) - 1/{\rm snr}$ are met, there is a stable extremum (a maximum) of the potential at $E=1$. The existence of this high-error
maximum thus requires $1/(2R\ln(2)) - 1/{\rm snr}<1$, and we thus define
the asymptotic $B\to\infty$ critical rate beyond which the state at $E=1$ is stable:
\begin{equation}
R_{\rm BP}^{\infty}\defeq [({1/{{\rm snr}}} + 1)2\ln(2)]^{-1}. \label{BPcrit}
\end{equation}
Since we initialize the recursion at $E=1$ when we attempt to
reconstruct the signal with AMP, we see that $R_{\rm BP}^{\infty}$ is a crucial
limit for the reconstruction abality by message-passing. See the right part of Fig.~\ref{fig:largeB} for an illustration of how this transition is separated from the Shannon capacity, leading to a computational gap that may be closed by spatial coupling.

In the low error case, the derivative of the potential is zero when:
\begin{equation}
  \frac 1{2R\ln(2)} \Big( \frac 1{1/{\rm snr}+E} - \frac{
    1/{\rm snr}+1}{(1/{\rm snr}+E)^2}\Big)=-\frac {1}{2R\ln(2)}
  \frac{1}{(1/{\rm snr} +E)^2},
\end{equation}
which happens when $E=0$. Hence, there is another maximum with
zero error. Let us determine which of these two is the global one. We
have
\begin{align}
\phi(0) &=-  \frac 1{2R\ln(2)} \( \ln(1/{\rm snr}) + {\rm snr} \) +
  \frac {\rm snr}{2R\ln(2)} = \frac{\log_2({\rm snr})}{2R},\\
\phi(1) &=-  \frac {\log_2(1/{\rm snr} + 1)}{2R} + 1,
\end{align}
so that the two are equal when $\log_2({\rm snr}) = 2R - \log_2(1+1/{\rm snr})$, or equivalently when $R = \log_2(1+{\rm snr})/2=C$, where we recognize the expression of the Shannon Capacity for the AWGN. These
results are confirming that, at large value of $B$, the optimal value of the section error rate vanishes. Therefore perfect
reconstruction is possible, at least as long as as the rate remains below the Shannon capacity after which, of course, this could not be true anymore. This confirms the results by \cite{barron2010sparse,barron2011analysis} that these codes are capacity achieving. 

These results are summarized by Fig.~\ref{fig:largeB}. The analysis of $\phi(E)$ have shown that the only possible maxima are at $E=0$ and $E=1$, which implies that \emph{the error floor vanishes as $B$ increases}. We have shown that if $R<R_{\rm BP}^\infty$, then $\phi(E)$ has a unique maximum at $E = 0$, meaning that AMP is optimal and leads to perfect decoding. Otherwise two minima coexist and AMP is sub-optimal. In this regime it is required to use spatial coupling or power allocation.
\subsection{Optimality of the AMP decoder with a proper power allocation} \label{sec:AMP-PowA}
In this section, we shall discuss a particular power allocation that
allows AMP to be capacity achieving in the large size limit, without
the need for spatial coupling. We shall work again in the large
section size $B$ limit.

We first divide the signal into $G$ blocks, see Fig.~\ref{fig_equivPowaSpc}. For our analysis, each of
these blocks has to be large enough and contains many sections, each of these sections being itself large so that $1\ll B\ll L_G$, $1\ll G$ where $L_G\defeq L/G$ is the number of sections per group. Now, in
each of these blocks, we use a different power allocation: the non-zero values of the sections inside block $g$ are all equal to $c_g$. This is
precisly the case which we have studied in sec.~\ref{sec:powA_SE}, so
we can apply the corresponding state evolution in a straightforward manner.

Our claim is that we can use the following power allocation: 
\begin{equation} \label{cg}
c_g=\frac{2^{-Cg/G}}{Z} \ \forall \ g\in\{1,\ldots,G\},
\end{equation}
where $C=\log_2{\(1+{\rm snr}\)}/2$ is the Shannon capacity. We choose $Z$ such that the power of the signal equals one, so that $\sum_{g}^G c_g^2/G=1$. With this definition,
we have
\begin{equation} \label{Z2}
Z^2 =  \frac{2^{-\frac{2C}G} \(1-2^{-2C}\)}{G(1-2^{-\frac{2C}G})}.
\end{equation}
This leads to the following useful identity:
\begin{equation}
\frac 1G \sum_{g}^{\tilde g} c_g^2= \frac {1-2^{-\frac{2C\tilde g}G}}{1-2^{-2C}}.
\label{mysum}
\end{equation}
Now, we want to show that, if we have decoded all sections until the
section $\tilde g$, then we will be able to decode section $\tilde g$
as well. If we can show this, then starting from $\tilde g=0$ we will
have a succession of decoding until all is decoded, and we would have
shown that this power allocation works. In this situation, using
(\ref{eq:SE_POWA_S}) and the expression of the rate $R$ (\ref{eq_alpha}), we have
for the section $\tilde g$ that
\begin{equation}  \label{eq41}
(\Sigma_{\tilde g}^{t+1})^2 =
R \ln(2) \frac{{1/{{\rm snr}}} +    {\cal E}_{\tilde g -1}}{c_{\tilde g}^2},
\end{equation}
with
\begin{equation} 
 {\cal      E}_{\tilde g} \defeq 1 - \frac 1G \sum_{g}^{\tilde g} c_g^2,  \label{eq_calE}
\end{equation}
where we have used our assumption of having already decoded until $\tilde g-1$ included: $E_g = \mathbb{I}(g \ge \tilde g)$. (\ref{eq_calE}) is the average ${\rm{MSE}}$ per section if all has been decoded until
$\tilde g$, given that the initial ${\rm{MSE}}$ is $E^0=1$ and that we have to remove what has been already decoded. We now ask if the block
${\tilde g}$ can be decoded as well. The evolution of the error in
this block is given by (\ref{eq:SE_POWA_E}), and we have seen, in
sec.~\ref{subsec_largeBrep}, that the condition for a perfect decoding
in the large $B$ limit is simply that $\Sigma^2<1/2$. Using \eqref{eq41}, we thus need the following to be satisfied (as long as $R<C$):
\begin{equation} 
R \ln(2) \frac{ 1/{\rm snr} + {\cal E}_{\tilde g-1}}{c_{\tilde g}^2} <\frac 12.
\label{condition}
\end{equation} 
If this condition is satisfied, there is no BP threshold to block
the AMP reconstruction in the block ${\tilde g}$, and thus the decoder
will move to the next block, etc. We thus need this condition to
be correct $\forall \ {\tilde g} \in \{1,\ldots,G\}$. Let us perform the large $G$ limit (remembering that $g/G$ stays however finite). We have from \eqref{cg} and \eqref{Z2} that
\begin{figure}[!t]
\centering
\includegraphics[width=0.45\textwidth]{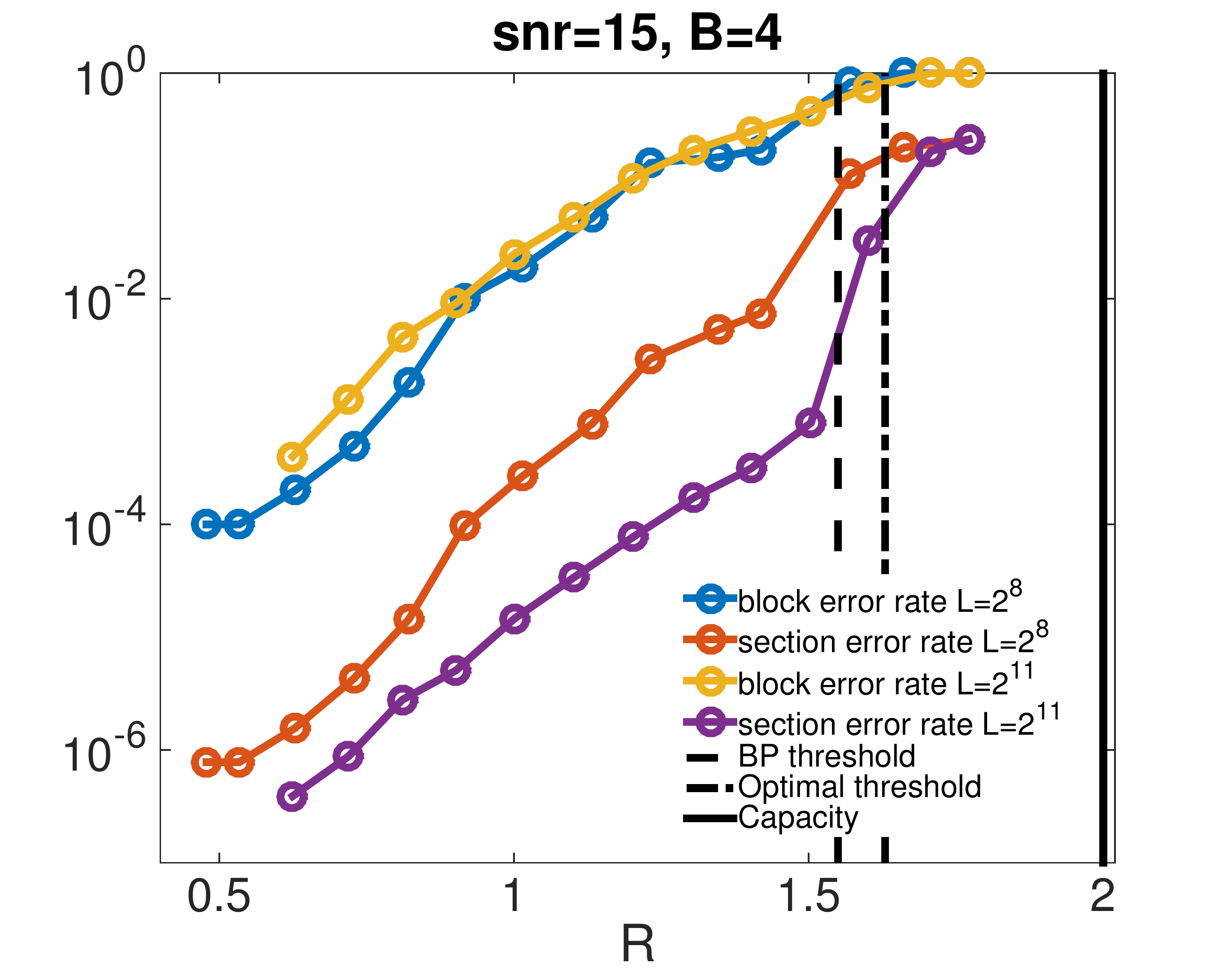}
\includegraphics[width=0.45\textwidth]{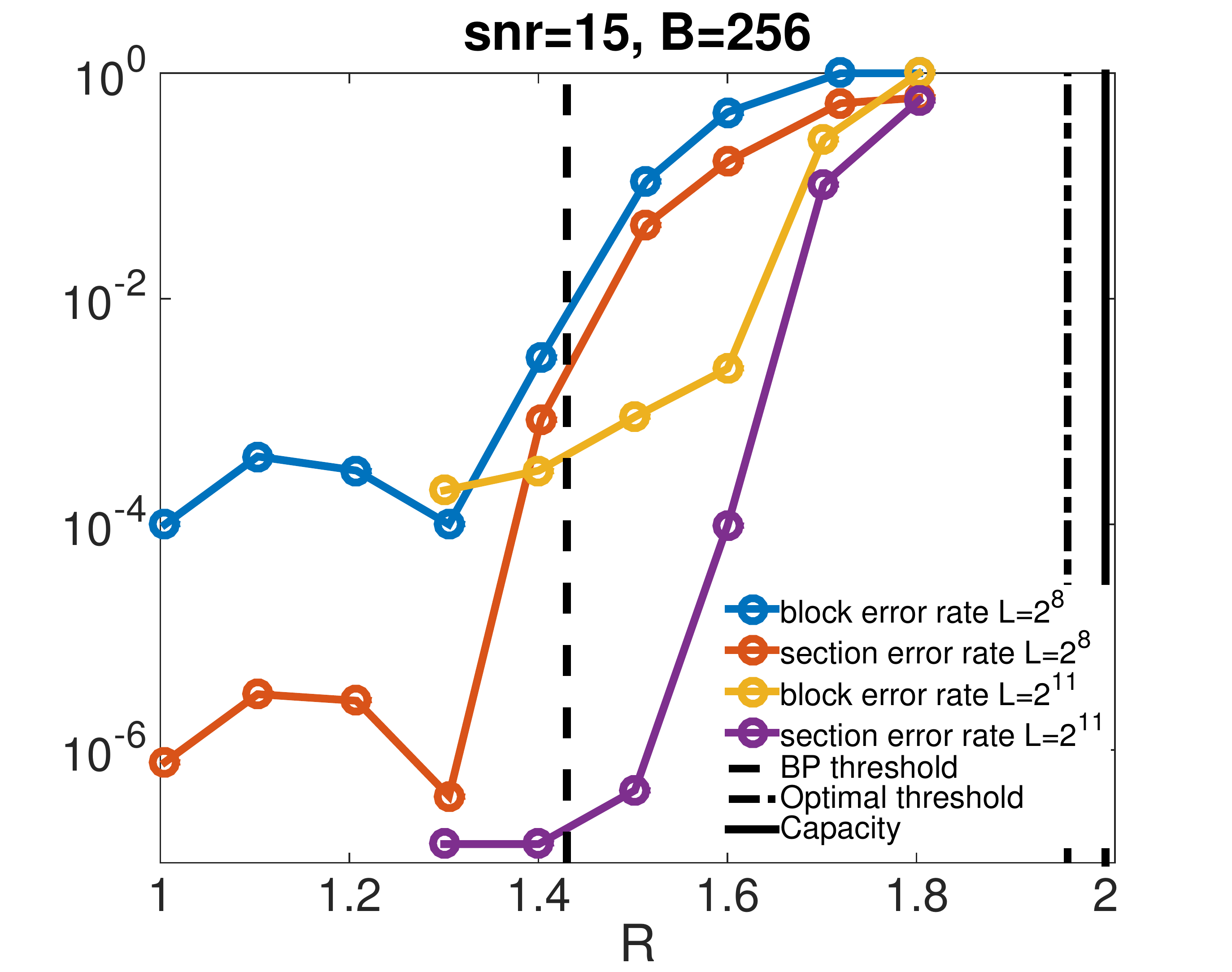}
\includegraphics[width=0.45\textwidth]{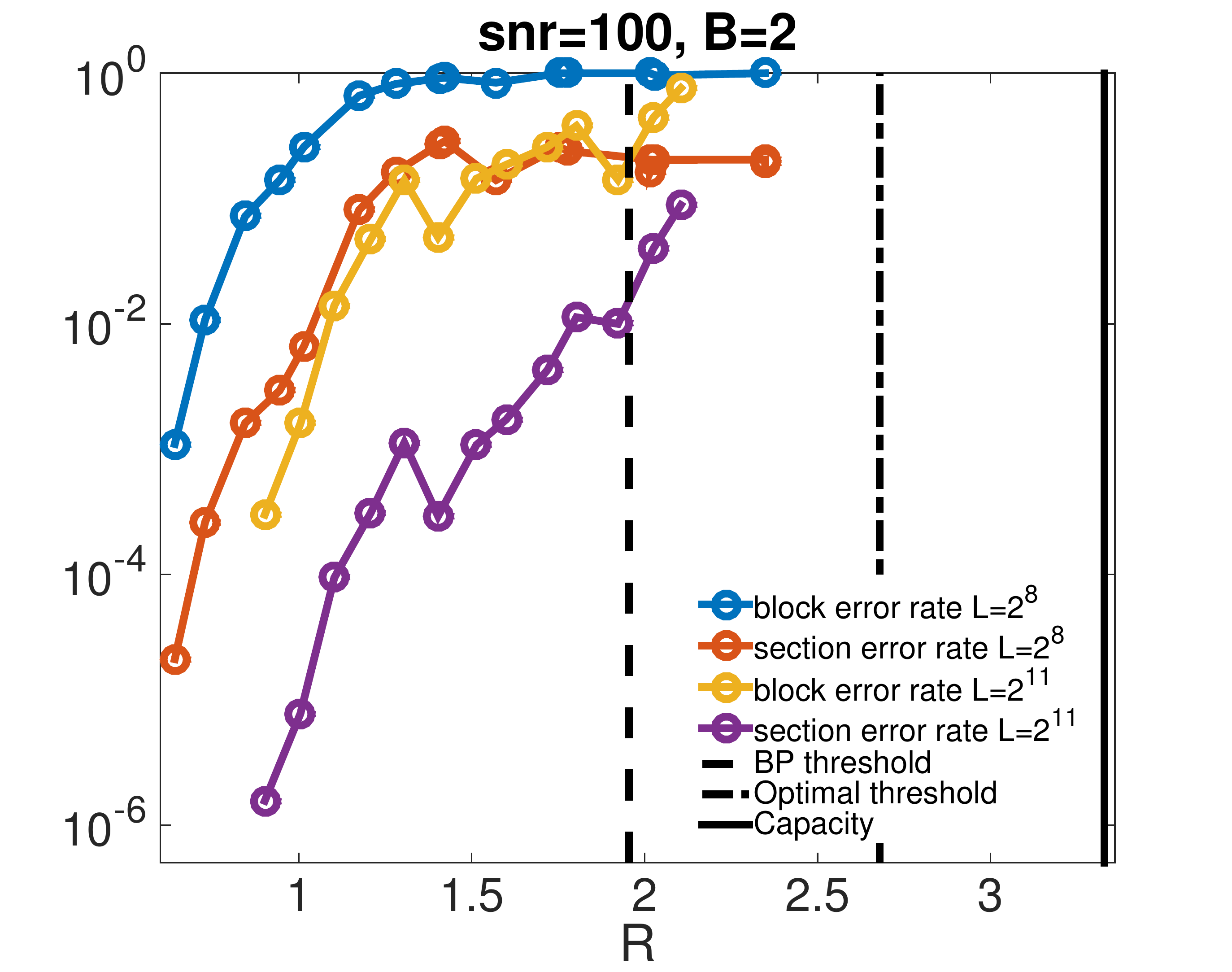}
\includegraphics[width=0.45\textwidth]{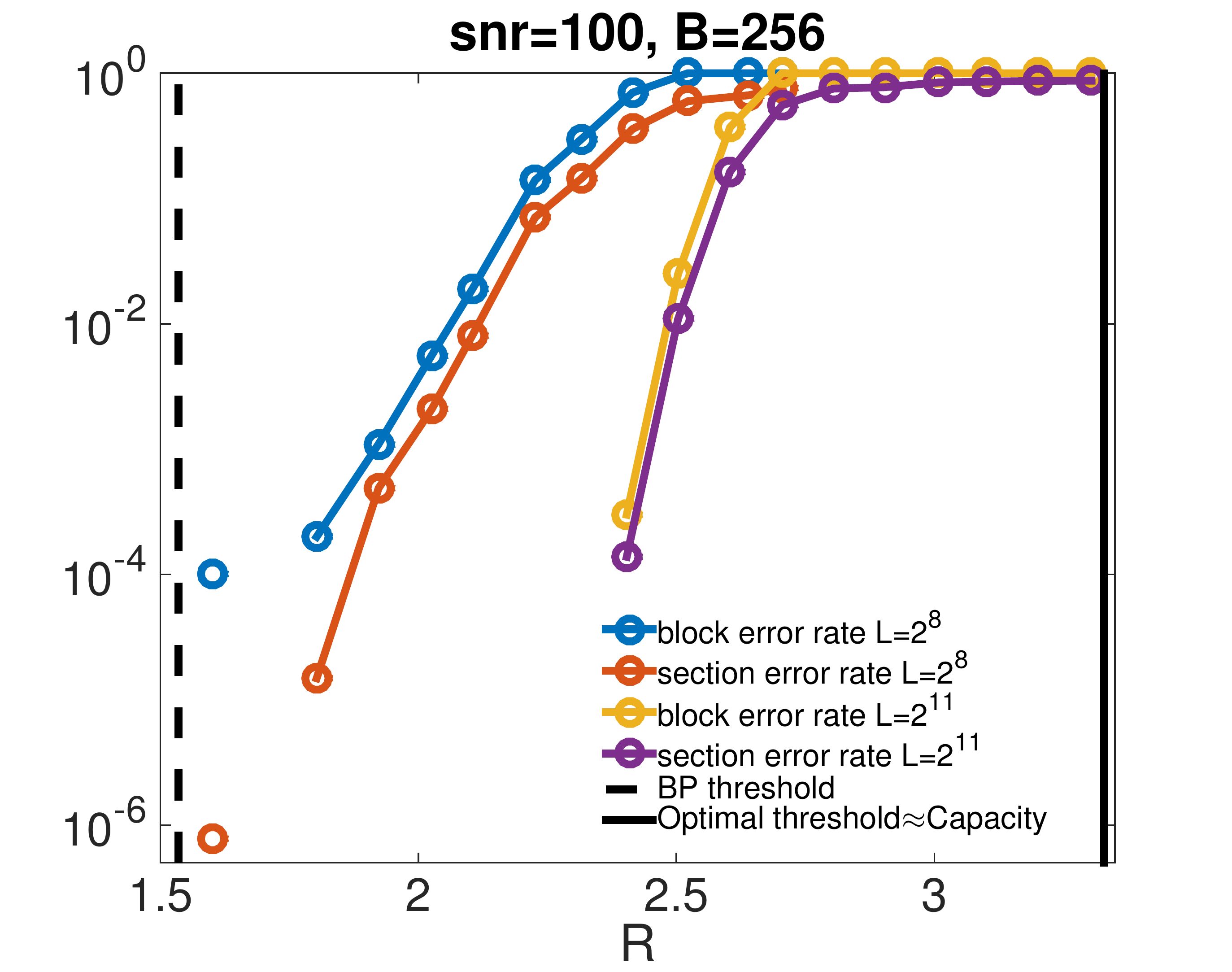} 
\caption{On this plot, we show the empirical block error rate and average section error rate of the
  superposition codes using the AMP decoder combined with spatially coupled
  Hadamard-based operators for two different ${{\rm snr}}$, signal sizes $L$
  and section sizes $B$. The block error rate is the fraction of the
  $10^4$ random instances we ran for each point that have not been
  perfectly reconstructed, i.e. in these instances at least one section
  has not been well recontructed and the final ${\rm SER} > 0$. The ${\rm SER}$ has been averaged over the $10^4$ random instances.
  The convergence
  criterion is that the mean change in the variables estimates between
  two consecutive iterations $\delta < 10^{-8}$ and the
  maximum number of iterations is $t_{\rm max}=3000$. The upper plots
  are for ${{\rm snr}}=15$, the lower for ${{\rm snr}}=100$ (notice the
  different $x$ axes). The first dashed black line is the BP transition
  obtained by state evolution analysis, the second one is the optimal
  transition obtained by the replica method from the free entropy (\ref{eq_freeEnt2}) and the solid black
  line is the capacity. In the (${{\rm snr}}=100, B=256$) case, the
  optimal transition is so close to the capacity that we plot a single
  line. For such sizes, the block error rate is $0$ for rates lower
  than the lowest represented one. The spatially coupled operators used for the experiments are drawn from the ensemble
  $(L_c=16,L_r=17,w=2,\sqrt{J}=0.4,R, \beta_{{{\rm seed}}}=1.8)$.}\label{fig_finiteSizeSeeded}
\end{figure}
\begin{align}
c_g^2 &= \frac{ G (1-2^{-\frac{2C}G})}{2^{-\frac{2C}G}
  \(1-2^{-2C}\)}2^{-\frac {2Cg}G}=
\frac{ G (1-2^{-\frac{2C}G})}{  \(1-2^{-2C}\)}2^{-\frac
  {2C(g-1)}G}\\
  &= G\frac{ 2^{-\frac
  {2C(g-1)}G} }{  \(1-2^{-2C}\)} \(\ln(2) 2C/G + O(1/G^2)\)= \frac {2C \ln(2)~2^{-\frac {2C(g-1)}G}}{1-2^{-2C}} + O(1/G). \label{eq_cgsq}
\end{align}
Now, we note from the expression of the Shannon capacity $C$ that the $\rm snr$ can be written as
\begin{equation}
\rm snr=2^{2C}-1=\frac {1-2^{-2C}}{2^{-2C}},
\end{equation}
so using (\ref{mysum}), \eqref{eq_calE} it leads to
\begin{align}
1/{\rm snr}+{\cal E}_{\tilde g-1} &= \frac {2^{-2C}}{1-2^{-2C}} + 1 - \frac {1-2^{-\frac{2C(\tilde   g-1)}G}}{1-2^{-2C}} =\frac{2^{-\frac{2C(\tilde   g-1)}G}}{1-2^{-2C}}.
\end{align}
Therefore to leading order, we have using (\ref{eq_cgsq}) that
\begin{equation}
\frac{1/{\rm snr} + {\cal E}_{\tilde g-1}}{c_{\tilde g}^2} \approx \frac 1{2C \ln(2)},
\end{equation}
so that the condition (\ref{condition}) becomes for large $G$
\begin{equation}
\frac {R \ln(2)}{2C \ln(2)} = \frac R{2C}<\frac 12,
\end{equation}
or equivalently $R<C$. This shows that, with a proper power allocation \eqref{cg} and as long as $R<C$, there aymptotically cannot exist a local maximum in the potential; or equivalently, that the AMP decoder cannot be stuck in such a spurious maximum and will reach the optimal solution with perfect reconstruction ${\rm SER}=0$.
\begin{figure}[t!]
\centering
  \includegraphics[width=0.5\textwidth]{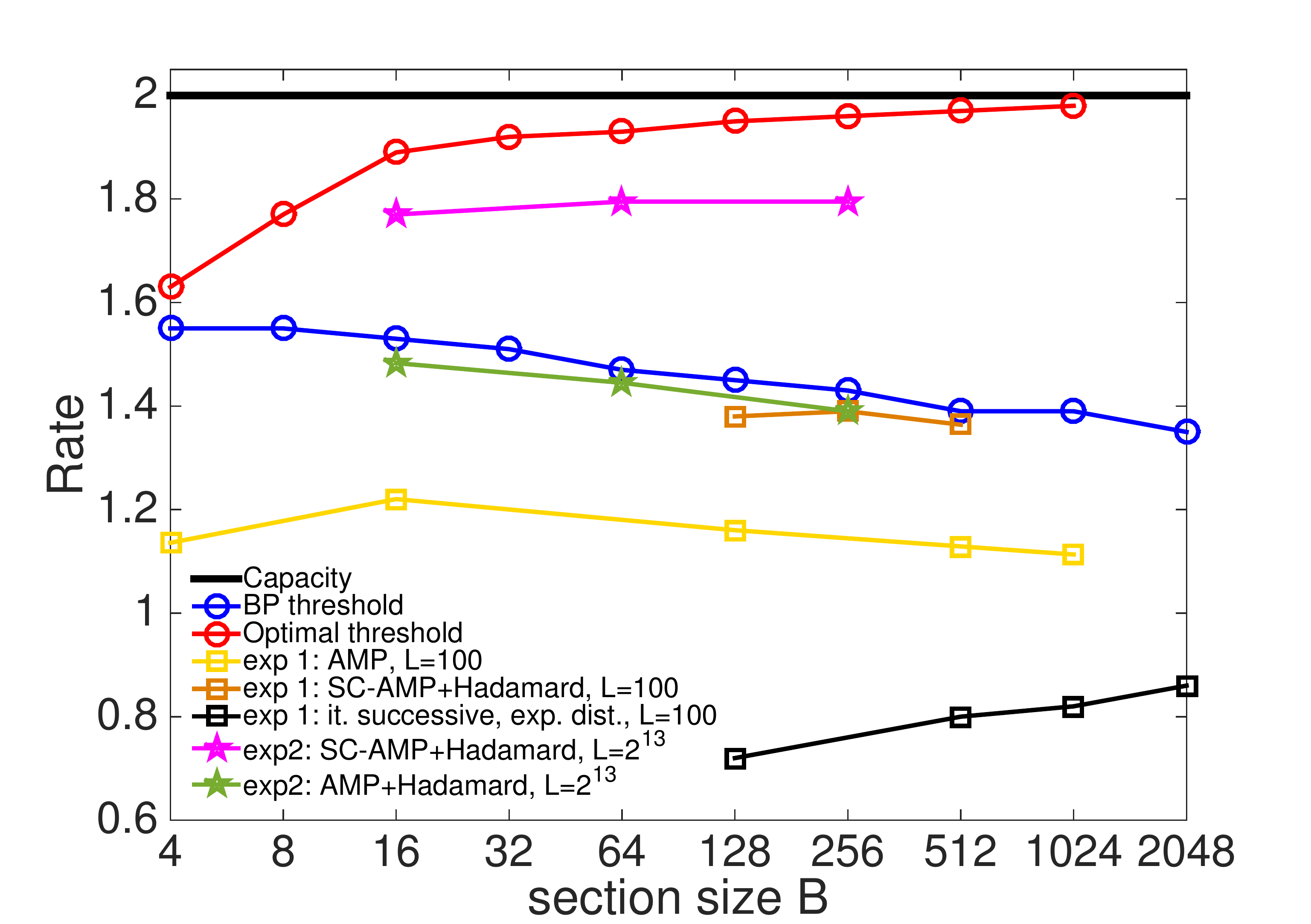}
  \caption{Phase diagram and experimental results for superposition
    codes at finite size $L$ for ${{\rm snr}}=15$,
    compared to the asymptotic transitions. The solid black line is the capacity which
    bounds the performance of any reconstruction algorithm for this
    ${{\rm snr}}$, the blue line is the BP transition $R_{\rm BP}({{\rm snr}}=15,B)$ obtained by state
    evolution analysis and the red line
    is the Bayesian optimal transition $R_{\rm opt}({{\rm snr}}=15,B)$ obtained from the potential (\ref{eq_freeEnt2}). The
    yellow, black and brown curves are results of the following
    experiment (exp 1): decode $10^4$ random instances and identify
    the empirical transition curve between a phase where the empirical probability
    $P\({\rm SER}>10^{-1}\)<10^{-3}$ (below the line) from a phase where
    $P\({\rm SER}>10^{-1}\)\ge10^{-3}$ (more than 9 instances have failed over the
    $10^4$ ones). The green and pink curves are the result of the
    second protocol (exp 2) which is a relaxed version of exp 1 with
    $10^2$ random instances and $P\({\rm SER}>10^{-1}\)<10^{-1}$ below the line,
    $P\({\rm SER}>10^{-1}\)\ge10^{-1}$ above. Note that in our experiments
    ${\rm SER}<10^{-1}$ essentially means ${\rm SER}=0$ at these
    sizes. The yellow curve compares our results with the iterative
    successive decoder (black curve) of
    \cite{barron2010sparse,barron2011analysis} where the number of
    sections $L=100$. Note that these data, taken from
    \cite{barron2010sparse,barron2011analysis}, have been generated
    with an exponential power allocation rather than the constant one
    we used. Compared with the yellow curve (AMP with the same
    value of $L$) the improvement of AMP is
    clear. The green and pink curves are the empirical transitions for Hadamard-based operators with AMP with (pink curve) or without
    (green curve) spatial coupling. For the experimental results, the
    maximum number of iterations of the algorithm is arbitrarily fixed
    to $t_{\rm max}=500$. The parameters used for the spatially coupled
    operators are $(L_c=16, L_r=17, w=2, \sqrt{J}=0.3,R, \beta_{\rm seed}=1.2)$.}\label{fig_phaseDiagSC}
\end{figure}
\section{Numerical experiments for finite size signals}
\label{sec:numerics}
We now present a number of numerical experiments testing the performance and behavior of the AMP decoder in different practical scenarios with finite size signals. The first experiment Fig.~\ref{fig_finiteSizeSeeded} quantifies the influence of the finite size effects over the superposition codes scheme with spatially coupled Hadamard-based operators, decoded by AMP. For each plot, we fix the ${\rm snr}$ and the alphabet size $B$ and repeat $10^4$ decoding experiments with each time a different signal and operator drawn from the ensemble $(L_c=16,L_r=17,w=2,\sqrt{J}=0.4,R, \beta_{{\rm{seed}}}=1.8)$. The curves present the empirical block error rate (blue and yellow curves) which is the fraction of instances that have not been perfectly decoded (i.e such that the final ${\rm SER} >0$) end the ${\rm SER}$ (red and purple curves). This is done for two different sizes $L=2^8$ and $L=2^{11}$. When the curves stop, it means that the empirical block error rate (and thus the section error rate as well) is actually $0$. The dashed lines are the BP threshold $R_{\rm BP}$ and optimal threshold $R_{{\rm opt}}$ extracted respectively from the state evolution analysis and potential (\ref{eq_freeEnt2}) and the solid black line is the capacity $C$. Thanks to the fact that at large enough section size $B$, the gap between the BP threshold and capacity is consequent, it leaves room for the spatially coupled ensemble with AMP decoding to beat the transition, allowing to decode at $R>R_{\rm BP}$ as in LDPC codes. For small section size $B$, the gap is too small to get real improvement over the full operators. We also note the previsible fact that as the signal size $L$ increases, the results are improving: one can decode closer to the asymptotic transitions and reach a lower error floor. For $B=256$, the sharp phase transition between the phases where decoding is possible/impossible by AMP with spatial coupling is clear and gets sharper as $L$ increases.

The next experiment Fig.~\ref{fig_phaseDiagSC} is the phase diagram 
for superposition codes at fixed ${{\rm snr}}=15$ like on Fig.~\ref{fig_diagsDist} but where we added on top finite size results. The asymptotic rates that can be reached
are shown as a function of $B$ (blue line for the BP threshold, red one for the optimal rate). The solid black line is the capacity. Comparing the black and yellow curves, it is clear that even without spatial coupling, AMP outperforms the iterative successive decoder of \cite{barron2010sparse} for practical
$B$ values. With the Hadamard-based spatially coupled operators and the AMP decoder, this is true for \emph{any} $B$ and is even more
pronounced (brown curve). The green (resp. pink) curve shows that the homogeneous (resp. spatially coupled) Hadamard-based operator has very good performances for reasonably large signals, corresponding here to a
blocklength $M<64000$ (the blocklength is the size of the transmitted vector $\tilde \by$).

\begin{figure}[!t]
\centering
\includegraphics[width=0.6\textwidth]{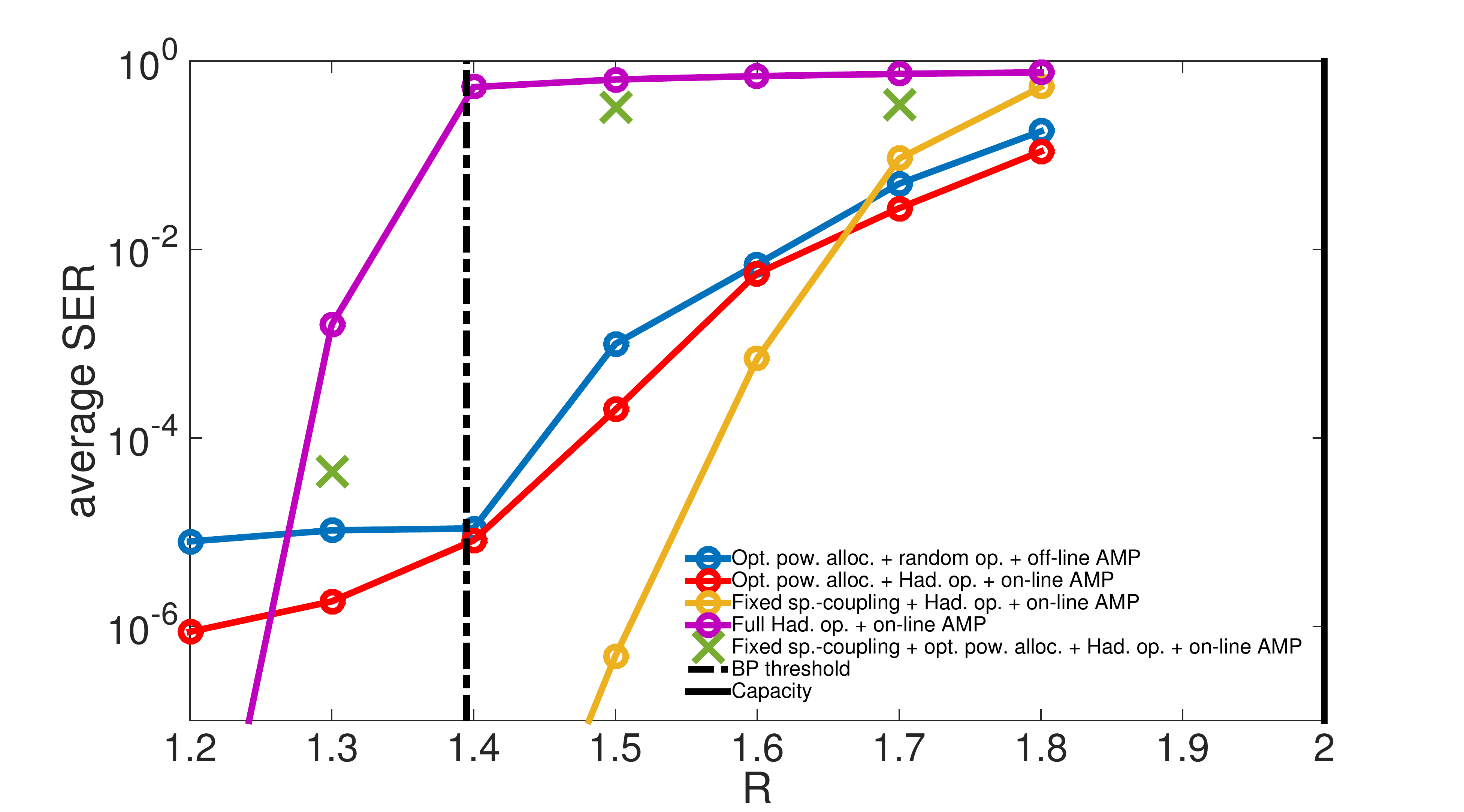}
\caption{The average section error rate ${\rm SER}$ in logarithmic scale as a function of the rate $R$ for different settings, all at fixed (${{\rm snr}}=15, B=512, L=1024$). The black dashed curve identifies the BP transition, the highest rate until which AMP can asymptotically perform well {\it{without}} spatial coupling nor non constant power allocation, the black solid line is the Shannon capacity. \textbf{Blue curve:} It corresponds to the results of Fig.~3 of \cite{rush2015capacity}: the points are averaged over $10^3$ random experiments using i.i.d Gaussian matrices and an optimized power allocation scheme where the parameters defining the power allocation are optimized for each rate. The values of the parameters and the associated power allocation scheme can be found in \cite{rush2015capacity}. The denomination off-line AMP refers to the AMP decoder update rules of \cite{rush2015capacity} that are different than ours and which require an off-line pre-processing part (as opposed to our on-line AMP implementation where all the quantities are computed without any need of pre-processing). \textbf{Red curve:} We reproduced exactly the same experiment (with same power allocation and parameters) as for the blue curve with two important differences: $i)$ we used our on-line AMP decoder instead of their off-line implementation and $ii)$ we used an Hadamard-based homogeneous operator instead of a random i.i.d Gaussian one. In addition, we runned $10^4$ instances instead of $10^3$ as we obtained an average ${\rm SER}$ equals to $0$ for the two first points. \textbf{Purple curve:} This experiment with $10^4$ instances per point is with an Hadamard-based homogeneous operator with on-line AMP decoding of constant power allocated signals. As it should, the decoder does not work anymore for $R>R_{\rm BP}$. \textbf{Yellow curve:} The points of this experiment have been averaged over $10^4$ instances. In this setting, we used our on-line AMP decoder and generated the signals with constant power allocation. We replaced the homogeneous operator by a spatially coupled Hadamard-based operator, described in Fig.~\ref{fig_seededHadamard}. The parameters defining the ensemble from which the operator is randomly generated are fixed and identical for all the points, as opposed to the power allocation curves (blue and red) where parameters have been optimized for each point. The ensemble is here given by $(L_c=16,L_r=17,w=2,\sqrt{J}=0.4,R, \beta_{\rm seed}=1.4)$. \textbf{Green crosses:} These points have been averaged over $10^4$ instances. We used the same spatially coupled Hadamard-based operator ensemble as for the yellow curve for decoding power allocated signals with same power allocation scheme as the blue and red curves. When the purple and yellow curves fall, it means that the points values are $0$. The codeword size for all these curves is between $5\times10^3$ to $9\times10^3$.\label{fig_compPowaVSspC}}
\end{figure}

Finally, the last experiment Fig.~\ref{fig_compPowaVSspC} is a comparison of the efficiency of the AMP decoder combined with spatial coupling or an optimized power allocation. The optimized power allocation used here comes from \cite{rush2015capacity}. We repeated their experiments and compared the results to a spatial coupling strategy. Comparing the results with Hadamard-based operators, given by the red and yellow curves for power allocation and spatial coupling respectively, it is clear that spatial coupling (despite not being optimized for each rate) greatly outperforms a (per rate) optimized power allocation scheme. 

In addition, we see that our red curve corresponding to the optimized
power allocation homogeneously outperforms the blue curve of
\cite{rush2015capacity} with exactly the same parameters. As we have numerically shown that Hadamard-based operators
gets same final performances as Gaussian i.i.d ones as used in
\cite{rush2015capacity} (see \cite{barbier2013compressed} and
Fig.~\ref{fig_distToRbp}), the difference in performance must come
from the AMP implementation: in our decoder implementation (that we denote by on-line decoder), there is no need of pre-processing but in the decoder of \cite{rush2015capacity} (denoted by off-line), quantities need to be computed by state evolution before running.

The advantage of spatial coupling (yellow) over power allocation (red) is \emph{independent of the AMP decoder implementation and the fact that we use Hadamard-based operators}, as it outperforms the red curve obtained with our on-line decoder and Hadamard-based operators as well. This is true at any rate except at very high values where spatial coupling does not allow to decode at all, while the very first components of the signal are decoded using power allocation as their power is very large. But it is not a
really useful regime as only a small part of the signal is decoded
anyway, even with power allocation. The green points show that a mixed
strategy of spatial coupling with optimized power allocation does not
perform well compared to individual strategies. This is easily
understood from the Fig.~\ref{fig_equivPowaSpc}: a power allocation
modify the spatial coupling and worsen its original design. In
addition we notice that at low rates, a power allocation strategy
performs worst that constant power allocation without spatial coupling (purple curve).
\section{Conclusion and future works}
We have derived and studied the approximate message-passing decoder,
combined with spatial coupling or power allocation, for the sparse
superposition codes over the additive white Gaussian
noise channel. Clear links have been established between the present
problem and compressed sensing with structured sparsity.

On the theoretical side, we have computed the potential of the scheme thanks to the heuristic replica method and have shown that the code is capacity achieving under AMP decoding in a proper limit. The analysis shows that there exists a sharp phase transition
blocking the decoding by message-passing before the capacity. The analysis also shows that the optimal Bayesian limit, that can be reached by AMP combined with spatial coupling or power allocation, tends to the Shannon capacity as the section size (i.e the input alphabet size) increases. We have also derived the state
evolution recursions associated to the AMP decoder, with
or without spatial coupling and power allocation. The replica and
state evolution analysies have been shown to be perfectly equivalent
for predicting the various phase transitions as the state evolution can be derived as fixed point equations of the replica
potential. The optimal asymptotic performances have been studied and it appeared
that the error floor decay and the rates of convergence of the various
transitions to their asymptotic values (in the section size) empirically follow power laws.

On the more practical and experimental side, we have presented an
efficient and capacity achieving solver based on spatially coupled
fast Hadamard-based operators. It allows to deal with very large instances
and performs as well as random coding operators. Intensive numerical
experiments show that a well designed spatial coupling performs way
better than an optimized power allocation of the signal, both in terms
of reconstruction error and robustness to noise. Finite size
performances of the decoder under spatial coupling have been studied
and it appeared that even for small signals, spatial coupling allows
to obtain very good perfomances. In addition, we have shown that the
AMP decoder (even without spatial coupling) beats the iterative
succesive decoder of Barron and Joseph for any manageable size. Futhermore, its performances with spatial coupling are way better for any section size.

The scheme should be now compared in a systematic way to other
state-of-the-art error correction schemes over the additive white
Gaussian noise channel. On the application side, from the structure of
the reconstructed signal itself in superpostion codes, we can also
interpret the problem as a structured group testing problem where one
is looking for the only individual that has some property (for example
infected) in each group. Finally, the link with the random energy
model and the similarity of sparse superposition codes with Shannon's random code suggest that these codes should be capacity achieving for a
much larger class of channels. We plan to look at these questions in
future works\footnote{Since the publication of the first version of this paper, one of the author has extended the study of sparse superposition codes and have shown that \emph{this code achieves capacity on any memoryless channel} with spatial coupling and under generalized approximate message-passing decoding, see \cite{barbier2016threshold}.}.
\appendices
\section{Derivation of the approximate message-passing decoder from belief-propagation} \label{app:BPtoAMP}
The following generic derivation is very close to that of
\cite{KrzakalaMezard12} albeit in the present
case it is done in a framework where the variables for which we
know the prior are $B$-d (the sections) instead of the $1$-d ones (for which we want to derive closed equations). For readibility purpose, we drop the time index from the equations and add it back at the end.
\subsection{Gaussian approximation of belief-propagation for dense linear estimation: relaxed belief-propagation} \label{app:relaxedBP}
It starts from the usual loopy BP
equations. We write them in the continuous framework despite the variables we want to infer are discrete. In this way the messages are densities, that can be expanded later on, an essential step in the derivation. Recall that $\bF_{\mu l}\defeq [F_{\mu i} : i \in l]$ is the vector of entries of the $\mu^{th}$ row of the matrix $\bF$ that act on $\bx_l$, see Fig.~\ref{fig_1dOp}. Furthermore, vectors are column vectors and thus terms of the form $\ba^\intercal \bF_{\mu l}=\bF_{\mu l}^\intercal \ba$ for some $\ba$ are scalar products. Then BP reads
\begin{align}
&\hat m_{\mu l}(\hat \bx_l) = \frac{1}{\hat z_{\mu l}} \int \Big[\prod_{k \neq l}^{L-1} d\hat \bx_k m_{k \mu}(\hat \bx_k)\Big] e^{-\frac{{{\rm snr}}}{2} \left(\sum_{k\neq l}^{L-1} \bF_{\mu k}^{\intercal} \hat \bx_k + \bF_{\mu l}^\intercal \hat \bx_l - y_\mu \right)^2 }, \label{eq_bp1} \\
&m_{l\mu}(\hat \bx_l) = \frac{1}{z_{l\mu}} P_0^l(\hat \bx_l) \prod_{\gamma \neq \mu}^{M-1} \hat m_{\gamma l}(\hat \bx_l) \label{eq_bp2},
\end{align}
where $d\hat \bx_k \defeq \prod_{i\in k}^B dx_i$ and we write with some abuse of notation $i\in k$ to refer to the set of indices of the scalar signal components composing the $k^{th}$ section. $P_0^l(\hat \bx_l)$ is the prior (\ref{eq_prior}) that strongly correlates the components inside the same section by enforcing that there is a single non-zero in it, with value given by the power allocation. 

These intractable equations must be simplified. Recall that we fix the power to $1$ which implies a scaling $F_{\mu i} = O(1/\sqrt{L})$ for the coding matrix i.i.d entries. Thus $F_{\mu i}\to 0$ as $L\to \infty$ which allows to expand the previous equations. We will need the following transform $\exp(-w^2{{\rm snr}}/2) = \sqrt{{{\rm snr}}/(2\pi)} \int d\lambda \exp(-\lambda^2{{\rm snr}}/2 + i{{\rm snr}}\lambda w)$. Using this for $w\defeq \sum_{k\neq l}^{L-1} \bF_{\mu k}^\intercal \hat \bx_k$, we express (\ref{eq_bp1}) as
\begin{align}
\hat m_{\mu l}(\hat \bx_l) &= \frac{\sqrt{{{\rm snr}}}}{\sqrt{2\pi}\hat z_{\mu l}} e^{-\frac{{{\rm snr}}}{2}\left(\bF_{\mu l}^\intercal \hat \bx_l - y_\mu\right)^2} \int d\lambda e^{-\frac{\lambda^2{{\rm snr}}}{2}} \prod_{k\neq l}^{L-1} \Big[\int d\hat \bx_k m_{k\mu}(\hat \bx_k) e^{{{\rm snr}}\,\bF_{\mu k}^\intercal \hat \bx_k\left(y_\mu - \bF_{\mu l}^\intercal \hat \bx_l + i\lambda\right)}\Big]. \label{eq_bp3}
\end{align}
In order to define the approximate messages using a Gaussian parametrization (that is using the first and second moments), we need the following vector definitions
\begin{align}
\ba_{u} \defeq \int \hat \bx\ \! m_{u}(\hat \bx) \ \! d\hat \bx,\ \bv_{u} \defeq \int \hat \bx^2 m_{u}(\hat \bx) \ \! d\hat \bx - \ba_{u}^2,
\end{align}
where $u$ represents a generic index and the square operation $\hat \bx^2$ is componentwise. Expanding (\ref{eq_bp3}), keeping only the terms not smaller than $O(1/L)$ and approximating the result by an exponential we obtain
\begin{align}
\hat m_{\mu l}(\hat \bx_l) \approx \frac{\sqrt{{{\rm snr}}}}{\sqrt{2\pi}\hat z_{\mu l}} e^{-\frac{{{\rm snr}}}{2}\left(\bF_{\mu l}^\intercal\hat \bx_l - y_\mu\right)^2} \int d\lambda e^{-\frac{\lambda^2{{\rm snr}}}{2}}  \prod_{k\neq l}^{L-1} \Big[e^{{{\rm snr}}\,\ba_{k\mu}^\intercal \bF_{\mu k} \left(y_\mu - \bF_{\mu l}^\intercal \hat \bx_l + i\lambda\right) + \frac{{{\rm snr}}^2}{2}\bv_{k\mu}^\intercal \bF_{\mu k}^2 \left(y_\mu - \bF_{\mu l}^\intercal \hat \bx_l + i\lambda\right)^2}\Big].
\end{align}
The symbol $\approx$ means equality up to terms of lower order than $O(1/L)$. The Gaussian integration with respect to $\lambda$ can now be performed. Putting all the $\hat \bx_l$-independent terms in the normalization constant $\hat z_{\mu l}$, we obtain
\begin{align}
\hat m_{\mu l}(\hat \bx_l) \approx \frac{1}{\hat z_{\mu l}} e^{-\frac{1}{2} \bA_{\mu l}^\intercal \hat \bx_l^2  +\bB_{\mu l}^\intercal \hat \bx_l},& \ \hat z_{\mu l}=\prod_{i\in  l}^{B} \sqrt{\frac{2\pi}{A_{\mu i}}} e^{\frac{B_{\mu i}^2}{2A_{\mu i}}},\label{eq_mmul} \\
\bA_{\mu l} \defeq \frac{\bF_{\mu l}^2}{1/{{\rm snr}} + \Theta_\mu - (\bF_{\mu l}^2)^\intercal \bv_{l\mu}},& \ \bB_{\mu l} \defeq \frac{\bF_{\mu l}(y_\mu - w_\mu + \bF_{\mu l}^\intercal \ba_{l \mu} ) }{1/{{\rm snr}} + \Theta_\mu - (\bF_{\mu l}^2)^\intercal \bv_{l\mu}}, \label{eq_bA}
%
% \bA_{\mu l} &\defeq \frac{\bF_{\mu l}^2}{1/{{\rm snr}} + \sum_{k\neq l}^{L-1}\bv_{k \mu}^\intercal \bF_{\mu k}^2}, \ \bB_{\mu l} \defeq \frac{\bF_{\mu l}(y_\mu - \sum_{k\neq l}^{L-1} \bF_{\mu k}^\intercal \ba_{k \mu} ) }{1/{{\rm snr}} + \sum_{k\neq l}^{L-1} \bv_{k \mu}^\intercal \bF_{\mu k}^2} \label{eq_bA} \\
\end{align}
where we introduced the shorthand notations
\begin{align}
w_\mu &\defeq \sum_{k}^L \bF_{\mu k}^\intercal \ba_{k\mu}, \ \Theta_\mu \defeq \sum_{k}^L (\bF_{\mu k}^2)^\intercal \bv_{k\mu}. \label{eq_wmu_thetamu}
\end{align}
It is noteworthy that at this stage, the joint distribution $\hat m_{\mu l}(\hat \bx_l)$ of the signal components inside the section $l$ is a multivariate Gaussian distribution with diagonal covariance. The independence of $\{\hat x_i : i\in l\}$ in the measure $\hat m_{\mu l}(\hat \bx_l)$ is \emph{not} an assumption, and it arises in the computation as $L\to\infty$ from the fact that the entries of the coding matrix $\bF$ are independently drawn, which makes the non diagonal terms of the covariance of a smaller order than the diagonal ones. We can thus safely neglect them as $L\to\infty$. This decouples in the factor-to-node messages all the components inside the same section, which simplifies a lot the equations. The strong correlations between the signal components inside a same section are purely due to the prior $P_0(\bx_l)$ (\ref{eq_prior}), and will be taken into account in the next step. Plugging (\ref{eq_mmul}) in (\ref{eq_bp2}), we deduce the node-to-factor messages
\begin{align}
m_{l\mu}(\hat \bx_l) \approx \frac{1}{z_{l\mu}} P_0^l(\hat \bx_l) e^{-([\hat \bx_l - \bR_{l\mu} ]^2)^\intercal (2\bxigma_{l\mu})^{-1} },& \ z_{l\mu} = \int d\hat \bx_l P_0^l(\hat \bx_l) e^{-([\hat \bx_l - \bR_{l\mu} ]^2)^\intercal (2\bxigma_{l\mu})^{-1} }, \label{eq64} \\
\bxigma_{l\mu}^2 \defeq \frac{1}{\sum_{\gamma\not =\mu}^{M-1} \bA_{\gamma l}},& \ \bR_{l\mu} \defeq \frac{\sum_{\gamma\not =\mu}^{M-1} \bB_{\gamma l}}{\sum_{\gamma\not =\mu}^{M-1} \bA_{\gamma l}}. \label{eq_sigma2cav}
%
% \end{align}
% %
% where we define
% % 
% \begin{align}
% m_{l\mu}(\hat \bx_l) = \frac{1}{z_{l\mu}} P_0^l(\hat \bx_l) e^{-\frac{1}{2} (\hat \bx_l^2)^\intercal \sum_{\gamma \neq \mu}^{M-1}\bA_{\gamma l}  + \hat \bx_l^\intercal \sum_{\gamma \neq \mu}^{M-1}\bB_{\gamma l} }, \ z_{l\mu} = \int d\hat \bx_l P_0^l(\hat \bx_l) e^{-\frac{1}{2} (\hat \bx_l^2)^\intercal \sum_{\gamma \neq \mu}^{M-1}\bA_{\gamma l}  + \hat \bx_l^\intercal \sum_{\gamma \neq \mu}^{M-1}\bB_{\gamma l} },
\end{align}
Here sums of the form $\sum_{\gamma}\boldsymbol{\Gamma}_{\gamma l} \defeq \big[\sum_{\gamma }\Gamma_{\gamma i}:i\in l\big]$ are vectors of size $B$ and the inverse operation $\bx^{-1}$ for a vector is a componentwise operation, similarly as $\bx^2$. Each message is now expressed as a Gaussian distribution, fully parametrized by its first and second moment. Thus the algorithm can be expressed only with these moments, instead of the messages.

We define $l_i$ as the section index to which the $i^{th}$ component belongs to. Depending on the context that should be clear, it may also be the set of indices of the components of the section to which belongs component $i$. We now introduce a generic probability measure $m_B(\hat\bx_l|\bxigma_{l}^2,\bR_{l})$ for a section. It is a joint probability distribution over the $B$ components composing a given section.
% , whereas $m_1$ are the marginals of these $1$-d components. 
%
\begin{align}
m_B(\hat\bx_l|\bxigma_{l}^2,\bR_{l}) &\defeq \frac{1}{z(\bxigma_{l}^2,\bR_{l})} P_0^l(\hat \bx_{l}) e^{-([\hat \bx_{l}-\bR_{l}]^2)^\intercal(2\bxigma_{l}^2)^{-1}}, \, z(\bxigma_{l}^2,\bR_{l}) = \int d\hat \bx_l P_0^l(\hat \bx_{l}) e^{-([\hat \bx_{l}-\bR_{l}]^2)^\intercal(2\bxigma_{l}^2)^{-1}}. \label{eq_marginal}
%
% m_1(\hat x_i|(\bxigma_{l_i})^2,\bR_{l_i}) &\defeq \int \!\!\!\prod_{j\in l_i : j\neq i}^{B-1} \!\!\!d\hat x_j m_B(\hat\bx_{l_i}|(\bxigma_{l_i})^2,\bR_{l_i}).
\end{align}
Here $z(\bxigma_{l}^2,\bR_{l})$ is a normalization. We define the denoisers $f_{a_i}, f_{c_i}$ as the marginal mean and variance with respect to the measure $m_B$ of the $i^{th}$ signal component 
% If the component $k\in l_i$, i.e the $k^{th}$ and $i^{th}$ components are in the same section, then
%
\begin{align}
f_{a_i}(\bxigma_{l_i}^2,\bR_{l_i}) &\defeq \Big[\int d\hat \bx\, m_B(\hat \bx|\bxigma_{l_i}^2,\bR_{l_i})\, \hat \bx\Big]_i, \label{eq:fai}\\
f_{c_i}(\bxigma_{l_i}^2,\bR_{l_i}) &\defeq \Big[\int d\hat \bx \, m_B(\hat \bx|\bxigma_{l_i}^2,\bR_{l_i})\, \hat \bx^2\Big]_i - f_{a_i}(\bxigma_{l_i}^2,\bR_{l_i})^2,\label{eq:fci}
\end{align}
where the notation $[\bx]_i$ means presently that the component of the vector $\bx \in \mathbb{R}^B$ associated to the $i^{th}$ component of the signal is selected, where $i\in\{1,\ldots,N\}$. For example in (\ref{eq:fai}), $\bx$ is the $B$-d vector of marginal means with respect to the measure $m_B(\hat \bx|\bxigma_{l_i}^2,\bR_{l_i})$. The denoisers are thus taking $B$-d vectors as input and output a single scalar. The interpretation of these functions is the following. The so-called ``AMP fields'' $\bR_{l_i}^t,(\bxigma_{l_i}^t)^2$ which are the mean and variance of the signal component $i$ with respect to the likelihood at a given time are computed. 
These quantities summarize the overall influence of the other variables on the $i^{th}$ one. Then in order to estimate the posterior marginal mean and associated variance, the prior has to be taken into account. This is the role of the denoisers to do so. One may also interpret the denoiser $f_{a_i}(\bxigma_{l_i}^2,\bR_{l_i})$ as the MMSE estimator associated with an effective AWGN channel of noise variance $\bxigma_{l_i}^2$ and channel observation $\bR_{l_i}$.

Now note that the marginal (\ref{eq_marginal}) appearing in the denoisers definitions verifies $m_B(\hat \bx_l|\bxigma_{l\mu}^2,\bR_{l\mu})=m_{l\mu}(\hat \bx_l)$ given by \eqref{eq64}. Then the Gaussian approximation of the BP equations is
\begin{align}
\ba_{l\mu} &= \big[f_{a_i}\big(\bxigma_{l\mu}^2, \bR_{l\mu}\big):i\in l\big], \ \bv_{l\mu} = \big[f_{c_i}\big(\bxigma_{l\mu}^2, \bR_{l\mu}\big):i\in l\big].\label{eq:relaxedCav}
\end{align}
At this stage, after indexing with the time, the algorithm defined by the set of equations (\ref{eq_bA}), (\ref{eq_wmu_thetamu}), (\ref{eq_sigma2cav}), (\ref{eq:relaxedCav}) together with the definition of the denoisers is usually referred to as relaxed-BP \cite{KrzakalaPRX2012}. After convergence, the final posterior estimates and variances $\{a_i,v_i\}$ of the signal components are obtained from
\begin{align}
\bxigma_l^2 \defeq \frac{1}{\sum_{\mu}^M \bA_{\mu l}},& \ \bR_l \defeq \frac{\sum_{\mu}^M \bB_{\mu l}}{\sum_{\mu}^M \bA_{\mu l}} \label{eq_sigma2marg},\\
a_i = f_{a_i}\big(\bxigma_{l_i}^2,\bR_{l_i}\big),& \ v_i = f_{c_i}\big(\bxigma_{l_i}^2, \bR_{l_i}\big).\label{eq:relaxedMarg}
\end{align}
In compressive sensing and more generally for linear estimation problems with AWGN and defined on dense factor graphs, this algorithm is asymptotically exact (as the number of sections $L\to \infty$) in the sense that it is perfectly equivalent to the BP algorithm. This means that the two algorithms would provide the same estimation of the signal up to corrections that asymptotically vanish. This equivalence is a direct consequence of the fact that the coding matrix has i.i.d entries and is dense. With this in mind, from ``central-limit like arguments'', the Gaussian expansions that we have introduced are very natural and asymptotically well justified.
\subsection{Reducing the number of messages: the approximate message-passing algorithm}
We can simplify further the equations, going from the relaxed-BP algorithm with $2MN$ messages ($2$ per edges on the factor graph Fig.~\ref{fig_factorSC}) to the AMP algorithm where only $M+N$ messages are computed. The following expansion is called the Thouless-Anderson-Palmer (TAP) equations in statistical physics \cite{MezardMontanari09}, which is again exact in the same sense as before: the estimation provided by AMP is asymptotically the same as the BP one in the limit $L\to\infty$. 

Deriving AMP starts by noticing that in the $L\to \infty$ limit (and thus the number $M$ of factors diverges as well, while the ratio $\alpha = M/(LB)$ is kept fixed), the quantities (\ref{eq_sigma2cav}), (\ref{eq:relaxedCav}) become almost independent of the index $\mu$. This is equivalent to say that each factor's influence becomes infinitely weak as there are so many. We can thus rewrite (\ref{eq_sigma2cav}), (\ref{eq:relaxedCav}) as marginal quantities, i.e that depend on a single index (as opposed to the cavity quantities that depend on both a variable and factor indices, i.e an edge index), while keeping the proper first order corrections in $F_{\mu i}$. These correcting terms, called the Onsager reaction terms in statistical physics \cite{MezardMontanari09,KrzakalaMezard12}, are essential for the performance of AMP and they make the resulting AMP algorithm different with respect to a naive mean-field approach \cite{variationalF_krzakala14}.

Recall that all the operations such as $1/\bx$ or the dot product $\bx_1 \bx_2$ applied to vectors or matrices are componentwise, whereas $\bx_1^\intercal \bx_2$ is the usual inner product between vectors. Furthermore, keep in mind that we always consider the section size $B$ finite as $L\to\infty$ for the derivation, so that if $B$ terms of order $O(1/\sqrt{L})$ are summed, the result remains of order $O(1/\sqrt{L})$. The aim now is to compute the corrections of the cavity quantities around their associated marginal approximation. Let us start with the corrections to the posterior average given by (\ref{eq:relaxedCav}). Denoting $\bff_{a_l}(\bxigma^2,\bR) \defeq [f_{a_i}(\bxigma^2,\bR) : i\in l]$ (and similarly for $\bff_{c_l}$), we obtain
\begin{align}
\ba_{l\mu} &= \bff_{a_l}(\bxigma_{l\mu}^2,\bR_{l\mu})\approx \bff_{a_l}(\bxigma_{l}^2,\bR_{l}) + (\bxigma_{l\mu}^2 - \bxigma_{l}^2) \nabla_{\bxigma_{l}^2} \bff_{a_l}(\bxigma_{l}^2,\bR_{l}) + (\bR_{l\mu} - \bR_{l}) \nabla_{\bR_{l}} \bff_{a_l}(\bxigma_{l}^2,\bR_{l}) \nonumber\\
&=\ba_l + \frac{\bA_{\mu l}}{(\sum_\gamma^M \bA_{\gamma l})(\sum_{\gamma}^M \bA_{\gamma l} - \bA_{\mu l})} \nabla_{\bxigma_{l}^2} \bff_{a_l}(\bxigma_{l}^2,\bR_{l}) + \frac{\bB_{\mu l}(\sum_{\gamma}^M \bA_{\gamma l}) - \bA_{\mu l}(\sum_{\gamma}^M \bB_{\gamma l})}{(\sum_{\gamma}^M \bA_{\gamma l})(\sum_{\gamma}^M \bA_{\gamma l} - \bA_{\mu l})} \nabla_{\bR_{l}} \bff_{a_l}(\bxigma_{l}^2,\bR_{l}).
\end{align}
Now we use the fact that $\bA_{\gamma l} = O(1/L)$ is a strictly positive term and $\bB_{\gamma l} = O(1/\sqrt{L})$ can be of both signs, see (\ref{eq_bA}), thus
$\sum_\gamma \bA_{\gamma l}$ and $\sum_\gamma \bB_{\gamma l}$ are both $= O(1)$. After some algebra, we obtain the first order corrections to $\ba_l$ and similarly, the corrections to $\bv_l$ as well
\begin{align}
&\ba_{l\mu} \approx \ba_l - \underbrace{\bxigma_{l}^2\bB_{\mu l} \nabla_{\bR_{l}} \bff_{a_l}(\bxigma_{l}^2,\bR_{l})}_{\defeq \beps_{\ba_{l \mu}}}, \ \bv_{l\mu} \approx \bv_l - \underbrace{\bxigma_{l}^2\bB_{\mu l} \nabla_{\bR_{l}} \bff_{c_l}(\bxigma_{l}^2,\bR_{l})}_{\defeq \beps_{\bv_{l \mu}}}, \label{eq_appCorrections}
\end{align}
where $\bxigma_{l}^2,\bR_{l}$ are defined by (\ref{eq_sigma2marg}). We introduced $\beps_{\ba_{l\mu}}\defeq [\epsilon_{a_{i\mu}}:i\in l ]$, the vector of $O(1/\sqrt{L})$ corrections linking the cavity quantity $\ba_{l\mu}$ to the marginal one $\ba_l$, and similarly $\beps_{\bv_{l\mu}}$ for $\bv_l$. Their components can be of both signs. To go further, we thus need to express $\bxigma_{l}^2,\bR_{l}$ in function of marginal quantities only. Keeping only the $O(1)$ dominant terms, we obtain
\begin{align}
\bxigma_{l}^2 & \approx \Big[\sum_{\mu}^M \frac{\bF_{\mu {l}}^2}{1/{{\rm snr}} + \Theta_{\mu} } \Big]^{-1}, \\
\bR_{l} &\approx \bxigma_{l}^2 \Big[\sum_{\mu}^M \frac{\bF_{\mu {l}} \left(y_\mu - w_\mu\right)}{1/{{\rm snr}} + \Theta_{\mu}} + \underbrace{\sum_{\mu}^M \bF_{\mu {l}}\frac{(\bF_{\mu {l}})^\intercal \ba_{l}}{1/{{\rm snr}} + \Theta_{\mu}}}_{=\bxigma_{l}
^{-2} \ba_{l} + O(1/\sqrt{L})} - \underbrace{\sum_{\mu}^M \bF_{\mu {l}}\frac{(\bF_{\mu {l}})^\intercal \beps_{\ba_{l \mu}}}{1/{{\rm snr}} + \Theta_{\mu}}}_{= O(1/L)}\Big] \approx \ba_{l} + \bxigma_{l}^2 \sum_{\mu}^M \frac{\bF_{\mu l}(y_\mu - w_\mu)}{1/{{\rm snr}} + \Theta_\mu}.
\end{align}
As these quantities depend on (\ref{eq_wmu_thetamu}) that depend themselves on cavity quantities, we need to expand them as well. Using (\ref{eq_appCorrections}), we obtain
\begin{align}
\Theta_{\mu} &\approx \sum_{k}^L (\bF_{\mu k}^2)^\intercal\bv_k - \underbrace{\sum_{k}^L (\bF_{\mu k}^2)^\intercal\beps_{\bv_{k\mu}}}_{= O(1/L)} \approx \sum_{k}^L (\bF_{\mu k}^2)^\intercal\bv_k, \label{eq_appThetamu} \\
% \end{align}
% %
% \begin{align}
w_\mu &\approx\sum_{k}^L \bF_{\mu k}^\intercal\ba_k - \sum_{k}^L \bF_{\mu k}^\intercal\beps_{\ba_{k\mu}} \approx \sum_{k}^L \bF_{\mu k}^\intercal\ba_k - \frac{y_\mu - w_\mu}{1/{{\rm snr}} + \Theta_\mu} \sum_{k}^L (\bF_{\mu k}^2)^\intercal \bv_k. \label{eq_appW}
\end{align}
The last equality is obtained neglecting $o(1)$ terms and combining the relation
\begin{align}
&f_{c_i}(\bxigma_{l_i}^2,\bR_{l_i})=v_i=\Sigma_{i}^2 \partial_{R_{i} }f_{a_i}(\bxigma_{l_i}^2,\bR_{l_i}) \Rightarrow \bff_{c_l}(\bxigma_{l}^2,\bR_{l})=\bv_l = \bxigma_{l}^2 \nabla_{\bR_{l} }\bff_{a_l}(\bxigma_{l}^2,\bR_{l}) \label{eq_propertyfafc}
\end{align}
with (\ref{eq_appCorrections}) and (\ref{eq_bA}). We thus now have a closed set of coupled equations on the marginal quantities $\{\bxigma^2_l, \bR_l, \Theta_\mu, w_\mu, \ba_l, \bv_l\}$. Adding back the time index to these equations to get an iterative algorithm, we obtain the AMP algorithm of Fig.~\ref{algo_AMP}.
\subsection{Taking into account the prior for sparse superposition codes}
The only problem-dependent objects in AMP are the denoisers $f_{a_i}, f_{c_i}$ given by (\ref{eq:fai}), (\ref{eq:fci}) that both depend on the prior \eqref{eq_prior}. Let us derive them for any power allocation $\{c_l > 0 : l\in \{1,\ldots,L\}\}$. Using \eqref{eq_prior}, we obtain after simple algebra the posterior estimate $a_i^t$ and variance $v_i^t$ at time $t$
\begin{align}
a_i^t &= f_{a_i}((\bxigma_{l_i}^t)^2,\bR_{l_i}^t) = c_{l_i}\frac{\exp\Big(-\frac{c_{l_i}(c_{l_i}-2R_i^t)}{2(\Sigma_i^t)^2}\Big)}{\sum_{j \in l}^B \exp\Big({-\frac{c_{l_i}(c_{l_i}-2R_j^t)}{2(\Sigma_j^t)^2}}\Big)}, \label{eq_meani}\\
v_i^t &= f_{c_i}((\bxigma_{l_i}^t)^2,\bR_{l_i}^t) =  a_i^t (c_{l_i} - a_i^t ), \label{eq_vari}
\end{align}
where $(\bxigma_{l_i}^t)^2,\bR_{l_i}^t$ are the AMP fields of the section $l_i$, to which the $i^{th}$ component of the signal belongs to. This closes the derivation of the AMP decoder for sparse superposition codes.
\subsection{Further simplifications for random matrices with zero mean and homogeneous variance}
\label{susec:fullTap}
We can go further in the simplification of some quantities computed by the decoder Fig.~\ref{algo_AMP} by considering the elements $F_{\mu i}^2$ equal to their variance. We start considering the general case where the matrix can have a block structure encoded through the $L_r\times L_c$ variance matrix $[J_{r,c}]$, and thus we assume $F_{\mu i}^2=J_{r_\mu,c_i}/L$, see Fig~\ref{fig_seededHadamard} (the $1/L$ factor makes the codeword fluctuations of $O(1)$, such that the power has proper scaling. Note that in order to have a codeword power strictly equal to one, the matrix elements are also multiplied by a proper $O(1)$ constant). Here the notation $r_\mu$ ($c_i$) means the block index $r\in\{1,\ldots,L_r\}$ (resp. $c\in\{1,\ldots,L_c\}$) to which the factor index $\mu$ (resp. component $i$) belongs to. 

This step is justified by the fact that the average with respect to the coding matrices ensemble of all the objects appearing in the decoder that depend on such squared elements (such as $\Theta_\mu$) are $O(1)$ whilst their variance are $O(1/N)$, see \cite{KrzakalaMezard12}. Thus in the large signal limit, we can neglect their fluctuations by simply replacing the matrix squared elements by their variance. Considering the most general version of AMP (the left version of Fig.~\ref{algo_AMP}) where it is written in terms of the operators, the dependency is just in the $\tilde O_\mu$ and $\tilde O_i$ operators that now depend only on the block indices and can thus be approximated as
\begin{align}
&\tilde O_r(\textbf{e}_c) \defeq \frac{J_{r,c}}{L}\sum_{i\in c}^{N/L_c} e_i,\  \tilde O_c(\textbf{f}_r) \defeq \frac{J_{r,c} }{L}\sum_{\mu\in r}^{\alpha_rN/L_c} f_\mu.
\end{align}
Thanks to this simplification, we obtain the simpler decoder of Fig.~\ref{algo_AMP2}. It is asymptotically equivalent to AMP as it provides the same estimation when $L\to\infty$ and for matrices which have zero mean entries with homogeneous variance per block $J_{r,c}$. 

Let us show how to go from this simplified AMP decoder to the equivalent notations of \cite{montanari2012graphical} in the case of a fully homogeneous matrix, i.e that does not have a block structure. We start defining the \emph{residual} $\btau_r^{t}\defeq [\tau_\mu^{t}:\mu\in r]$ 
\begin{align}
\btau_r^t &\defeq \textbf{y}_r - \textbf{w}^{t+1}_r = \textbf{y}_r - \Big[\sum_{c}^{L_c}O_\mu(\textbf{a}_c^t) : \mu \in r \Big] + \boldsymbol{\Theta}^{t+1}_{r}\frac{\by_r-\textbf{w}^t_r}{{1/{{\rm snr}}} + \boldsymbol{\Theta}^t_{r}} \\
&= \textbf{y}_r - \Big[\sum_{c}^{L_c}O_\mu(\textbf{a}_c^t) : \mu \in r \Big] + \btau_r^{t-1}\frac{\boldsymbol{\Theta}^{t+1}_{r}}{{1/{{\rm snr}}} + \boldsymbol{\Theta}^t_{r}}. \label{eq:residual_monta}
\end{align}
Then plugging the $R_i^{t+1}$ expression of Fig.~\ref{algo_AMP2} into the denoiser, one obtains
\begin{align}
a_{i}^{t+1} &= f_{a_i}\Big((\Sigma_{c_i}^{t+1})^2, \underbrace{\Big[a^t_j + (\Sigma^{t+1}_{c_i})^2 \sum_{r}^{L_r} O_i\Big(\frac{\btau_r^t}{{1/{{\rm snr}}} + \boldsymbol{\Theta}^{t+1}_r}\Big):j\in l_i\Big]}_{\defeq \bR_{l_i}^{t+1}}\Big). \label{eq:ai_monta}
\end{align}
It is naturally considered that the blocks are such that all the components inside a same section are part of the same block. We define the shorthand notations
\begin{align}
\langle f_c^t\rangle_k \defeq \frac{L_c}{N}\sum_{i\in k}^{N/L_c} f_{c_i}((\Sigma_{k}^{t})^2,\bR_{l_i}^{t}), \ \langle (f_a^t)'\rangle_k \defeq \frac{L_c}{N}\sum_{i\in k}^{N/L_c} \frac{\partial f_{a_i}(x,\textbf{y})}{\partial y_i}\bigg|_{(\Sigma_{k}^{t})^2,\bR_{l_i}^{t}}.
\end{align}
\begin{figure}[!t]
\centering
\begin{minipage}{.45\textwidth}
\centering
\begin{algorithmic}[1]
\State $t\gets 0$
\State $\delta \gets \epsilon + 1$
\While{$t<t_{\rm max} \ \textbf
{and} \ \delta>\epsilon$} 
\State $\Theta^{t+1}_r \gets  \sum_{c}^{L_c}\tilde O_r(\textbf{v}_c^t)$
\State $w^{t+1}_\mu \gets \sum_{c}^{L_c}O_\mu(\textbf{a}_c^t) - \Theta^{t+1}_{r_\mu}\frac{y_\mu-w^t_\mu}{{1/{{\rm snr}}} + \Theta^t_{r_\mu}}$
\State $\Sigma^{t+1}_c \gets \left[\sum_{r}^{L_r}\tilde O_c\left([{1/{{\rm snr}}} + \Theta_r^{t+1}]^{-1}\right)\right]^{-1/2}$
\State $R^{t+1}_i \gets a^t_i + (\Sigma^{t+1}_{c_i})^2 \sum_{r}^{L_r} O_i\left(\frac{\textbf{y}_r - \textbf{w}^{t+1}_r}{{1/{{\rm snr}}} + \Theta^{t+1}_r}\right)$
\State $v^{t+1}_i \gets f_{c_i}\left((\Sigma^{t+1}_{c_i})^2,\bR_{l_i}^{t+1}\right)$
\State $a^{t+1}_i \gets f_{a_i}\left((\Sigma^{t+1}_{c_i})^2,\bR_{l_i}^{t+1}\right)$
\State $t \gets t+1$
\State $\delta \gets 1/N\sum_i^N(a_i^t - a_i^{t-1})^2$
\EndWhile
\State \textbf{return} $\{a_i\}$
\end{algorithmic}
\end{minipage}
\caption{The simplified (with respect to Fig.~\ref{algo_AMP}) AMP decoder for sparse superposition codes, where we have approximated the squared elements of the matrix by their variance.}  
\label{algo_AMP2}
\end{figure}
Now using the definition of $\boldsymbol{\Theta}^{t+1}_{r}\eqdef \boldsymbol{1}_{\alpha_rN/L_c} \Theta^{t+1}_{r}$ where $\boldsymbol{1}_{u}$ is a vector full of ones of size $u$, we obtain
\begin{align}
\Theta^{t+1}_{r} = \frac{B}{L_c}\sum_k^{L_c} J_{r,k}\langle f_c^t\rangle_k \, = \,\frac{B}{L_c}\sum_k^{L_c} J_{r,k} (\Sigma_k^{t})^2 \langle (f_a^t)'\rangle_k \, , \label{eq_defThetaAMP2}
\end{align}
where we have used the property (\ref{eq_propertyfafc}) of the denoising function for the last equality. (\ref{eq:residual_monta}), (\ref{eq:ai_monta}) and (\ref{eq_defThetaAMP2}) together with the iteration of $\Sigma_c^t$ from Fig.~\ref{algo_AMP2} forms a set of closed equations, which is just AMP written in term of the residual. We now show that this form gives back the one of \cite{montanari2012graphical} in the full operator case, i.e when $L_c=L_r=J_{r,c}=1$. In this case, the quantities in the algorithm become
\begin{align}
\Theta^{t+1} &= B (\Sigma^{t})^2 \langle (f_a^t)'\rangle , \label{eq87}\\
(\Sigma^{t+1})^2 &= (\Theta^{t+1} + 1/{\rm snr})/(B\alpha),\label{eq88}\\
\end{align}
\begin{align}
a_{i}^{t+1} &= f_{a_i}\Big((\Sigma^{t+1})^2, \Big[a^t_j + (\Sigma^{t+1})^2 \sum_{\mu}^{M} F_{\mu j} \tau_\mu^t \frac{1}{\Theta^{t+1} + 1/{\rm snr}} : j\in l_i\Big] \Big)\nonumber\\
&=f_{a_i}\Big((\Sigma^{t+1})^2, \Big[a^t_j + \frac{1}{B\alpha} \sum_{\mu}^{M} F_{\mu j} \tau_\mu^t: j\in l_i\Big] \Big),  \\
\tau_\mu^t &= y_\mu - \sum_{i}^N F_{\mu i}a_i^t + \tau_\mu^{t-1}\frac{\Theta^{t+1}}{{1/{{\rm snr}}} + \Theta^t} \\
&=y_\mu - \sum_{i}^N F_{\mu i}a_i^t + \tau_\mu^{t-1}\frac{\langle (f_a^t)'\rangle }{\alpha},
\end{align}
using \eqref{eq87}, \eqref{eq88} for the last equality. The last step is to rescale the coding matrix by dividing its elements by $B\alpha: \tilde \bF\defeq \bF / (B\alpha)$, that implies that the codeword $\tilde \by$ is similarly rescaled. We finally obtain the more classical form of AMP for homogeneous matrices
\begin{align}
\tilde \tau_\mu^t &= \tilde y_\mu - \sum_{i}^N \tilde F_{\mu i}a_i^t + \tilde \tau_\mu^{t-1}\langle (f_a^t)'\rangle /\alpha,\\
(\Sigma^{t+1})^2 &= \Big((\Sigma^{t})^2 \langle (f_a^t)'\rangle  + 1/(B{\rm snr})\Big)/\alpha,\\
a_{i}^{t+1} &= f_{a_i}\Big((\Sigma^{t+1})^2, \Big[a^t_j + \sum_{\mu}^{M} \tilde F_{\mu j} \tilde \tau_\mu^t: j\in l_i\Big] \Big),
\end{align}
where $\tilde \btau^t$ is the rescaled residual, and
\begin{align}
\langle (f_a^t)'\rangle  &\defeq \frac{1}{N}\sum_{i}^{N} \frac{\partial f_{a_i}(x,\textbf{y})}{\partial y_i}\Big|_{(\Sigma^{t})^2,\big[a^{t-1}_j + \sum_{\mu}^{M} \tilde F_{\mu j} \tilde \tau_\mu^{t-1}:j\in l_i\big]} \\
&= \frac{1}{N(\Sigma^{t})^2}\sum_{i}^{N} f_{c_i}\Big((\Sigma^{t})^2,\Big[a^{t-1}_j + \sum_{\mu}^{M} \tilde F_{\mu j} \tilde \tau_\mu^{t-1}:j\in l_i\Big]\Big).
\end{align} 
\section{State evolution analysis} \label{app:SE}
We consider for the present analysis that the matrix $\bF$ has i.i.d Gausian entries with zero mean and a variance scaling as $O(1/L)$, such that the codeword has fluctuations (and thus a power) of $O(1)$.
\subsection{The Bayesian optimal setting and the Nishimori identity} \label{sec:NishimoriConditions}
Before deriving the state evolution analysis, let us show how the perfect knowledge of the statistical properties of the communication channel and of the signal prior imply great simplifications and useful identities for the state evolution and replica analysies. This perfect knowledge of all the problem parameters is refered as the \emph{Bayesian optimal setting}. The induced simplifications are refered as the \emph{Nishimori identities} in statistical physics
\cite{KrzakalaMezard12,MezardMontanari09}. While these were first derived as a concequence of a gauge symmetry in spin systems \cite{nishimori2001statistical}, they are actually a direct concequences of Bayes rule in the present continuous framework. We shall here derive the Nishimori
identities similarly to \cite{KrzakalaMezard12}. 

Assume the signal $\bx$ is distributed according to some known prior $P_0(\bx)$ and its observation $\by$ is drawn from some known conditional distribution $P(\by|\bx)$. Furthermore, assume $\hat \bx_1$ drawn from the posterior $P(\hat\bx_1|\by) = P_0(\hat\bx_1)P(\by|\hat\bx_1)/P(\by)$. Then for any function $g(\hat\bx_1, \bx)$, and using the Bayes formula, we obtain the following relation
\begin{align}
&\mathbb{E}_{\bx,\by}\{\mathbb{E}_{\hat\bx_1|\by} \{g(\hat\bx_1, \bx)\}\} = \mathbb{E}_{\bx}\{\mathbb{E}_{\by|\bx}\{\mathbb{E}_{\hat\bx_1|\by} \{g(\hat\bx_1, \bx)\}\}\} = \mathbb{E}_{\by} \{\mathbb{E}_{\hat\bx_2|\by}\{\mathbb{E}_{\hat\bx_1|\by} \{g(\hat\bx_1, \hat\bx_2)\}\}\}. \label{eq:Nishimori}
\end{align}
We have renamed $\hat\bx_2=\bx$ to emphasize that $\hat\bx_2\sim P(\hat\bx_2|\by)$ is drawn from the posterior (independently of $\hat\bx_1$) and thus play the same role as $\hat\bx_1$: we speak about two ``replicas''. The Nishimori identity \eqref{eq:Nishimori} says that if a function depending on both the signal and a replica is averaged with respect to both the quenched disorder and the posterior, then one may replace in it the signal by a new i.i.d replica distributed according to the posterior (the two replicas being distributed according to the product measure). In particular, this implies
\begin{align}
&\mathbb{E}_{\bx,\by}\{\mathbb{E}_{\hat\bx_1|\by} \{g(\hat\bx_1)\}\} = \mathbb{E}_{\bx} \{g(\bx)\}. \label{eq:Nishimori_signalOnly}
\end{align}

In statistical physics, macroscopic \emph{intensive} observables (i.e divided by the system size), such as the MMSE or ${\rm SER}$ in the present setting, are assumed to concentrate around their expectation with respect to the problem realization (that is with respect to the posterior and quenched disorder) as the system size diverges. This assumption implies that we can replace the asymptotic value of such observables by their expectation\footnote{See for example \cite{KoradaMacris_CDMA} for such concentration proofs.}.

Let us use this concentration assumption combined with the Nishimori identity to show a number of useful sub-identities. Define the following \emph{overlaps}
\begin{align}
&m  \defeq  \frac{1}{L}\sum_{l}^L \hat \bx_l^\intercal \bx_l,\ Q  \defeq  \frac{1}{L}\sum_{l}^L\hat\bx_l^\intercal \hat\bx_l, \ q \defeq \frac{1}{L} \sum_{l}^L {\hat\bx_l}^\intercal \hat\bx_l', \label{eq_repOrderParam_0}
\end{align}
where $\hat \bx, \hat \bx'$ are i.i.d replicas distributed according to \eqref{eq:posterior}. Recall that $\ba = \mathbb{E}_{\hat\bx|\by}\{\hat\bx\}$ is the MMSE estimate \eqref{eq_aiTrue}. The concentration of these overlaps together with the Nishimori identity straightforwardly implies that in the limit $L\to\infty$
\begin{align}
m &= q = \mathbb{E}_{\bx,\by}\{\mathbb{E}_{\hat\bx|\by} \{\mathbb{E}_{\hat\bx'|\by} \{q\}\}\} = \mathbb{E}_{\bx,\by}\Big\{\frac{1}{L} \sum_{l}^L \ba_l^\intercal \ba_l \Big\}, \label{eq101}\\
Q &= \mathbb{E}_{\bx,\by}\{\mathbb{E}_{\hat\bx|\by} \{Q\}\} = \mathbb{E}_{\bx,\by}\Big\{\mathbb{E}_{\hat\bx|\by} \Big\{\frac{1}{L} \sum_{l}^L\hat \bx_l^\intercal \hat \bx_l\Big\}\Big\}=\mathbb{E}_\bx \Big\{ \frac{1}{L} \sum_{l}^L \bx_l^\intercal \bx_l\Big\} = 1 ,\label{eq102}
\end{align}
where the last equality is valid for sparse superposition codes.

These identities imply various expressions of the MMSE $E$. Let us show that it equal to the posterior variance of the MMSE estimate. The MMSE is
\begin{align}
E &= \mathbb{E}_{\by,\bx}\Big\{\frac{1}{L} \sum_{l}^{L} (\ba_l - \bx_l)^\intercal(\ba_l - \bx_l) \Big\} = \mathbb{E}_{\by,\bx}\Big\{\mathbb{E}_{\hat \bx|\by}\Big\{\mathbb{E}_{\hat \bx'|\by}\Big\{\frac{1}{L} \sum_{l}^{L} (\hat \bx_l - \bx_l)^\intercal(\hat \bx_l' - \bx_l) \Big\}\Big\}\Big\}.
\end{align}
From the previous identities \eqref{eq101}, \eqref{eq102}, it can be written in the limit $L\to\infty$ as
\begin{align}
E &= q - 2m + Q = Q - m = 1 - m. \label{eq103}
\end{align}
Let us link this to the posterior expected variance $V$. By concentration and the Nishimori identity
\begin{align}
V &= \mathbb{E}_{\by,\bx}\Big\{\frac{1}{L}\sum_l^L\Big(\mathbb{E}_{\hat \bx|\by}\{\hat \bx_l^\intercal \hat \bx_l \} - \ba_l^\intercal \ba_l\Big)\Big\}= \mathbb{E}_{\by,\bx}\Big\{\mathbb{E}_{\hat \bx|\by}\Big\{\mathbb{E}_{\hat \bx'|\by}\Big\{\frac{1}{L}\sum_l^L\Big(\hat \bx_l^\intercal \hat \bx_l  - \hat \bx_l^\intercal \hat \bx_l'\Big)\Big\}\Big\}\Big\} \\
&=Q - q = 1 - m = E.
\end{align}
Thus the posterior expected variance $V$ and MMSE are asymptotically the equal.
\subsection{Coding matrices with homogeneous variance} \label{appSec:SE_Nonseeded}
We start deriving the state evolution analysis for coding matrices with homogeneous variance for all its entries. We consider constant power allocation which drastically simplifies the analysis due to the symmetry between all the sections. Non constant power allocation will follow as a special case of the analysis for structured matrices (next section) thanks to the discussion of sec.~\ref{sec:powA_SE}. The state evolution analysis may start from the cavity quantities that appeared in the derivation of AMP as in \cite{KrzakalaMezard12}, but here we will follow another path starting from the decoder Fig.\ref{algo_AMP} directly. 

The aim is to evaluate the AMP estimate of the signal components $\ba^t$ in order to compute the asymptotic ${\rm{MSE}}$ per section of AMP $E^t$. Once this is done, we can deduce the asymptotic ${\rm{SER}}^t$. The posterior estimate $\ba^t$ is given by (\ref{eq_meani}), thus it is a deterministic function of the AMP fields $(\bxigma_l^t, \bR_l^t)$. We thus need to access the asymptotic of $\bR_l^t$ as $L\to\infty$. We define
\begin{align}
\Lambda_\mu^t &\defeq \Theta_\mu^{t+1} \frac{y_\mu - w_\mu^t}{1/{{\rm snr}} + \Theta_\mu^{t}}, \label{eq:deflambdamu_0}\\
\br_l^{t+1} &\defeq \sum_{\mu}^M \bF_{\mu l} \Big[\sum_{k\neq l}^{L-1} \bF_{\mu k}^\intercal (\bx_k - \ba_k^t) + \xi_\mu  + \Lambda_\mu^t \Big]. \label{eq:deflambdamu}
\end{align}
Recall $\boldsymbol{1}_B$ is a vector of ones of size $B$. As $L\to\infty$, $\Theta_\mu^t$ becomes asymptotically independent of $\mu$ as we can replace the $F_{\mu i}^2$ elements by the matrix variance $1/L$ (see appendix~\ref{susec:fullTap} for a justification of this)
\begin{equation}
\Theta_\mu^t \approx \Theta^t \defeq \frac{1}{L}\sum_i^N v_i^t \Rightarrow (\bxigma^{t+1}_l)^2 \approx \frac{1/{{\rm snr}}+ \Theta^{t+1}}{B\alpha}\boldsymbol{1}_B, \label{eq:thetaSimplifies}
\end{equation}
where $\approx$ means equality at the leading order in $L$. Injecting the definitions given by Fig.~\ref{algo_AMP} of the quantities appearing in the $\bR_l^t$ term, recalling $B\alpha\defeq M/L$, using (\ref{eq_yfx}) to replace $\by$ and (\ref{eq:thetaSimplifies}), we obtain
\begin{align}
\bR_l^{t+1} &= \ba_l^t + (\bxigma^{t+1}_l)^2\sum_{\mu}^M \frac{\bF_{\mu l}}{1/{{\rm snr}}+ \Theta_\mu^{t+1}} \Big[\sum_k^L \bF_{\mu k}^\intercal(\bx_k - \ba_k^t) + \xi_\mu + \Lambda_\mu^t \Big] \approx \bx_l + \frac{\br_l^{t+1}}{B\alpha}. \label{eq_rl}
\end{align}
From (\ref{eq_rl}), we understand that $\br_l^{t+1}$ are the fluctuations of $\bR_l^{t+1}$ around the signal section $\bx_l$, due to the precense of noise. The aim is thus to compute the statistical properties of these fluctuations. The last step in (\ref{eq_rl}) uses that the second term of the right hand side of the following equality can be safely neglected, as we keep only the leading $O(1)$ terms when evaluating the moments of $\br_l^{t+1}$:
\begin{equation}
\sum_\mu^M\bF_{\mu l} \Big[\bF_{\mu l}^\intercal(\bx_l-\ba_l^t)\Big]=\underbrace{\Big[\sum_\mu^M F_{\mu i}^2(x_i-a_i^t):i\in l \Big]}_{=B\alpha(\bx_l - \ba_l^t)}+\Big[\underbrace{\sum_\mu^M \sum_{j\in  l:j\neq i}^{B-1} F_{\mu i}F_{\mu j}(x_j-a_j^t)}_{= O(1/\sqrt{L})} :i\in l \Big].
\label{eq_forgetSecPart}
\end{equation}

Recall that the $\bF$ entries are i.i.d, so one could ``naively'' think about applying the central limit theorem to $\br_l^{t+1}$ \eqref{eq:deflambdamu} in order to evaluate its distribution. Unfortunately, this is not justified as $\ba^{t}$ and $\boldsymbol{\Lambda}^t$ \eqref{eq:deflambdamu_0} are correlated with $\bF$ through the iterations of AMP. Nevertheless, in AMP, the Onsager reaction terms (the second terms in the iterations of $\bw^{t+1}$ and $\bR^{t+1}$ in Fig.~\ref{algo_AMP}) are precisely removing these correlations from iteration to iteration. We refer to \cite{BayatiMontanari10} for intuition on this fact and rigorous statements. So despite not being a priori justified, this naive application of the central limit theorem brings the good result, and suggests that $\br_l^{t+1}$ is Gaussian distributed with moments that we evaluate now. Note that the derivation starting from the relaxed-BP algorithm of appendix~\ref{app:relaxedBP} would not require this assumption, see \cite{KrzakalaPRX2012} for this derivation.

We remind the reader that the noise has zero mean. Furthermore, in the large size limit the ${\rm{MSE}}$ per component is assumed to concentrate on its expectation with respect to the quenched disorder
\begin{equation}
\tilde E^{t+1} = \frac{1}{N}\sum_k^N [x_k - a_k^{t+1}]^2 \underset{L\to\infty}{\to} \mathbb{E}_{\bF,\bx,\bxi}\Big\{\frac{1}{N}\sum_k^N[x_k - a_k^{t+1}]^2\Big\}.
\end{equation}
The matrix $\bF$ being of $0$ mean, only the terms with even power of the matrix entries will survive in the following equations, because of the quenched average. $\bzero_B$ is a vector of zeros of size $B$. Let us start by computing the mean of the fluctuations $\br_l^{t+1}$. First notice from (\ref{eq:deflambdamu}) combined with (\ref{eq_appThetamu}), (\ref{eq_appW}) that we can identify 
\begin{align}
&\Lambda^t_\mu \approx \sum_k^L \bF_{\mu k}^\intercal\beps_{a_{k\mu}}, \label{eq_lambdaMu}
\end{align}
where $\beps_{a_{k\mu}} = O(1/\sqrt{L})$ is given by (\ref{eq_appCorrections}), (\ref{eq_appW}). From this and \eqref{eq:deflambdamu} we obtain
\begin{align}
\mathbb{E}_{\bF,\bxi,\bx} \{\br_l^{t+1}\} &= \underbrace{\mathbb{E}_{\bF,\bxi,\bx} \Big\{\sum_\mu^M \bF_{\mu l}\sum_{k\neq l}^{L-1} \bF_{\mu k}^\intercal (\bx_k - \ba_k^t) + \sum_{\mu}^M \bF_{\mu l}\xi_{\mu}\Big\} }_{=\bzero_B} + \mathbb{E}_{\bF,\bxi,\bx}\Big\{\sum_{\mu}^M\bF_{\mu l} \sum_{k}^L \bF_{\mu k}^\intercal \beps_{a_{k \mu}}\Big\} \label{eq_rFirstMom1}\\
&\approx \underbrace{\mathbb{E}_{\bF,\bxi,\bx} \Big\{\sum_{\mu}^M (\bF_{\mu l}^3)^\intercal \bv_l\frac{y_\mu - w_\mu^t}{1/{{\rm snr}} + \Theta^t}\Big\}}_{= O(1/L)} \approx \bzero_B,  \label{eq_rFirstMom2}\\
\end{align}
We now turn our attention to the cross terms. If $l'\neq l$ then
\begin{align}
\mathbb{E}_{\bF,\bxi,\bx}\{{\br_l^{t+1} \br_{l'}^{t+1}}\} &= \mathbb{E}_{\bF,\bxi,\bx}\Big\{\sum_{\mu,\nu}^{M,M} \bF_{\mu l} \bF_{\nu l'}\Big[\sum_{k\neq l}^{L-1} \bF_{\mu k}^\intercal (\bx_k - \ba_k^t) + \xi_\mu  + \Lambda_\mu^t  \Big] \Big[\sum_{k'\neq l'}^{L-1} \bF_{\nu k'}^\intercal (\bx_{k'} - \ba_{k'}^t) + \xi_\nu  + \Lambda_\nu^t  \Big]\Big\} \label{eq_rSecMomCross1}\\
&=\mathbb{E}_{\bF,\bxi,\bx}\Big\{\sum_{\mu}^M \bF_{\mu l} \bF_{\mu l'}\Big[\bF_{\mu l'}^\intercal (\bx_{l'} - \ba_{l'}^t) + \xi_\mu  + \Lambda_\mu^t  \Big] \Big[\bF_{\mu l}^\intercal (\bx_{l} - \ba_{l}^t) + \xi_\mu  + \Lambda_\mu^t  \Big] \Big\} \label{eq_rSecMomCross2}\\
&=\underbrace{\mathbb{E}_{\bF,\bxi,\bx}\Big\{\sum_{\mu}^M \bF_{\mu l}^2 \bF_{\mu l'}^2(\bx_{l'} - \ba_{l'}^t)(\bx_{l} - \ba_{l}^t) \Big\}}_{= O(1/L)} \approx \bzero_B .\label{eq_rSecMomCross3}
\end{align}
We then compute the variance
\begin{align}
\mathbb{E}_{\bF,\bxi,\bx}\{(\br_l^{t+1})^2\} &= \mathbb{E}_{\bF,\bxi,\bx}\Big\{\sum_{\mu,\nu}^{M,M} \bF_{\mu l}\bF_{\nu l} \Big[\sum_{k\neq l}^{L-1} \bF_{\mu k}^\intercal (\bx_k - \ba_k^t) + \xi_\mu  + \Lambda_\mu^t  \Big] \Big[\sum_{k'\neq l}^{L-1} \bF_{\nu k'}^\intercal (\bx_{k'} - \ba_{k'}^t) + \xi_\nu  + \Lambda_\nu^t  \Big]\Big\} \label{eq_rVar1} \\
&=\mathbb{E}_{\bF,\bxi,\bx}\Big\{\sum_{\mu}^M \bF_{\mu l}^2 \Big[\sum_{k\neq l}^{L-1} \bF_{\mu k}^\intercal (\bx_k - \ba_k^t) + \xi_\mu  + \Lambda_\mu^t  \Big] \Big[\sum_{k'\neq l}^{L-1} \bF_{\mu k'}^\intercal (\bx_{k'} - \ba_{k'}^t) + \xi_\mu  + \Lambda_\mu^t  \Big]\Big\}\label{eq_rVar2} \\
&=\mathbb{E}_{\bF,\bxi,\bx}\Big\{\sum_{\mu}^M \bF_{\mu l}^2 \Big[\sum_{k\neq l}^{L-1} \bF_{\mu k}^\intercal (\bx_k - \ba_k^t) \Big] \Big[\sum_{k'\neq l}^{L-1} \bF_{\mu k'}^\intercal (\bx_{k'} - \ba_{k'}^t) \Big]\Big\}\nonumber\\ 
&\ \ \ + \frac{\alpha B}{{{\rm snr}}}\boldsymbol{1}_B + \underbrace{\mathbb{E}_{\bF,\bxi,\bx}\Big\{\sum_\mu^M \bF_{\mu l}^2 \Lambda_\mu^2\Big\}}_{= O(L^{-3/2})} + 2\underbrace{\mathbb{E}_{\bF,\bxi,\bx} \Big\{\sum_{\mu}^M \bF_{\mu l}^2 \Lambda_\mu \Big[\sum_{k\neq l}^{L-1} \bF_{\mu k}^\intercal (\bx_k - \ba_k^t)\Big] \Big\}}_{=\boldsymbol{0}_B}\label{eq_rVar3}
\end{align}
\begin{align}
&\approx \mathbb{E}_{\bF,\bxi,\bx}\Big\{\sum_{\mu}^M \bF_{\mu l}^2 \Big[\sum_{k\neq l}^{L-1} (\bF_{\mu k}^2)^\intercal (\bx_k - \ba_k^t)^2 \Big] \Big\} + \frac{\alpha B}{{{\rm snr}}}\boldsymbol{1}_B\\
&\approx \Big(\frac{M}{L^2}\sum_{k}^L\mathbb{E}_{\bF,\bxi,\bx}\{(\bx_k - \ba_k^t)^\intercal(\bx_k - \ba_k^t)\}+ \frac{\alpha B}{{{\rm snr}}}\Big) \boldsymbol{1}_B =B\alpha \Big({1/{{\rm snr}}} + B \tilde E^t\Big) \boldsymbol{1}_B.
\end{align}
Now that we have computed the moments of the Gaussian fluctuation $\br_l^{t+1}$, from \eqref{eq_rl} we can write $R_i^t$ as a random Gaussian variable as well:
\begin{align}
r_i^{t+1} \sim \mathcal{N}\Big(r_i^{t+1}\Big|0,B\alpha ({1/{{\rm snr}}} + B \tilde E^t)\Big) \Rightarrow R_i^{t+1} \sim \mathcal{N}\Big(R_i^{t+1}\Big|x_i,\frac{{1/({{\rm snr}}}B)+\tilde E^t}{\alpha}\Big).
\end{align}
It remains to perform the average with respect to the signal $\bx_l \sim P_0(\bx_l)$ as it is the mean of $R_i^{t+1}$. We can focus a single section as the ${\rm{MSE}}$ is asymptotically homogeneous over all sections, the power allocation being constant. Thus the state evolution recursion for the ${\rm MSE}$ per component of AMP reads
\begin{align}
\tilde E^{t+1} &= \frac{1}{B} \sum_{i\in  l}^B\int d\bx_l \ \!\mathcal{D}\bz\ \! P_0(\bx_l) \left[f_{a_i}\left((\tilde\Sigma^{t+1})^2, \bR^{t+1}(\bz,\bx_l)\right) - x_i\right]^2, \label{eq_E1} \\
\tilde \Sigma^{t+1}(\tilde E^t) &\defeq \sqrt{\frac{1/({{\rm snr}}B) + \tilde E^t}{\alpha} }= \sqrt{\frac{RB({1/({{\rm snr}}}B) + \tilde E^t)}{\log_2 B} }, \label{eq_tildeS2def} \\ 
\bR^{t+1}(\bz,\bx_l) &\defeq \bx_l + \bz \tilde\Sigma^{t+1}, \label{eq128}
\end{align}
where $\mathcal{D} \bz \defeq \prod_{i}^{B} \mathcal{D} z_i = \prod_{i}^{B} \mathcal{N}(z_i|0,1) dz_i $ is a $B$-d standardized Gaussian measure.
% Now using the prior for constant power allocation $P_0(\bx_l) \defeq1/B \sum_{i\in  l}\delta(x_i-1)\prod_{j\in  l : j\neq i}\delta(x_j)$, we obtain
%
% \begin{align}
% \tilde E^{t+1} &= \frac{1}{B} \int \mathcal{D} \bz \left[f_{a_1}\left((\tilde\Sigma^{t+1})^2,\bR^{(1),t+1}(\bz) \right)-1\right]^2 + \frac{B-1}{B} \int \mathcal{D} \bz \ f_{a_1}\left((\tilde\Sigma^{t+1})^2, \bR^{(2),t+1}(\bz)\right)^2, \label{eq_E2}\\
% %
% R^{(i),t+1}_j(\bz, \tilde E^t) &\defeq \delta_{i,j} + z_j \tilde \Sigma^{t+1}(\tilde E^t), \label{eq_tildeR}
% \end{align}
% %
%where $f_{a_i}$ is given by (\ref{eq_meani}) and $\delta_{i,j}$ being the kroneker symbol. 
Let us define quantities that do not scale with $B$
\begin{align}
E^t \defeq B\tilde E^t, ~ \Sigma^{t+1}(E^{t}) \defeq \tilde \Sigma^{t+1}(E^t) \sqrt{\ln(B)} = \sqrt{R({1/{{\rm snr}}} + E^t)\ln(2)}. \label{eq_defS2}
\end{align}
Now using the prior for sparse superposition codes with constant power allocation $P_0(\bx_l) \defeq1/B \sum_{i\in  l}\delta(x_i-1)\prod_{j\in  l : j\neq i}\delta(x_j)$ and after some algebra, one obtains the final form of the state evolution given by (\ref{eq_SE_MSE}).

The Nishimori identity \eqref{eq:Nishimori} implies another way of expressing the ${\rm{MSE}}$ that will be useful later on in appendix~\ref{subsec:repIsSE} to show the equivalence between the replica and state evolution analysies. In appendix~\ref{sec:NishimoriConditions}, we have shown in full generality that in the Bayes optimal setting, the MMSE associated with some posterior can be written in the last form of \eqref{eq103}. As already noticed in appendix~\ref{app:relaxedBP} (see also \eqref{eq_marginal}, \eqref{eq:fai}, \eqref{eq_meani}), the denoiser $f_{a_i}((\tilde\Sigma^{t+1})^2, \bR^{t+1}(\bz,\bx_l))$ is the Bayes optimal MMSE estimator for a $B$-d AWGN channel with noise variance $(\tilde\Sigma^{t+1})^2$ and channel observation $\bR^{t+1}(\bz,\bx_l)$ given by \eqref{eq128}. Thus from \eqref{eq_E1}, one sees that $\tilde E^{t+1}$ actually corresponds to the (averaged) MMSE associated with this simple $B$-d vectorial AWGN channel and thus the Nishimori identity is valid. It implies that \eqref{eq103} is true, where the overlap $m$ is given by \eqref{eq_repOrderParam_0}. It leads to another equivalent form of the state evolution
\begin{align}
E^{t+1} &= 1 - \sum_{i\in  l}^B\int d\bx_l \ \!\mathcal{D}\bz\ \! P_0(\bx_l) f_{a_i}\big((\tilde\Sigma^{t+1})^2, \bR^{t+1}(\bz,\bx_l)\big)x_i = 1 - \int \mathcal{D}\bz\ \! f_{a_{1|1}}\big((\Sigma^{t+1})^2,\bz \big), \label{SE_MSE2}
\end{align}
where we have used (\ref{eq_tildeS2def}), (\ref{eq_defS2}) and the last equality is obtained using the prior (\ref{eq_prior}), (\ref{eq_fa1fun}) and integrating $\bx_l$. We have used here that the overlap $m$ concentrates on its expectation $\mathbb{E}_{\bx,\by}\{\mathbb{E}_{\hat\bx|\by}\{\hat \bx\}\bx\}$ (thus the precense of the averages with respect to the prior and the noise $\bz$ of the effective AWGN channel \eqref{eq128}).

This equivalent form of the state evolution is computationally easier and faster to compute. But it can be more cumbersome to use than (\ref{eq_SE_MSE}) as it may become negative if the difference is really small, due to finite numerical precision. 
\subsection{Coding matrices with a block structure} \label{appSec:SE_seeded}
The derivation of the state evolution in the block structured case is
very similar to the homogeneous one. The difference is that now each block of the matrix can have a different variance. We give here the main steps, the details being similar to the previous section. All the computations are done keeping in mind the limit $L\gg L_c, L_r$ in wich AMP is valid with structured matrices. As before, we start from the algorithm Fig.~\ref{algo_AMP} and the operators definitions (\ref{eq_fastOpDefs2}). Note also that the arguments on the use of the central limit theorem, and the decorrelations happening due to the Onsager reaction terms discussed in the previous section are valid here. 

Let us study the fluctuations of the AMP field $\bR_l^{t+1}$. We define 
\begin{equation}
\Lambda_\mu^t \defeq \Theta_\mu^{t+1} \frac{y_\mu - w_\mu^t}{1/{{\rm snr}} + \Theta_\mu^{t}}. 
\end{equation}
Then similarly as before, but making the block structure explicit, one gets
\begin{align}
\bR_l^{t+1} &= \ba_l^t + (\bxigma^{t+1}_l)^2\sum_r^{L_r}\sum_{\mu\in r}^{\alpha_r N/L_c} \frac{\bF_{\mu l}}{1/{{\rm snr}}+ \Theta_\mu^{t+1}} \Big[\sum_c^{L_c} \sum_{k\in c}^{L/L_c} \bF_{\mu k}^\intercal(\bx_k - \ba_k^t) + \xi_\mu +  \Lambda_\mu^t \Big].
\end{align}
Using that the variance of the entries of $\bF$ depends only on the block indices, we obtain
\begin{align}
\Theta_\mu &= \sum_c^{L_c}\sum_{l\in c}^{L/L_c}(\bF_{\mu l}^2)^\intercal\bv_l \approx \sum_c^{L_c} \frac{J_{r_\mu,c}}{L}\sum_{l\in c}^{L/L_c}\sum_{i\in  l}^B v_i \eqdef \Theta_{r_\mu}, \label{eq_thetar}\\
\Rightarrow \Lambda_{\mu}^t &= \Theta_{r_\mu}^{t+1}\frac{y_\mu - w_\mu^t}{1/{{\rm snr}} + \Theta_{r_\mu}^{t}},
\end{align}
where $J_{r,c}/L$ is the variance of the entries of $\bF$ composing the block with indices $(r,c)$, see Fig.~\ref{fig_seededHadamard}. Recall the notation $r_\mu$ ($c_l$) means the block index $r\in\{1,\ldots,L_r\}$ (resp. $c\in\{1,\ldots,L_c\}$) to which the factor index $\mu$ (resp. section index $l$) belongs to. The last relation allows to simplify $(\bxigma_l^{t+1})^2$ appearing in Fig.~\ref{algo_AMP} as
\begin{equation}
(\bxigma_l^{t+1})^2 = \frac{L_c}{B} \Big(\sum_r^{L_r} \frac{J_{r,c_l}\alpha_r}{1/{{\rm snr}} + \Theta_{r}^{t+1}}\Big)^{-1}\boldsymbol{1}_B \eqdef (\Sigma_{c_l}^{t+1})^2 \boldsymbol{1}_B. \label{eq_sigmalSq}
\end{equation}
The parameters $(\alpha_r,L_c,L_r)$ are defined in sec.~\ref{sec:AMP}. We deduce the expression of $\bR_l^{t+1}$
\begin{align}
\bR_l^{t+1} &\approx \ba_l + (\bxigma^{t+1}_l)^2 \sum_r^{L_r}\frac{1}{1/{{\rm snr}} + \Theta_{r}^{t+1}} \sum_{\mu\in r}^{\alpha_rN/L_c} \bF_{\mu l} \Big[\sum_c^{L_c}\sum_{k\in c:k\neq l}^{L/L_c} \bF_{\mu k}^\intercal (\bx_k - \ba_k^t) + \xi_\mu  + \Lambda_{\mu}^t  \Big] \nonumber \\ 
&\ + (\bxigma^{t+1}_l)^2 \underbrace{\sum_r^{L_r}\frac{1}{1/{{\rm snr}} + \Theta_{r}^{t+1}}\sum_{\mu\in r}^{\alpha_rN/L_c} \bF_{\mu l} \Big[\bF_{\mu l}^\intercal(\bx_l-\ba_l)\Big]}_{\defeq U}.
\end{align}
We now notice that
\begin{align}
&U=(\bxigma^{t+1})^{-2} (\bx_l - \ba_l) + O(1/\sqrt{L}).
\end{align}
This allows to obtain, using a simplification similar to (\ref{eq_forgetSecPart})
\begin{align}
\bR_l^{t+1} &\approx \bx_l + (\bxigma^{t+1}_l)^2 \sum_r^{L_r}\frac{1}{1/{{\rm snr}} + \Theta_{r}^{t+1}} \sum_{\mu\in r}^{\alpha_rN/L_c} \bF_{\mu l} \Big[\sum_c^{L_c}\sum_{k\in c:k\neq l}^{L/L_c} \bF_{\mu k}^\intercal (\bx_k - \ba_k^t) + \xi_\mu  + \Lambda_{\mu}^t  \Big]. \label{eq_Rltp1}
\end{align}
We define
\begin{align}
\br_{r l}^{t+1} &\defeq \sum_{\mu\in r}^{\alpha_rN/L_c} \bF_{\mu l} \Big[\sum_c^{L_c}\sum_{k\in c:k\neq l}^{L/L_c} \bF_{\mu k}^\intercal (\bx_k - \ba_k^t) + \xi_\mu  + \Lambda_{\mu}^t  \Big], \\
\br_{l}^{t+1} &\defeq (\bxigma^{t+1}_l)^2 \sum_r^{L_r} \frac{\br_{r l}^{t+1}}{1/{{\rm snr}} + \Theta_r^{t+1} }, \label{eq:defrl}\\
\Rightarrow \bR_l^{t+1} &\approx \bx_l + \br_l^{t+1}. \label{eq:largeRfuncr_sc}
\end{align}
We can now compute the moments of the Gaussian distributed variables $\br_{r l}^{t+1}$ in order to deduce the distribution of $\br_{l}^{t+1}$ (see the previous section on why these variables are indeed Gaussian distributed). As before, we only keep the $O(1)$ terms. We can actually identify $\br_{r l}^{t+1}$ with $\br^{t+1}$ of (\ref{eq_rl}) and thus the computations are exactly the same as in the previous section, except that the variance in now a function of the block indices. Using (\ref{eq_appW}), the equation (\ref{eq_lambdaMu}) remains valid, so that we get a similar result to (\ref{eq_rFirstMom1}) and (\ref{eq_rFirstMom2})
\begin{equation}
\mathbb{E}_{\bF,\bxi,\bx} \{\br_{rl}^{t+1}\} \approx \bzero_B \Rightarrow \mathbb{E}_{\bF,\bxi,\bx} \{\br_{l}^{t+1}\} \approx \bzero_B. \label{eq_rlSeeded}
\end{equation}
The cross terms cancel as well at the dominant order. Indeed, if $l'\neq l$ then
\begin{align}
\mathbb{E}_{\bF,\bxi,\bx}\{{\br_l^{t+1} \br_{l'}^{t+1}}\} = (\bxigma^{t+1}_l)^2(\bxigma^{t+1}_{l'})^2 \sum_{r,r'}^{L_r,L_r} \frac{\mathbb{E}_{\bF,\bxi,\bx} \{\br_{rl}^{t+1}\}\mathbb{E}_{\bF,\bxi,\bx}\{\br_{r'l'}^{t+1}\}}{(1/{{\rm snr}} + \Theta_{r}^{t+1})(1/{{\rm snr}} + \Theta_{r'}^{t+1})} \approx \bzero_B.
\end{align}
The only moment that changes is the variance. Skipping some steps similar to (\ref{eq_rVar1}), (\ref{eq_rVar2}), we get
\begin{align}
&\mathbb{E}_{\bF,\bxi,\bx}\{(\br_{rl}^{t+1})^2\} =\mathbb{E}_{\bF,\bxi,\bx}\Big\{\sum_{\mu\in r}^{\alpha_rN/L_c} \bF_{\mu l}^2 \Big[\sum_c^{L_c}\sum_{k\in c:k\neq l}^{L/L_c} \bF_{\mu k}^\intercal (\bx_k - \ba_k^t) \Big] \Big[\sum_{c'}^{L_c}\sum_{k'\in c':k'\neq l}^{L/L_c} \bF_{\mu k'}^\intercal (\bx_{k'} - \ba_{k'}^t) \Big]\Big\}\nonumber\\ 
&+ \mathbb{E}_{\bF,\bxi,\bx}\Big\{\sum_{\mu\in r}^{\alpha_rN/L_c} \bF_{\mu l}^2 \xi_\mu^2\Big\} + \underbrace{\mathbb{E}_{\bF,\bxi,\bx}\Big\{\sum_{\mu\in r}^{\alpha_rN/L_c} \bF_{\mu l}^2 \Lambda_\mu^2\Big\}}_{= O(L^{-3/2})} + 2\underbrace{\mathbb{E}_{\bF,\bxi,\bx} \Big\{\sum_{\mu\in r}^{\alpha_rN/L_c} \bF_{\mu l}^2 \Lambda_\mu \Big[\sum_c^{L_c}\sum_{k\in c:k\neq l}^{L/L_c} \bF_{\mu k}^\intercal (\bx_k - \ba_k^t)\Big] \Big\}}_{=\boldsymbol{0}_B}.
\end{align}
Let us define the asymptotic ${\rm{MSE}}$ per block: $\tilde E_c$ is the ${\rm{MSE}}$ of the block $c$ of the signal, where the block structure of the signal is induced by the design of the spatially coupled operator (see Fig.~\ref{fig_seededHadamard}):
\begin{align}
\tilde E_c\defeq\frac{L_c}{N}\sum_{k\in c}^{L/L_c} \mathbb{E}_{\bF,\bxi,\bx}\{(\bx_k - \ba_k^t)^\intercal(\bx_k - \ba_k^t)\}. \label{eq_tildeEc}
\end{align}
With this new definition, we obtain
\begin{align}
\mathbb{E}_{\bF,\bxi,\bx}\{(\br_{rl}^{t+1})^2\} &\approx \frac{\alpha_r B J_{r,c_l}}{{{\rm snr}}L_c} \boldsymbol{1}_B + \mathbb{E}_{\bF,\bxi,\bx}\Big\{\sum_{\mu\in r}^{\alpha_rN/L_c} \bF_{\mu l}^2 \Big[\sum_c^{L_c}\sum_{k\in c:k\neq l}^{L/L_c} (\bF_{\mu k}^2)^\intercal (\bx_k - \ba_k^t)^2 \Big] \Big\}\\
&\approx \Big(\frac{\alpha_r B J_{r,c_l}}{{{\rm snr}}L_c} + \sum_{\mu\in r}^{\alpha_rN/L_c} \frac{J_{r,c_l}}{L}\Big[\sum_c^{L_c} \frac{J_{r,c}}{L} \sum_{k\in c:k\neq l}^{L/L_c} \mathbb{E}_{\bF,\bxi,\bx}\{(\bx_k - \ba_k^t)^\intercal(\bx_k - \ba_k^t)\} \Big]\Big)\boldsymbol{1}_B\\ 
&\approx \frac{\alpha_r B J_{r,c_l}}{L_c} \Big(\frac{1}{{{\rm snr}}} + \frac{B}{L_c}\sum_c^{L_c} J_{r,c}\tilde E_c\Big)\boldsymbol{1}_B. \label{eq_rl2last}
\end{align}
The variance of $\br_{r}^{t+1}$ is deduced from (\ref{eq:defrl}), (\ref{eq_rlSeeded}) using again the independence of the matrix elements
\begin{align}
\mathbb{E}_{\bF,\bxi,\bx}\{(\br_{l}^{t+1})^2\} = (\bxigma_l^{t+1})^4\sum_{r,r'}^{L_r,L_r} \frac{\mathbb{E}_{\bF,\bxi,\bx}\{\br_{rl}^{t+1}\br_{r'l}^{t+1}\}}{(1/{{\rm snr}} + \Theta_{r}^{t+1})(1/{{\rm snr}} + \Theta_{r'}^{t+1})} =(\bxigma_l^{t+1})^4\sum_{r}^{L_r} \frac{\mathbb{E}_{\bF,\bxi,\bx}\{(\br_{rl}^{t+1})^2\}}{(1/{{\rm snr}} + \Theta_{r}^{t+1})^2}. \label{eq_meanrl2}
\end{align}
We define the average variance of the estimates inside the block $c$ of the estimated signal as 
\begin{equation}
\tilde V_c\defeq \frac{L_c}{N} \sum_{l\in c}^{L/L_c}\sum_{i \in  l}^{B} v_i.
\end{equation}
The Nishimori identities derived in appendix~\ref{sec:NishimoriConditions} allow to write $\tilde V_c=\tilde E_c \ \forall \ c \in \{1,\ldots,L_c\}$. From this and (\ref{eq_thetar}), we can rewrite $\Theta_{r}$ as
\begin{equation}
\Theta_r = \frac{B}{L_c} \sum_c^{L_c} J_{r,c}\Tilde V_c= \frac{B}{L_c} \sum_c^{L_c} J_{r,c}\Tilde E_c.
\end{equation}
We plug this expression into (\ref{eq_rl2last}) and using (\ref{eq_meanrl2}), (\ref{eq_sigmalSq}) we obtain
\begin{align}
\mathbb{E}_{\bF,\bxi,\bx}\{(\br_{l}^{t+1})^2\} = (\bxigma_l^{t+1})^4 \frac{B}{L_c} \sum_{r}^{L_r} \frac{\alpha_rJ_{r,c_l}(1/{{\rm snr}} + \Theta_{r}^{t+1})}{(1/{{\rm snr}} + \Theta_{r}^{t+1})^2} = (\bxigma_l^{t+1})^2.
\end{align}
We now know the distribution of $R_i^{t+1}$. From (\ref{eq:largeRfuncr_sc}) and (\ref{eq_sigmalSq}) one obtains
\begin{align}
&r_i^{t+1} \sim \mathcal{N}\big(r_i^{t+1}\big|0,(\tilde \Sigma_{c_i}^{t+1})^2\big) \Rightarrow R_i^{t+1}\sim \mathcal{N}\big( R_i^{t+1}\big|x_i,(\tilde \Sigma_{c_i}^{t+1})^2\big),\\
&\tilde \Sigma_c^{t+1}(\{E_{c'}^t\}) = \Big[B\sum_{r}^{L_r} \frac{\alpha_{r} J_{r,c}}{{L_c/{{\rm snr}}} + B\sum_{c'}^{L_c} J_{r,c'}\tilde E_{c'}^t}\Big]^{-1/2}.
\end{align}
Defining $E_c^t \defeq B\tilde E_c^t$ and $\Sigma_c^{t+1} \defeq \tilde \Sigma_c^{t+1} \sqrt{\ln(B)}$, from the same arguments as in the previous section, we finally obtain the state evolution for block structured coding operators (\ref{eq_SESeeded}), (\ref{eq_SEsigmaSeeded}).
\section{Replica analysis}
\label{app:replicaAnalysis}
The following derivation is similar to the free entropy calculation of \cite{KrzakalaMezard12} with the difference that in the present case, computations are made considering $B$-d i.i.d variables, that is the sections. We place ourselves in the constant power allocation case. As for the state evolution analysis, we consider $\bF$ to be drawn from the ensemble of random matrices with i.i.d zero mean Gaussian entries, and variance scaling as $O(1/L)$.
\subsection{Derivation of the replica symmetric free entropy for constant power allocation by the replica method} \label{app:replicaAnalysis_A}
We start from the definition of the potential at fixed section size $B$, or replica free entropy, that can be expressed using the replica trick \cite{MezardMontanari09} as
\begin{equation}
\Phi_B \defeq \lim_{L\to \infty} \frac{1}{L}\mathbb{E}_{\bF,\bxi,\bx}\{\ln Z\} = \lim_{L\to \infty}\lim_{n\to0} \frac{\mathbb{E}_{\bF,\bxi,\bx}\{Z^n\}-1}{Ln},\label{eq_replicaTrick}
\end{equation}
where $\mathbb{E}_{\bF,\bxi,\bx}$ is the average over the quenched disorder and $Z$ is the partition function \eqref{eq_fullZ} (a random variable of the channel observation, itself a function of the quenched disorder). $Z^n$ is the so-called replicated partition function as it can be interpreted as the partition function associated with $n$ ``replicas'' $\{\hat \bx^a\}$ indexed by $a \in \{1,\ldots,n\}$, all independently drawn from the posterior \eqref{eq:posterior}. We define 
\begin{equation}
v_\mu^a \defeq \sum_{l}^L\bF_{\mu l}^\intercal (\bx_l - \hat\bx_l^a) + \xi_\mu, \quad X_\mu \defeq \mathbb{E}_{\bF,\bxi}\Big\{e^{-\frac{{{\rm snr}}}{2}\sum_{a}^n (v_\mu^a)^2}\Big\}.
\end{equation}
Then the replicated partition function is
\begin{align}
\mathbb{E}_{\bF,\bxi,\bx}\{Z^n\} &= ({{\rm snr}/2\pi})^{\frac{Mn}{2}} \mathbb{E}_\bx \Big\{\int \Big[\prod_{l,a}^{L,n} d\hat\bx_l^a P_0(\hat\bx_l^a) \Big]\prod_{\mu}^M X_\mu\Big\}. \label{eq:replicatedZ_before}
\end{align}
In order to compute $X_\mu$, we apply the central limit theorem to $v_\mu^a$ (the sum appearing in it is a sum of i.i.d terms). We thus need its first two moments to define its associated Gaussian distribution. It has zero mean $\mathbb{E}_{\bF,\bxi} \{v_\mu^a\} = 0$ and a variance
\begin{align}
\mathbb{E}_{\bF,\bxi} \{(v_\mu^a)^2 \} &= \mathbb{E}_{\bF,\bxi} \Big\{\sum_{l,k}^{L,L} [\bF_{\mu l}^\intercal (\bx_l - \hat\bx^a_l)]^\intercal \bF_{\mu k}^\intercal (\bx_k - \hat\bx^a_k) + 2\xi_\mu \sum_{l}^L \bF_{\mu l}^\intercal (\bx_l - \hat\bx^a_l) + \xi_\mu^2\Big\}\nonumber\\
&= \sum_{l,k}^{L,L} \Big[(\bx_l - \hat\bx^a_l)^\intercal \mathbb{E}_{\bF}\{\bF_{\mu l} \bF_{\mu k}^\intercal\} (\bx_k - \hat\bx^a_k) \Big] + {1/{{\rm snr}}}.
\end{align}
Using the fact that each element of the matrix is i.i.d, we find that only the diagonal elements of the matrix $\mathbb{E}_{\bF}\{\bF_{\mu l} \bF_{\mu k}^\intercal\}$ are not vanishing 
\begin{equation}
\mathbb{E}_{\bF}\{\bF_{\mu l} \bF_{\mu k}^\intercal\} = \frac{\delta_{k,l}}{L} \boldsymbol{I}_{B,B} \Rightarrow \mathbb{E}_{\bF,\bxi} \{(v_\mu^a)^2\} = \frac{1}{L} \sum_{l}^L(\bx_l - \hat\bx^a_l)^\intercal(\bx_l - \hat\bx^a_l) + {1/{{\rm snr}}}, \label{eq_vmu2}
\end{equation}
where $\boldsymbol{I}_{B,B}$ is the identity matrix of dimension $B\times B$. Now we define new parameters, referred as overlaps (similarly as \eqref{eq_repOrderParam_0})
\begin{align}
&m_a  \defeq  \frac{1}{L}\sum_{l}^L (\hat\bx_l^a)^\intercal \bx_l,\ Q_a  \defeq  \frac{1}{L}\sum_{l}^L(\hat\bx_l^a)^\intercal \hat\bx_l^a, \ q_{ab} \defeq \frac{1}{L} \sum_{l}^L (\hat\bx_l^a)^\intercal \hat\bx_l^b. \label{eq_repOrderParam}
\end{align}
$m_a$ is the overlap between the replica $\hat\bx^a$ and the signal $\bx$, $Q_a$ is the self overlap of $a$ and $q_{ab}$ is the overlap between replicas $a$ and $b$. These overlaps will serve as a way to re-parametrize the averaged replicated partition function \eqref{eq:replicatedZ_before}, which makes its evaluation easier. The MMSE $E$ is linked to the overlaps by 
\begin{equation} \label{eq151}
E \defeq \langle x^2 \rangle_L -2m +q, 
\end{equation}
where $\langle x^2 \rangle_L \defeq \sum_{l}^L \bx_l^\intercal\bx_l/L = 1$ for sparse superposition codes. Using these, the variance becomes
\begin{align}
\mathbb{E}_{\bF,\bxi} \{(v_\mu^a)^2 \} = 1 - 2m_a + Q_a + {1/{{\rm snr}}}.
\end{align}
Exactly in the same way, we get the cross terms $\forall\ a\neq b$
\begin{equation}
\mathbb{E}_{\bF,\bxi} \{v_\mu^a v_\mu^b\} = 1 - (m_a + m_b) + q_{ab} + {1/{{\rm snr}}}.
\end{equation}
We now introduce the replica symmetric ansatz, that is known to be valid for inference problems as long as the Nishimori identity \eqref{eq:Nishimori} is verified (that is in the Bayes optimal setting where all the problem parameters are known, see appendix~\ref{sec:NishimoriConditions}) and the factor graph associated to the problem is locally tree-like or dense such as in the present case, see Fig.~\ref{fig_factorSC}. See \cite{MezardMontanari09,DBLP:journals/corr/KabashimaKMSZ14} for more details on this ansatz and its justification. This ansatz states that all the replicas are statistically equivalent, and thus the overlaps are independent of the replica indices. It reads
\begin{align}
q_{ab}=q \ \forall \ a,b : a\neq b, \ Q_a = Q \ \forall \ a, \ m_a = m \ \forall \ a.
\end{align}
The covariance matrix $\bG$ of $\{v_\mu^a\}$ under this ansatz reads $\forall\ a,b$
\begin{align}
&G_{ab} \defeq  \mathbb{E}_{\bF,\bxi} \{v_\mu^a v_\mu^b\} = 1 - 2m + {1/{{\rm snr}}} + q + (Q - q)\delta_{a,b}, \\
\Rightarrow ~ &\bG = \left(1 - 2m + {1/{{\rm snr}}} + q\right)\boldsymbol{1}_{n,n} + (Q - q)\boldsymbol{I}_{n,n},
\end{align}
where $\boldsymbol{1}_{n,n}$ is a matrix full of ones of dimension $n\times n$. We thus obtain
\begin{align}
X_\mu &= \mathbb{E}_{\bF,\bxi}\{e^{-\frac{{{\rm snr}}}{2}\sum_{a}^n (v_\mu^a)^2}\} = \mathbb{E}_{\bv} \{e^{-\frac{{{\rm snr}}}{2} \bv^\intercal \bv}\}, \\
P(\bv) &= [(2\pi)^n {\rm det}(\bG)]^{-1/2} e^{-\frac{1}{2}\bv^\intercal \bG^{-1} \bv}, \\
\Rightarrow X_\mu &= [(2\pi)^n {\rm det}(\bG)]^{-1/2} \int d\bv e^{-\frac{1}{2}\bv^\intercal (\bG^{-1} + {{\rm snr}} \boldsymbol{I}_n) \bv} = {\rm det}\left(\boldsymbol{I}_{n,n} + {{\rm snr}} \, \bG \right)^{-1/2},
\end{align}
where the last equality is obtained by Gaussian integration. The eigenvectors of $\bG$ are one eigenvector $[1,1,\ldots,1]$ with associated eigenvalue $Q - q + n \left(1 - 2m + {1/{{\rm snr}}} + q \right)$ and $n-1$ eigenvectors of the type $[0,\ldots,0,-1,1,0,\ldots,0]$ with degenerated eigenvalue $Q-q$. Therefore
\begin{align}
{\rm det}\left(\boldsymbol{I}_{n,n} + {{\rm snr}}\, \bG \right) &= \big(1 + {{\rm snr}} \left[Q-q + n \left(1 - 2m + {1/{{\rm snr}}} + q \right) \right] \big)\left[1 + {{\rm snr}} (Q-q)\right]^{n-1}, \\
\Rightarrow \lim_{n\to0} X_\mu &= \exp\Big(-\frac{n}{2} \Big[ \frac{q - 2m + 1 + {1/{{\rm snr}}}}{Q - q + {1/{{\rm snr}}}}+ \ln({1/{{\rm snr}}} + Q - q) - \ln({1/{{\rm snr}}}) \Big] \Big). \label{eq:Xmu_simplified}
\end{align}
Now we know $X_\mu$, let us come back to \eqref{eq:replicatedZ_before}. We re-parametrize this expression thanks to the overlaps, so that we can then evaluate the integral by the saddle-point method over these macroscopic parameters. We need to enforce in \eqref{eq:replicatedZ_before} the overlaps to satisfy their definitions (\ref{eq_repOrderParam}). This is done by plugging the inverse Fourier transform of the Dirac delta function
\begin{align}
&1 = \int \Big[\prod_{a}^n dQ_ad\hat Q_a dm_a d\hat m_a\Big] \!\Big[\prod_{b,a<b}^{n(n-1)/2} dq_{ab}d\hat q_{ab}\Big] \nonumber\\
&\exp\Big[{-\sum_a^n \hat m_a(m_a L - \sum_{l}^L (\hat\bx_l^a)^\intercal \bx_l) + \sum_a^n \hat Q_a (Q_aL/2 - 1/2 \sum_l^L(\hat\bx_l^a)^\intercal \hat\bx_l^a) - \sum_{b,a<b}^{n(n-1)/2} \hat q_{ab}(q_{ab}L - \sum_l^L (\hat\bx_l^a)^\intercal \hat\bx_l^b)}\Big] \label{eq_1Fourier}.
\end{align}
The new conjugated parameters $\{\hat Q_a, \hat q_{ab}, \hat m_a\}$ are here to enforce the consistency conditions \eqref{eq_repOrderParam}. Plugging this expression and (\ref{eq:Xmu_simplified}) in (\ref{eq:replicatedZ_before}) leads
\begin{align}
&\mathbb{E}_{\bF,\bxi,\bx}\{Z^n\} = \int \Big[\prod_{a}^n dQ_a d\hat Q_a dm_a d\hat m_a \Big]\Big[\prod_{a,b<a}^{n(n-1)/2} dq_{ab}d\hat q_{ab}\Big] e^{L \left( {1\over2} \sum_a^n\hat Q_a Q_a - {1\over2} \sum_{a,b\neq a}^{n(n-1)}\hat q_{ab} q_{ab} - \sum_{a}^n \hat m_a m_a \right)} \nonumber \\
&\Big[\prod_{\mu}^M X_\mu \Big]\Big(\underbrace{ \int d\bx P_0(\bx) \Big[\prod_{a}^n d\hat\bx^a P_0(\hat\bx^a) \Big] e^{-\frac{1}{2} \sum_{a}^n \hat Q_a (\hat\bx^a)^\intercal \hat\bx^a + \frac{1}{2} \sum_{a,b\neq a}^{n(n-1)} \hat q_{ab} (\hat\bx^a)^\intercal \hat\bx^b  + \sum_{a}^n \hat m_a (\hat\bx^a)^\intercal \bx}}_{\defeq\Gamma}\Big)^L \left({{\rm snr}/2\pi} \right)^{\frac{Mn}{2}}.\label{eq_avZ}
\end{align}
We now need to evaluate this last expression. To do so, we define
\begin{align}
f(\bz) &\defeq \int d\hat\bx P_0(\hat\bx) \exp\Big(-\frac{1}{2} (\hat Q+\hat q) \hat\bx^\intercal\hat\bx + \hat m \hat\bx^\intercal \bx + \bz^\intercal \hat\bx \sqrt{\hat q} \Big), \\
\tilde\Gamma &\defeq \int d\bx \mathcal{D}\bz P_0( \bx) {f(\bz)}^n. \label{eq_Y}
\end{align}
Using the following transformation
\begin{align}
&e^{\frac{\hat q}{2}\sum_{a,b\neq a}^{n(n-1)} \hat\bx_a^\intercal \hat\bx_b} = \prod_{i}^B e^{\frac{\hat q}{2} \sum_{a,b\neq a}^{n(n-1)} \hat x_{a,i} \hat x_{b,i}} = \prod_i^B \int \mathcal{D}z_i \ \! e^{\sqrt{\hat q}\ \! z_i \sum_a^n \hat x_{a,i} - \frac{\hat q}{2} \sum_a^n {\hat x_{a,i}}^2}= \int \mathcal{D}\bz\ \! e^{\sqrt{\hat q}\ \!\bz^\intercal\sum_a^n \bx_a} e^{-\frac{\hat q}{2} \sum_a^n \hat\bx_a^\intercal \hat\bx_a},
\end{align}
we obtain $\tilde \Gamma = \Gamma$. In addition
\begin{align}
&\int \mathcal{D}\bz f(\bz)^n \underset{n\to 0}{\approx} \exp\Big(n \int \mathcal{D} \bz \ln f(\bz)\Big) \Rightarrow \tilde \Gamma \underset{n\to 0}{\approx} \exp\Big(n \int d \bx P_0(\bx) \int \mathcal{D}\bz \ln f(\bz)\Big).  \label{eq_fz}
\end{align}
Combining (\ref{eq_fz}) and (\ref{eq_avZ}), we reach the expression of the averaged replicated partition function under the replica symmetric ansatz (assumed for the conjugated variables as well)
\begin{align}
\mathbb{E}_{\bF,\bxi,\bx}\{Z^n\} &= \int dQ d\hat Q dm d\hat m dq d\hat q \, \exp\Big(nL \tilde \Phi_B(m,\hat m,q,\hat q, Q, \hat Q)\Big), \label{eq_saddle}\\
\tilde \Phi_B(m,\hat m,q,\hat q, Q, \hat Q) &= \frac{1}{2} \Big(\hat Q Q + \hat q q - 2 \hat m m \Big) - \frac{\alpha B}{2} \Big(\frac{q - 2m + \langle x^2 \rangle_L + 1/{{\rm snr}}}{Q - q + 1/{{\rm snr}}} + \ln(1/{{\rm snr}} + Q - q) \Big)\nonumber \\
&\ \ \ + \int d\bx P_0(\bx) \mathcal{D}\bz \ln\Big( \int d\hat\bx P_0(\hat\bx) \exp\Big(\hat m \bx^\intercal \hat\bx + \sqrt{\hat q} \bz^\intercal \bhx - \frac{1}{2}(\hat q + \hat Q)\hat\bx^\intercal\hat\bx\Big) \Big). \label{eq_freeEnt}
\end{align}
For the replica trick (\ref{eq_replicaTrick}) to be formally valid, the limit $n\to 0$ should be taken before $L\to\infty$. But we need to estimate the integral (\ref{eq_saddle}) by the saddle point method, which is justified only if the limit $L\to\infty$ is performed first. We thus assume that the limits commute, which is not rigorous, but heuristically verified in many models including in inference \cite{KrzakalaMezard12,DBLP:journals/corr/KabashimaKMSZ14,MezardMontanari09}. The saddle point estimate of (\ref{eq_saddle}) is performed by taking the extremum of the potential with respect to the free parameters 
\begin{equation}
\Phi_B\defeq {\rm extr}[\tilde \Phi_B(m,\hat m,q,\hat q, Q, \hat Q)] =\tilde\Phi_B(m^*,\hat m^*,q^*,\hat q^*, Q^*, \hat Q^*),
\end{equation}
where the extremum values are denoted with stars. Letting $n\to 0$ after the saddle point estimate of (\ref{eq_saddle}), the resulting expression corresponds to the replica free entropy as seen from (\ref{eq_replicaTrick}). The extremization gives
\begin{align}
&\frac{\partial\tilde \Phi_B}{\partial m}=0 \Rightarrow \hat m^* = \alpha B\frac{1}{Q^*-q^*+{1/{{\rm snr}}} },\\
&\frac{\partial\tilde \Phi_B}{\partial q}=0 \Rightarrow \hat q^* = \alpha B\frac{1/{{\rm snr}} + \langle x^2 \rangle_L - 2m^* + q^*}{(Q^*-q^*+{1/{{\rm snr}}})^2},\\
&\frac{\partial\tilde \Phi_B}{\partial Q}=0 \Rightarrow \hat Q^* = \alpha B\frac{2m^* - \langle x^2 \rangle_L - 2q^* + Q^*}{(Q^*-q^*+{1/{{\rm snr}}})^2}.
\end{align}
It is explained in appendix~\ref{sec:NishimoriConditions} that the Nishimori identity (\ref{eq:Nishimori}) together with the assumed concentration of the overlaps (\ref{eq_repOrderParam}) (a canonical assumption in statistical physics) imply that in the limit $L\to\infty$ the overlaps verify
\begin{align}
&q^* = m^*, \ Q^* = \langle x^2 \rangle_L = 1 \Rightarrow E = 1 - m^* \label{eq_nishi1},
\end{align}
where the last implication follows from \eqref{eq151}, $\Sigma(E)$ is given by \eqref{eqSIGMA2} and we have used (\ref{eq_alpha}) to get rid of $\alpha$. Note that this last equality has already been derived in appendix~\ref{sec:NishimoriConditions}, see \eqref{eq103}. These simplifications imply
\begin{align}
&\hat q^* = \hat m^*={\alpha B \over E + 1/{{\rm snr}}}={\ln(B) \over \Sigma(E)^2}, \ \hat Q^* = 0. \label{eq:extremumHatVars}
\end{align}
Thus due to the Nishimori identity, a single free parameter survives in the potential expression ($m^*$ or equivalently the MMSE $E$ due to \eqref{eq_nishi1}). Combining all, plugging the prior (\ref{eq_prior}) in (\ref{eq_freeEnt}) and simplifying the expression by integrating $\bx, \hat\bx$ we obtain the final expression of the potential, or replica symmetric free entropy (\ref{eq_freeEnt2}).
\subsection{The link between replica and state evolution analysies}
\label{subsec:repIsSE}
We now show that the extrema of the potential (\ref{eq_freeEnt2}) correspond to the stationary points of state evolution \eqref{eq_SE_MSE}, \eqref{eq_SE_var}. We restrict ourselves to constant power allocation but the derivation for generic power allocation is similar. We start from the potential (\ref{eq_freeEnt}). Thanks to (\ref{eq_nishi1}), (\ref{eq:extremumHatVars}) the potential only depends on a single variable $m^*$ (or $E$). The fixed point condition leading to $m^*$, when (\ref{eq:extremumHatVars}) is verified, is
\begin{align}
0=\frac{\partial\tilde \Phi_B}{\partial \hat m}\bigg|_{\hat m^*, \hat q^*, \hat Q^*} \Rightarrow m^*(E) &= \int d\bx \mathcal{D}\bz P_0(\bx) \int d\bhx P_0(\bhx) {1\over Z(\bx, \bz, E)} e^{{\ln(B) \over 2\Sigma(E)^2} \left(2\bhx^\intercal \left[\bx+\bz\Sigma(E)/ \sqrt{\ln(B)} \right] - 1\right)}\bx^\intercal\bhx, \\
Z(\bx, \bz, E) &\defeq \int d\bhx P_0(\bhx) e^{{\ln(B) \over 2\Sigma(E)^2} \left(2\bhx^\intercal \left[\bx+\bz\Sigma(E)/ \sqrt{\ln(B)} \right] - 1\right)}.
\end{align}
After integrating this expression with respect to $\bhx, \bx$ using (\ref{eq_prior}) and simple algebra, one obtains
\begin{align}
&m^*(E)=\int \mathcal{D}\bz \ \! f_{a_{1|1}}(\Sigma(E)^2,\bz ),
\end{align}
where $f_{a_{1|1}}$ is given by (\ref{eq_fa1fun}) and $\Sigma(E)$ by \eqref{eqSIGMA2}. Using the last equality of (\ref{eq_nishi1}), we see that the fixed point conditions of the replica potential give back the state evolution recursion (\ref{SE_MSE2}) at its stationary point (when the time index is dropped). We can thus assert that using the state evolution analysis to compute the typical (i.e averaged over the quenched disorder) mean-square error of the fixed points of AMP or extracting this information from the potential is equivalent. In addition, this strenghten further the claim that the replica analysis is exact for computing the potential of the problem, despite not rigorous\footnote{See \cite{barbier2016mutual,reeves2016replica} for recent rigorous results on the validity of the replica analysis in linear estimation.}.
\subsection{Alternative derivation of the large section limit of the potential via the replica method} \label{subsec:largeB_rep}
We now re-derive the results of sec.~\ref{subsec_largeBrep}, that the superposition codes are capacity achieving, using the replica method to compute \eqref{eq_logKB}. The computation is performed at fixed $\Sigma$ which plays again the role of a temperature. The replica method is appropriate because we have to average the logarithm of a partition function over some disorder $\bz$. Starting from \eqref{eq_logKB}, we can re-write this partition function $K_B$ (defined as what appears inside the logarithm in \eqref{eq_logKB}) as
\begin{align}
K_B(\bz) &\defeq \exp\Big(\frac{\ln(B)}{2\Sigma^2} + \frac{\sqrt{\ln(B)}z_1}{\Sigma} \Big) + \sum_{i = 2}^B \exp\Big(-\frac{\ln(B)}{2\Sigma^2} + \frac{\sqrt{\ln(B)}z_i}{\Sigma} \Big) \\
&= \sum_{i }^B \exp\Big(-\frac{1}{\Sigma}\Big(\frac{\ln(B)}{2\Sigma}(1 - 2\delta_{i,1}) - \sqrt{\ln(B)}z_i \Big)\Big) \eqdef \sum_{i}^B \exp\Big(-\frac{h_i(z_i)}{\Sigma}\Big). \label{eq_KB}
\end{align}
Meanwhile $Z$ given by (\ref{eq_fullZ}) is the (random) partition function of the overall signal, $K_B$ can be interpreted as the partition function of one single section of size $B$.
An important difference with the random energy model is that here there is a favored section state distinct from the other ones (noted state $1$), corresponding to the actual transmitted section in the original signal. It has been treated apart in sec.~\ref{subsec_largeBrep} but we keep it here in the "energy states" $\{h_i\}_i^B$. From the statistic of $z_i$ we get the one of $h_i$:
\begin{align}
z_i &\sim \mathcal{N}(z_i|0,1) \\
\Rightarrow h_i &\sim \mathcal{N}\Big(h_i\Big|\frac{(1 - 2\delta_{i,1})\ln(B)}{2\Sigma}, \ln(B)\Big).
\end{align}
The average of $K_B$ with respect to $\bz$ can thus be replaced by the average over $\bh$, the vector of independent energy states (independent because the $\{z_i\}_{i}^B$ are). We use again the replica trick for computing $I_B=\mathbb{E}_{\bh}\{\ln\(K_B(\bh)\)\}$ as $B$ diverges. We thus need the average replicated partition function as in the section appendix.~\ref{app:replicaAnalysis_A}:
\begin{align}
I&\defeq\lim_{B\to \infty}\mathbb{E}_{\bh}\{\ln K_B(\bh)\} \\
&= \lim_{B\to \infty}\lim_{n\to 0}\frac{\mathbb{E}_{\bh}\{K_B^n\}-1}{n}, \label{eq_repTrickK}\\
% \end{align}
% %
% \begin{align}
%
\mathbb{E}_{\bh}\{K_B^n\} &= \mathbb{E}_{\bh} \bigg\{\sum_{i_1,..,i_n }^{B,..,B} \exp\Big(-\frac{1}{\Sigma}(h_{i_1} + \ldots + h_{i_n})\Big)\bigg\} \\
&= \mathbb{E}_{\bh} \bigg\{\sum_{i_1,..,i_n }^{B,..,B} \prod_{j}^B \exp\Big(-\frac{h_j}{\Sigma} \sum_{a}^n\delta_{j,i_a}\Big)\bigg\} \\
% \end{align}
% %
% \begin{align}
&= \sum_{i_1,..,i_n }^{B,..,B} \prod_{j}^B \mathbb{E}_{h_j} \bigg\{\exp\Big(-\frac{h_j}{\Sigma} \sum_{a}^n\delta_{j,i_a}\Big)\bigg\}\\
&= \sum_{i_1,..,i_n }^{B,..,B} \exp\Big(\frac{\ln(B)}{2\Sigma^2}\sum_{j}^B\Big(\sum_{a,b}^{n,n}\delta_{j,i_a}\delta_{j,i_b} - (1 - 2\delta_{j,1})\sum_{a}^n\delta_{j,i_a} \Big)\Big)\\
&= \sum_{i_1,..,i_n }^{B,..,B}  \exp\Big(\frac{\ln(B)}{2\Sigma^2}\Big(\sum_{a,b}^{n,n}\delta_{i_a,i_b} - \sum_{j}^B\sum_{a}^n\delta_{j,i_a} (1 - 2 \delta_{j,1}) \Big)\Big) \\
&= \sum_{i_1,..,i_n }^{B,..,B} \exp\Big(\frac{\ln(B)}{2\Sigma^2}\Big(\sum_{a,b}^{n,n}\delta_{i_a,i_b} - n + 2\sum_{a}^n\delta_{1,i_a} \Big)\Big). \label{eq_KB3}
\end{align}
We now define new macroscopic order parameters for re-parametrizing the replicated partition function: 
\begin{align}
q_{ab} &\defeq \delta_{i_a,i_b} \ \forall\ (a,b),\label{eq_repOrderParamBbig_q}\\ 
m_a &\defeq \delta_{i_a,1} \ \forall\ a .\label{eq_repOrderParamBbig}
\end{align}
The first one indicates if two replicas are in the same state or not, the second one if a given replica is in the favored state $1$. We now replace the sum over the single replica states by sums over all the authorized order parameters combinaisons which become the new free variables; the sums are restricted over the subspace matching the order parameters definitions (\ref{eq_repOrderParamBbig_q}), (\ref{eq_repOrderParamBbig}). In the appendix.~\ref{app:replicaAnalysis_A}, this condition was enforced by the introduction of Dirac delta functions in the integral through (\ref{eq_1Fourier}), here it is simpler because we are in a discrete case. We deduce from (\ref{eq_KB3})
\begin{equation}
\mathbb{E}_{\bh} \{K_B^n\} = \sum_{\bq,\bm} \exp\Big(\frac{\ln(B)}{2\Sigma^2} \Big(\sum_{a,b}^{n,n} q_{ab} + 2\sum_{a}^n m_a - n + 2\Sigma^2 s_{\bq,\bm}\Big)\Big),
\end{equation}
where we have introduced the entropy associated to these new order parameters: $s_{\bq,\bm} \defeq S_{\bq,\bm}/\ln(B)$ where $S_{\bq,\bm}$ is the logarithm of the number of states of the replicas (the number of terms in the sum \eqref{eq_KB3}) compatible with $\bq$ and $\bm$ at the same time, where $\bq \defeq \[q_{ab}\]_{a,b}^{n,n}$ and $\bm \defeq \[m_a\]_a^n$. We use the replica symmetric ansatz, where each replica is considered equivalent. This reads
\begin{align}
q_{ab} &= q + (1-q)\delta_{a,b} \ \forall\ (a,b),\\ 
m_a &= m \ \forall \ a.
\end{align}
It allows to simplify the average replicated partition function as
\begin{align}
\mathbb{E}_{\bh} \{K^n_B\}&= \sum_{q,m} \exp\Big(n\ln(B) \underbrace{\Big[\frac{(n-1)q + 2m + \frac{2\Sigma^2}{n} s_{q,m}}{2\Sigma^2}\Big]}_{\defeq \tilde I(q,m)}\Big) \\
&\eqdef \sum_{q,m} \exp\Big(n\ln(B)\tilde I(q,m)\Big). \label{eq_saddleK}
\end{align}
Looking at (\ref{eq_repOrderParamBbig_q}), (\ref{eq_repOrderParamBbig}), there are a priori four different possible ansatz, corresponding to four different states of the section: $(q=m=0), (q=m=1), (q=0, m=1)$ and $(q=1, m=0)$ but actually, only three possibilities remain as the state $(q=0, m=1)$ has no meaning: the replicas cannot be all in different states ($q=0$) and all in the favored one ($m=1$) at the same time. Thus it remains:\\
$~~~\bullet ~(q=m=0)$ : all the replicas are in different states but none of them are in the favored one 1. \label{eq_mq0}\\
$~~~\bullet ~(q=m=1)$ : all the replicas are in the favored state 1.\label{eq_mq1}\\
$~~~\bullet ~(q=1, m=0)$ : all the replicas are in the same state, which is not the favored one. \label{eq_thirdAnsatz}\\
The last ansatz can be forgotten as the computation shows that it always leads to lower free entropy than the two other ones. This is understandable as there should be a symmetry among all the ``wrong'' states (different from $1$) as none of them is special with respect to the other ones, so the replicated system should not choose a particular one spontaneously. It leaves two ansatz. The last sum $\sum_{q,m}$ is performed by the saddle point method as $B \to \infty$, assuming the commutativity of the limits in (\ref{eq_repTrickK}). From (\ref{eq_repTrickK}), the ``section potential'' is thus given by the maximizer of the sum among the two possible ansatz:
\begin{equation}
I/\ln(B)=\max_{(q^*,m^*)}\, \tilde I(q^*,m^*). \label{eq_maxTildeI}
\end{equation}
Let's compute the value of $\tilde I$ for the two remaining ansatz as $n\to 0$ in order to find the maximum:
\begin{align}
(q^*=m^*=0) &\Rightarrow s_{0,0} = \ln ((B-1)^n)/\ln(B) \approx n \\
&\Rightarrow \tilde I(E|q^*=m^*=0) \approx 1,\\
(q^*=m^*=1) &\Rightarrow s_{1,1} = \ln (1)/\ln(B) = 0 \\
&\Rightarrow \tilde I(E|q^*=m^*=1)= (2\Sigma(E)^2)^{-1},
\end{align}
where $\Sigma(E)^2$ is given by \eqref{eqSIGMA2} and the approximate equalities are up to vanishing terms with $B$. Thus \eqref{eq_maxTildeI} is the same as the second term of the right hand side of \eqref{Phi_largeB}; the result is consistent with appendix.~\ref{app:replicaAnalysis_A} and leads to the same potential $\phi(E)$.
\section*{Acknowledgments}
The research leading to these results has received funding from the
European Research Council under the European Union's $7^{th}$
Framework Programme (FP/2007-2013/ERC Grant Agreement 307087-SPARCS)
and from the French Ministry of defense/DGA. Part of this work was
revised during a visit to the Simons Institute for the Theory of
Computing, University of California, Berkeley. We also want to thank
Rüdiger Urbanke for useful discussions.  \ifCLASSOPTIONcaptionsoff
\fi
\bibliographystyle{IEEEtran}
\bibliography{IEEEabrv,refs}
\end{document}